\documentclass[preprint]{elsarticle}

\makeatletter
	\def\ps@pprintTitle{%
 	\let\@oddhead\@empty
	\let\@evenhead\@empty
	\def\@oddfoot{\centerline{\thepage}}%
	\let\@evenfoot\@oddfoot}
\makeatother

\usepackage{etoolbox}
\patchcmd{\MaketitleBox}{\footnotesize\itshape\elsaddress\par\vskip36pt}{\footnotesize\itshape\elsaddress\par\parbox[b][36pt]{\linewidth}{\vfill\hfill\textnormal{\today}\hfill\null\vfill}}{}{}%
\patchcmd{\pprintMaketitle}{\footnotesize\itshape\elsaddress\par\vskip36pt}{\footnotesize\itshape\elsaddress\par\parbox[b][36pt]{\linewidth}{\vfill\hfill\textnormal{\today}\hfill\null\vfill}}{}{}%

\usepackage[
            colorlinks=true,%
            breaklinks=true,%
            linkcolor=blue,
            urlcolor=blue,%
            citecolor=blue,%
            pdftitle={Bayesian Parameter Estimation for Dynamical Models in Systems Biology}, 
            pdfkeywords={Parameter Estimation, Systems Biology, Uncertainty Quantification, Bayesian Methods},
            pdfauthor={Nathaniel J. Linden, Boris Kramer, Padmini Rangamani},
            bookmarksopen=false,
            pdfpagemode=UseNone]{hyperref}

\usepackage[margin = 1.2in]{geometry}
\usepackage[T1]{fontenc}
\usepackage[english]{babel}
\usepackage[latin1]{inputenc}
\usepackage{mathtools}
\usepackage{amsmath}
\usepackage{amsfonts}
\usepackage{amsthm}
\usepackage{amssymb}
\usepackage{array}
\usepackage{color}
\usepackage{url}
\usepackage{cleveref}
\usepackage{graphicx}
\usepackage{ragged2e}
\usepackage{caption}
\usepackage{threeparttable}
\usepackage{breakcites}
\usepackage{soul}
\usepackage{enumitem}
\usepackage[title,titletoc,toc]{appendix}
\usepackage{tabularx}
\usepackage{natbib}
\newcommand{\subpanelref}[2]{\hyperref[#1]{\ref*{#1}.#2}}

\usepackage{lineno}

\usepackage{framed}

\newtheorem{theorem}{Theorem}

\newtheorem{remark}{Remark}
\newtheorem{problem}{Problem}
\newtheorem{definition}{Definition}
\newcommand {\real} {\mathbb{R}}
\newcommand {\nat} {\mathbb{N}}

\DeclareMathOperator*{\argmin}{arg\,min}%

\newcommand{\bA}{\ensuremath{\mathbf{A}}}

\newcommand{\bC}{\ensuremath{\mathbf{C}}}

\newcommand{\bH}{\ensuremath{\mathbf{H}}}
\newcommand{\bI}{\ensuremath{\mathbf{I}}}
\newcommand{\bK}{\ensuremath{\mathbf{K}}}
\newcommand{\bL}{\ensuremath{\mathbf{L}}}
\newcommand{\bP}{\ensuremath{\mathbf{P}}}
\newcommand{\bS}{\ensuremath{\mathbf{S}}}

\newcommand{\ba}{\ensuremath{\mathbf{a}}}
\newcommand{\bx}{\ensuremath{\mathbf{x}}}
\newcommand{\by}{\ensuremath{\mathbf{y}}}

\newcommand {\boldeta} {\mbox{\boldmath $\eta$}}
\newcommand {\bGamma} {\boldsymbol{\Gamma}}

\newcommand {\bxi} {\mbox{\boldmath $\xi$}}
\newcommand {\bXi} {\mbox{\boldmath $\Xi$}}
\newcommand {\btheta} {\mbox{\boldmath $\theta$}}
\newcommand{\bSigma}{\ensuremath{\boldsymbol{\Sigma}}}

\newcommand{\cL}{\ensuremath{\mathcal{L}}}
\newcommand{\cN}{\ensuremath{\mathcal{N}}}
\newcommand{\cU}{\ensuremath{\mathcal{U}}}
\newcommand{\cX}{\ensuremath{\mathcal{X}}}
\newcommand{\cY}{\ensuremath{\mathcal{Y}}}

\newcommand{\hbx}{\ensuremath{\mathbf{\hat x}}}
\newcommand{\bbE}{\ensuremath{\mathbb{E}}}

\newcommand{\ben}{\begin{enumerate}}
\newcommand{\een}{\end{enumerate}}

\begin{document}
	\begin{frontmatter}

		\title{Bayesian Parameter Estimation for Dynamical Models in Systems Biology}

		\author{Nathaniel J. Linden}

		\author{Boris Kramer\corref{cor1}}
		\ead{bmkramer@ucsd.edu; +1 858-246-5327}

		\author{Padmini Rangamani\corref{cor1}}
		\ead{prangamani@ucsd.edu}

		\cortext[cor1]{Corresponding authors}

		\address{Department of Mechanical and Aerospace Engineering, University of California San Diego, CA, United States}

		\begin{abstract}
		
		Dynamical systems modeling, particularly via systems of ordinary differential equations, has been used to effectively capture the temporal behavior of different biochemical components in signal transduction networks.
		Despite the recent advances in experimental measurements, including sensor development and `-omics' studies that have helped populate protein-protein interaction networks in great detail, modeling in systems biology lacks systematic methods to estimate kinetic parameters and quantify associated uncertainties.
		This is because of multiple reasons, including sparse and noisy experimental measurements, lack of detailed molecular mechanisms underlying the reactions, and missing biochemical interactions.
		Additionally, the inherent nonlinearities with respect to the states and parameters associated with the system of differential equations further compound the challenges of parameter estimation.
		In this study, we propose a comprehensive framework for Bayesian parameter estimation and complete quantification of the effects of uncertainties in the data and models.
		We apply these methods to a series of signaling models of increasing mathematical complexity. 
		Systematic analysis of these dynamical systems showed that parameter estimation depends on data sparsity, noise level, and model structure, including the existence of multiple steady states. 
	    These results highlight how focused uncertainty quantification can enrich systems biology modeling and enable additional quantitative analyses for parameter estimation.
		
		\end{abstract}

\end{frontmatter}
\thispagestyle{empty} 


\section*{Author Summary}
Mathematical models of biological signal transduction networks have been widely used to capture the temporal behavior of such systems.
Calibrating these models to increasingly available experimental data is essential to ensure that models accurately portray biological phenomena.
However, measurement noise, the inability to measure all biochemical species in a system, and the lack of detailed knowledge about all reactions make model calibration difficult and can introduce errors.
In this study, we propose a principled and complete computational framework to enable model calibration in the face of these challenges.
Therein, we quantify any uncertainties (potential errors) in the calibrated model.
We apply the framework to a series of example models demonstrating various dynamic regimes common in biology (limit cycles, steady states) to highlight how our method provided additional context and insights to modeling-based studies.


\section{Introduction} \label{sec:intro}
Mathematical modeling is an integral part of systems biology; indeed, the use of approaches from dynamical systems analyses resulted in a paradigmatic shift in our understanding of biochemical signal transduction and enabled the identification of the emergent properties of a signaling network~\cite{bhalla2001robustness, eungdamrong2004computational, lipshtat2010design, ma2005toward}.
Additionally, mathematical models allow us to investigate the dynamics of biological systems beyond what is experimentally possible~\cite{Song2021-qx, Chakraborty2017-oz, haney2010ultrasensitive, qiao2019network}. 
A classical approach to modeling the dynamics of signal transduction is the use of systems of ordinary differential equations (ODEs)~\cite{wolkenhauer2008modelling, zi2012tutorial}.
Often these equations include nonlinear functions to capture complex biochemical interactions using Michaelis-Menten kinetics and Hill functions for cooperative binding~\cite{keener2009mathematical}. 
One of the ongoing challenges in developing and constraining predictive models of signal transduction has been the estimation and identification of the kinetic parameters associated with these reactions and quantifying the associated uncertainty~\cite{Mitra2019-ky, Babtie2017-np, Yazdani2020systems}.
The use of rigorous, quantitative approaches to estimate kinetic parameters and their uncertainties is in its early stages in systems biology~\cite{Geris2016uncertainty, Mitra2019-ky, Raue2013lesons, Ashyraliyev2009-yh, Valderrama-Bahamondez2019mcmc} even though such methods are far more prevalent in the greater computational science community under the field of uncertainty quantification~(UQ)~\cite{smith2013uncertainty,oden2017predictive, stuart2010inverse, sullivan2015introduction}. 

There are many sources of uncertainty in dynamical systems modeling of signal transduction, including the model structure itself, the values of model parameters, and the quality of the data used for model calibration.
Uncertainty in the model equations, known as model form or topological uncertainty~\cite{kennedy2001bayesian,Babtie2014-nc, morrison2018representing, Galioto2020-gq} often arises during model development.
However, the reaction fluxes for many biochemical reactions (ODE formulations) can be established in terms of classical rate equations~\cite{wolkenhauer2008modelling, zi2012tutorial, keener2009mathematical}.
The more significant challenge is establishing suitable model parameters, including the kinetic rate constants, for various flux terms~\cite{Raue2013lesons}.
Direct measurement of these parameters often occurs in isolated reaction systems and does not capture the complexity of the entire network of reactions represented by the models. 
Estimating these biological parameters and identifying any remaining uncertainties requires selecting a statistical model and then learning the distribution of these parameters from available data~\cite{smith2013uncertainty, gelman2013bayesian}.
From this viewpoint, the biological parameters are random variables that either have parametric or nonparametric distributions.
However, parameter estimation is complicated by the noisy, sparse (few time points), and incomplete nature of data found in systems biology (few or select readouts due to experimental limitations) that introduce uncertainties in the biological parameters~\cite{Gutenkunst2007sloppy, Wieland2021structural, Yazdani2020systems, Lillacci2010estimation, schmiester2021petab}.
In the face of these complicating factors, there is a need for statistical modeling of parameters that enables uncertainty quantification.

A comprehensive parameter estimation and UQ framework should consider the impact of structural parameter identifiability and parameter sensitivity~\cite{Raue2013lesons, Wieland2021structural, Kent2013learn, Anstett-Collin2020-qn, Erguler2011-es}. 
Structural parameter identifiability analysis reveals which of the parameters can be estimated given a specific dynamical systems model and a set of measurable outputs~\cite{Gutenkunst2007sloppy, Wieland2021structural, Raue2009identifiability, Hong2020global}.
A parameter is globally structurally identifiable if there is only one unique model output for each value of that parameter~\cite{Wieland2021structural}.
Parameters that do not meet this criterion are deemed structurally nonidentifiable and cannot successfully be estimated from the specified model outputs.
Structural nonidentifiabilities can arise due to complex nonlinear equations and incomplete experimental data that only measures a subset of the system's states.
Additionally, parametric sensitivity analysis~\cite{smith2013uncertainty, Saltelli2007-ta, Varma1999parametric} quantifies how sensitive a model output is to variations in the model parameters.
Gutenkunst et al.~\cite{Gutenkunst2007sloppy} found that most models in systems biology contain parameters with a wide range of sensitivities, which they termed `sloppy'. 
Despite this challenge, sensitivity analysis helps rank the set of identifiable parameters by their contributions to specified model outputs,~\cite{Marino2008uncertsens} as was done in Mortlock \textit{et al}.~\cite{Mortlock2021-lk} for the prolactin-mediated JAK-STAT signaling pathway.
This analysis enabled them to select a subset of 33 out of 60 total parameters that significantly contribute to variations in the model outputs.

Commonly used methods to estimate parameters for systems biology models include frequentist~\cite{Ashyraliyev2009-yh} and Bayesian approaches~\cite{Valderrama-Bahamondez2019mcmc}.
In the frequentist setting, parameter estimation is formulated as an optimization problem, and the solution to the parameter estimation problem is the set of parameters that best recapitulates the data~\cite{smith2013uncertainty}.
Additionally, frequentist approaches quantify uncertainty via estimated confidence intervals around the optimal parameters~\cite{kay1993fundamentals, moon2000mathematical, smith2013uncertainty}.
In contrast, in the Bayesian perspective, parameters are assumed to be random variables whose unknown probability distributions, called \textit{posterior distributions}, quantify the probability of assuming any value in the parameter space~\cite{smith2013uncertainty, stuart2010inverse, gelman2013bayesian}.
The advantage of Bayesian approaches comes from their ability to characterize the entire posterior distribution and quantify the uncertainty in parameter estimates via credible intervals~\cite{gelman2013bayesian}.
Many methods have been developed for Bayesian parameter estimation~\cite{Valderrama-Bahamondez2019mcmc, Liepe2014abc, Toni2009approx, Golightly2011pmcmc, Ghasemi2011estimation, Bianconi2020application, Wilkinson2007-rt, Klinke2009-cc} that all aim to characterize the posterior distribution, by leveraging Bayes' rule~\cite{smith2013uncertainty, gelman2013bayesian}.
For example, Mortlock et al.~\cite{Mortlock2021-lk} successfully used Bayesian estimation to study the uncertainty in the model predictions and assess the statistical significance of their modeling results.

Despite the successes of Bayesian parameter estimation in systems biology~\cite{Mitra2019-ky, Geris2016uncertainty, Raue2013lesons, Valderrama-Bahamondez2019mcmc}, failure to account for all sources of uncertainty in a model can significantly inhibit parameter estimation and uncertainty quantification~\cite{Raue2013lesons, Galioto2020-gq, Gutenkunst2007sloppy}.
Thus, a comprehensive framework for UQ in systems biology should include rigorous accounting of uncertainties in the model structure, nonidentifiable parameters, mixed parameter sensitivities, and noisy, sparse, or incomplete experimental data.
While identifiability and sensitivity analyses are typically performed prior to parameter estimation~\cite{Raue2013lesons, Wieland2021structural}, accounting for model form uncertainty requires us to consider a stochastic model instead of a deterministic one~\cite{kennedy2001bayesian, Galioto2020-gq, Renardy2018-km, Lillacci2010estimation}.
One promising approach to account for model form uncertainty is the Unscented Kalman filter Markov chain Monte Carlo (UKF-MCMC) method~\cite{Galioto2020-gq, Erazo2017-ls}. 
This method includes statistical models for noisy data and model form uncertainty simultaneously; however, it has not been adapted for dealing with the unique challenges in system biology.
Similarly, the parameter estimation and model selection method in~\cite{Lillacci2010estimation} accounts for model form uncertainty with extended Kalman filtering but does not provide complete uncertainty estimates because it takes a frequentist approach for parameter estimation.
Thus, there is a need for a framework that combines structural identifiability analysis, global sensitivity analysis, a statistical model for data and model form uncertainty, and Bayesian parameter estimation for comprehensive UQ of dynamical models in systems biology.

The novelty of this work lies in the comprehensive workflow we have developed to advance the current state of the art for uncertainty quantification in systems biology.
Additionally, the proposed framework accounts for uncertainty in the data and model structure by building on the Bayesian framework and the UKF-MCMC method~\cite{Galioto2020-gq, Erazo2017-ls}.
Our proposed workflow begins with structural identifiability analysis~\cite{Hong2020global, Hong2019sian} and global sensitivity analysis~\cite{Saltelli2007-ta, Sobol2001gsa} to eliminate parameters that cannot be learned from the measurements or that are not affecting the model outputs and then extends UKF-MCMC for systems biology by leveraging the constrained interval unscented Kalman filter (CIUKF)~\cite{Teixeira2008-vy}.
Taken together, each of these steps quantitatively addresses uncertainties encountered during model development and calibration to improve predictive modeling in systems biology.

The remainder of this paper details our comprehensive workflow for uncertainty quantification in systems biology and presents several examples to highlight this analysis.
First, in Section~\ref{sec:methods} we introduce the proposed framework and provide the mathematical details needed to understand and apply the approach.
Next, in Section~\ref{sec:results} we apply this framework to three systems biology models of increasing complexity, including a simple two-state model~\cite{Villaverde2019-zc}, a model of the core mitogen-activated protein kinase (MAPK) signaling pathway~\cite{Nguyen2015dyvipac}, and a phenomenological model of synaptic plasticity to capture long-term potentiation/depression~\cite{Pi2008coupled}.
We found that even in simple models, estimation of parameters depends on the level of data noise and data sparsity.
Finally, the framework enables uncertainty quantification for model structures that include non-linearities and multistability.
In all of these cases, we leverage identifiability and sensitivity analyses to narrow the subset of parameters for estimation and then use Bayesian estimation to determine the role of model structure in parameter estimation.
These results establish an uncertainty quantification-focused approach to systems biology that can enable rigorous parameter estimation and analysis.
Lastly, in Section~\ref{sec:discussion} we discuss these findings in the context of the three examples, address challenges in applying Bayesian methods, and provide directions for future UQ in systems biology.


\section{Methods} \label{sec:methods}

\begin{figure}[ht!]
    \centering
    \includegraphics[width=\textwidth]{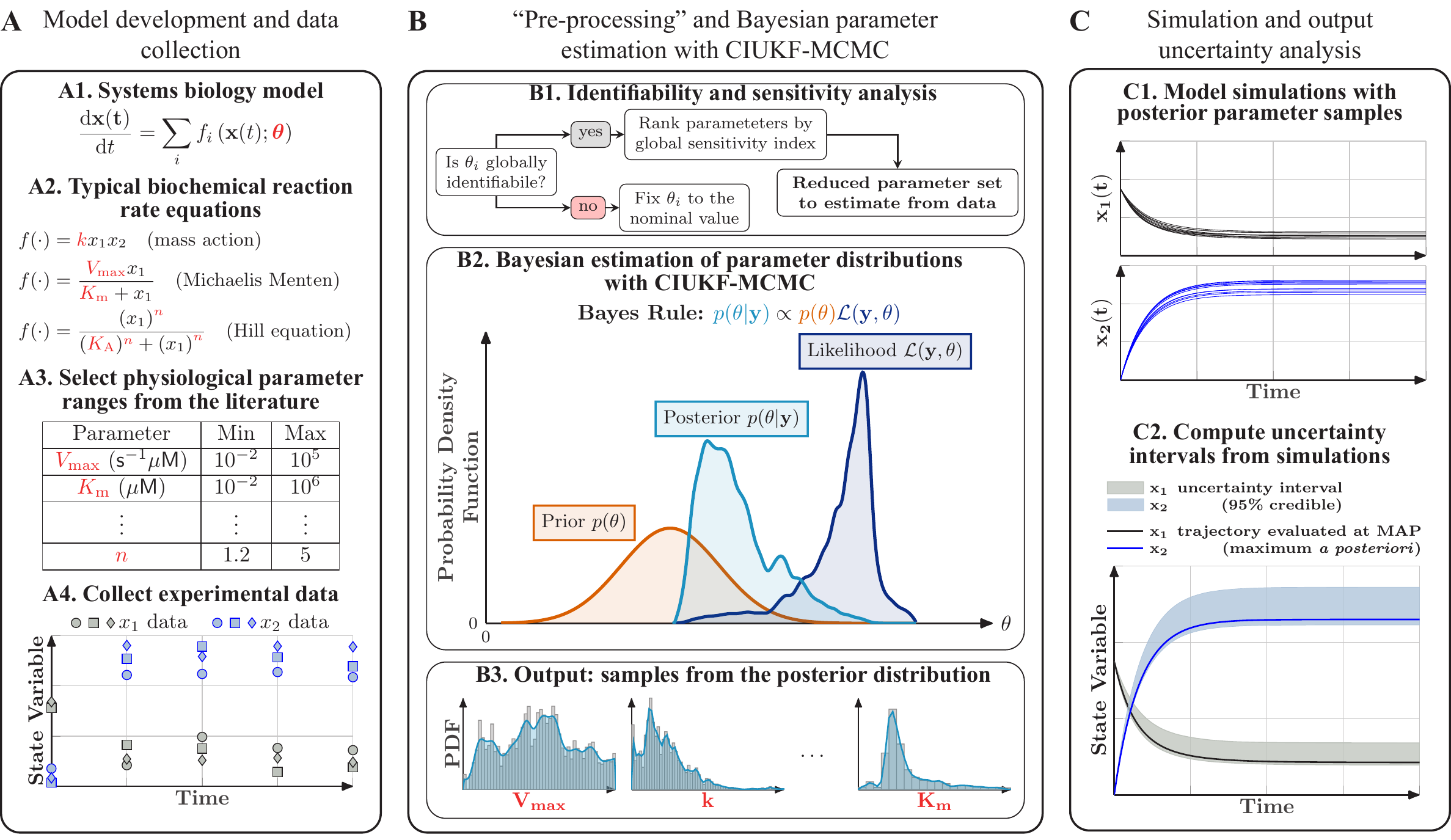}
    \caption{A comprehensive Bayesian parameter estimation and uncertainty quantification framework for dynamical models in systems biology.
    (\textbf{A}) 
    Model development in systems biology begins with model construction and data collection. 
    Dynamical models in systems biology typically involve a system of ODEs that capture the dynamics of the concentrations of different chemical species in the system (A1). 
    The reaction rates associated with these concentration changes are usually mass action, Michaelis Menten kinetics, or cooperative kinetics represented by the Hill equation (A2). 
    The free parameters in these models include kinetic rate constants, e.g. $k$, $V_{\text{max}}$, equilibrium constants, e.g. $K_m$, $K_A$, and Hill coefficients, e.g. $n$.
    These parameters are first constrained by \textit{best guess} values based on physiological ranges and typical values of model parameters from the literature (A3).
    Finally, the model needs experimental data for validation; this data can either be from published work or new experiments. 
    (\textbf{B}) 
    Parameter preprocessing and Bayesian parameter estimation with the CIUKF-MCMC algorithm.
    First structural identifiability and global sensitivity analyses on the entire parameter set reduce the set of free parameters that can be estimated (B1).
    Next, we perform Bayesian parameter estimation for this reduced set of parameters to learn their posterior distributions.
    The posterior distribution is the parameter distribution conditioned on the data (B2).
    Bayes' rule relates the posterior distribution to the product of the prior distribution and the likelihood function.
    The prior distribution encodes known information about the parameters and the likelihood function (which requires simulating the model) measures the misfit between predictions and the data.
    A state-constrained Unscented Kalman filter approximates the likelihood function to account for uncertainty in the model equations.
    Although Bayes' rule provides a means to evaluate the posterior distribution at specific points in the parameter space, we use Markov chain Monte Carlo (MCMC) sampling (B3) to characterize the entire distribution.
    (\textbf{C}) 
    The posterior distributions enable uncertainty analysis of model outputs through ensemble simulation.
    We perform simulations using the posterior parameter samples to propagate parameter uncertainty through the model (C1).
    Statistical analysis of the ensemble enables us to compute uncertainty intervals and study various system behaviors, for example, the statistics of the steady state values (C2).
    }
    \label{fig:uqOverview}
\end{figure}

This section describes the technical details of the comprehensive framework for uncertainty quantification (see Fig~\ref{fig:uqOverview}) proposed in Section~\ref{methods:framework}.
Next, Section~\ref{methods:partial-obs} introduces dynamical systems biology models and the parameter estimation problem.
Sections~\ref{methods:strucID}~and~\ref{methods:gsa} respectively overview structural identifiability and global sensitivity analyses to reduce the dimension of the parameter space.
Then Section~\ref{methods:bayes} introduces Bayesian estimation, Section~\ref{methods:CIUKF-MCMC} outlines the CIUKF-MCMC algorithm, and Section~\ref{methods:CIUKF} describes the constrained interval unscented Kalman filter in detail.
Following this, Section~\ref{methods:prior} discusses how to construct prior distributions and Section~\ref{methods:MCMC} details Markov chain Monte Carlo sampling.
Lastly, Section~\ref{methods:ensemble-model} discusses output uncertainty propagation with ensemble modeling, Section~\ref{methods:point-est} highlights choosing point estimators for the parameters, Section~\ref{method:synthetic-data} delineates synthetic data generation and Section~\ref{methods:limit-cycle} outlines limit cycle analysis.


\subsection{Framework for comprehensive uncertainty quantification for dynamical models in systems biology} \label{methods:framework}
This section previews the proposed comprehensive framework for parameter estimation and uncertainty quantification of dynamical models in systems biology.
Figure~\ref{fig:uqOverview} outlines the framework and its components, which are then described in much more detail in the subsequent sections. 
The proposed framework follows three main steps.
First, we assume that dynamical models of intracellular signal transduction (Fig~\subpanelref{fig:uqOverview}{A1}) use classical biochemical rate laws, such as mass action kinetics, Michaelis-Menten kinetics, and Hill functions~\cite{wolkenhauer2008modelling, zi2012tutorial, keener2009mathematical} (Fig~\subpanelref{fig:uqOverview}{A2}).
The key challenge to applying these models is estimating the associated parameters, such as the rate constants $k$ and $V_{\text{max}}$, equilibrium coefficients $K_m$ and $K_A$, and Hill coefficients $n$ in Fig~\subpanelref{fig:uqOverview}{A2}, from available experimental data (Fig~\subpanelref{fig:uqOverview}{A4}).
The comprehensive framework uses Bayesian inference to estimate a statistical model (a probability distribution; see Fig~\subpanelref{fig:uqOverview}{B}) for the model parameters given a set of noisy measurement data and a specific model form.

We argue that identifiability and sensitivity analysis are necessary steps to perform before parameter estimation (Fig~\subpanelref{fig:uqOverview}{B1}).
To eliminate uncertainty due to nonidentifiable parameters, we perform global structural identifiability analysis using the Structural Identifiability Analyzer (SIAN)~\cite{Hong2020global, Hong2019sian} (see Section~\ref{methods:strucID} for details).
The nonidentifiable parameters are fixed to their nominal values from the literature or based on their physiological ranges.
Next, variance-based global sensitivity analysis~\cite{smith2013uncertainty, Saltelli2007-ta} is performed to rank the identifiable parameters in order of their contributions to the variance of the model outputs (see section~\ref{methods:gsa} for details).
A subset of the identifiable parameters with the largest sensitivity indices is selected for parameter estimation. 
The remaining model parameters are then fixed to their nominal values in the same fashion as nonidentifiable parameters.

Bayesian parameter estimation completely characterizes uncertainty in the model parameters by estimating a nonparametric statistical model.
Bayes' rule (see Fig~\subpanelref{fig:uqOverview}{B2}) provides the \textit{best guess} distribution, called the \textit{posterior distribution}, for the parameters starting from an initial guess (the prior distribution) that is transformed by the available experimental data with the likelihood function.
The likelihood function measures the mismatch between the data and the model predictions and returns higher probabilities for parameter sets that produce outputs \textit{closer} to the data.
We use the CIUKF-MCMC algorithm~\cite{Galioto2020-gq, Erazo2017-ls} to approximate the likelihood function and account for uncertainty in the model formulation, data, and parameters.
Markov chain Monte Carlo sampling, with either delayed rejection adaptive Metropolis~\cite{Haario2006dram} or the affine invariant ensemble sampler~\cite{Goodman2010aies}, generates a set of samples that represents the posterior distribution (Fig~\subpanelref{fig:uqOverview}{B3}).

Lastly, we leverage the posterior distribution to quantify how uncertainty in the model parameters affects uncertainty in the model predictions (Fig~\subpanelref{fig:uqOverview}{C}).
An ensemble simulation with the parameter samples generates sets of trajectories (see Fig~\subpanelref{fig:uqOverview}{C1}) that capture the uncertainty in the predicted dynamics.
Computing uncertainty intervals such as the $95 \%$ credible intervals presented in Fig~\subpanelref{fig:uqOverview}{C2} provides a visualization of this uncertainty.
Notably, credible intervals are different from confidence intervals because credible intervals capture a specified percentage of the samples whereas confidence intervals are random variables that capture regions where estimators will lie at a specified probability level~\cite{gelman2013bayesian} (see Fig~\ref{fig:point-estimate} for an example of a credible interval).
Additionally, statistical analysis of the ensemble enables quantitative analysis of computational modeling results in the same way that running multiple experimental trials enables analysis of experimental results.
Before discussing the methods that enable comprehensive uncertainty quantification in the following sections, the next section formally introduces parameter estimation for dynamical models in systems biology.

\subsection{Parameter estimation for systems biology models in the form of partially-observed systems of ordinary differential equations} \label{methods:partial-obs}
We consider nonlinear ordinary differential equation models of the form
\begin{linenomath*}
    \begin{alignat}{2}
      \frac{{\rm d}\bx(t)}{{\rm d}t} &= \sum_i f_i\left(\bx(t); \ \btheta_f\right) \quad \label{eq:state-eq} \\
        \by(t) &= h\left(\bx(t); \ \btheta_h \right) + \boldeta(t), \quad && \boldeta(t) \sim \cN\left(0, \bGamma(\btheta_\Gamma)\right) \label{eq:obs-eq},
    \end{alignat}
\end{linenomath*}
where $\bx \in \real^d_{\geq 0}$ is the state vector of nonnegative species concentrations and $\by \in \real^m$ is the vector of potentially incomplete,  $m \leq d$, measurements of $\bx$.
The functions $f_i(\cdot; \cdot): \real^d \times \real^p \to \real^d$ govern the rates of the involved biochemical reactions and are derived using biochemical theory (see Fig~\subpanelref{fig:uqOverview}{A2} for example terms).
Further, $\btheta_f \in \real^{p_f}$ is the vector of biological model parameters, including but not limited to rate constants, binding coefficients, and equilibrium coefficients.
The function $h(\cdot; \cdot): \real^d \times \real^p \to \real^m$ is the measurement function that maps from the states to the set of observables (experimental data), where $\btheta_h$ is the vector of associated parameters.
Lastly, the measurements $\by(t)$ are corrupted by independently and identically distributed (iid) Gaussian measurement noise $\boldeta(t) \in \real^m$ with zero mean and covariance matrix $\bGamma \in \real^{m \times m}$.
The covariance matrix is parameterized by $\btheta_\Gamma \in \real^m$ such that $\bGamma \left(\btheta_\Gamma \right) = \text{diag}\left(\btheta_\Gamma \right)$.
The parameter space is then defined as the multidimensional space of all possible values of $\btheta =[\btheta_f$, $\btheta_\Gamma$], e.g., $\btheta \in \real^{p_f + m}$. 
    
In this work, we make several simplifying assumptions to the model in Eqs~\eqref{eq:state-eq}--\eqref{eq:obs-eq}.
First, we assume that the measurement function $h(\cdot; \cdot)$ is linear and that all the parameters in $\btheta_h$ are known, so Eq~\eqref{eq:obs-eq} becomes
\begin{linenomath*}
\begin{equation} \label{eq:linear-meas}
    \by(t) = \bH \bx(t) + \boldeta(t),
\end{equation}
\end{linenomath*}
with $\bH \in \real^{m \times d}$.
Second, we assume that the initial condition $\bx(t = 0) = \bx_0$ is known, so it is excluded from parameter estimation.

Although Eqs~\eqref{eq:state-eq}--\eqref{eq:obs-eq} define a continuous-time dynamical system, we mostly deal with discrete observations.
The set of $n$ measurements $\cY_n = \{\by_1, \ldots, \by_n\}$ denotes the experimental data taken at time instances $t_1, \leq t_2, \ldots, \leq t_n$, where $\by_k = \bH \bx_k + \boldeta_k$.
In this discrete setting, $\bx_k$ is the state vector at time $t_k$ and $\boldeta_k$ is an independent realization of the measurement noise. 
Additionally, the set of states at the discrete measurement times defined above is $\cX_n = \{\bx_1, \ldots, \bx_n \}$. 
Note that the full (internal) states may not be available for estimation, only the measurements.

We can now define the problem setting of the proposed parameter estimation framework.
This work assumes that the model form is known and seeks to estimate the model parameters and associated uncertainties by learning a probability distribution for the parameters.
The following problem statement formalizes the parameter estimation problem.
\begin{framed}
\begin{problem} \label{prob:problem-setting}
    Given a known model form as in Eq~\eqref{eq:state-eq} and a set of noisy, sparse and incomplete, experimental measurements $\cY_n = \{ \by_1,\ldots, \by_n \}$ at time instances $t_1 \leq t_2, \ldots, \leq t_n$, estimate the complete probability distribution, $p(\btheta|\cY_n)$, for the model parameters $\btheta = [\btheta_f, \ \btheta_\Gamma]^\top$.
\end{problem}
\end{framed}
The next section outlines the structural identifiability and global sensitivity analyses performed to reduce the dimension of $\btheta$ before estimating parameters.

\subsection{Global structural identifiability analysis with the structural identifiability analyzer (SIAN)} \label{methods:strucID}

Structural identifiability analysis determines if parameters can be uniquely estimated from the available measurement function~\cite{Wieland2021structural}.
Structural identifiability is a mathematical property of the model itself and does not consider the quality or quantity of the available experimental data.
The following definition from~\cite{norton2010identification} provides an intuitive condition for global structural identifiability and can be shown to be equivalent to alternative definitions~\cite{Anstett-Collin2020-qn, Hong2020global}.
\begin{definition}[Global structural identifiability~\cite{norton2010identification}] \label{def:globalID}
    A parameter $\theta_i$ is globally structurally identifiable if, for almost all $\btheta \in \real^p$ and for all times $t > 0$,
    \begin{linenomath*}
    \begin{equation*}
        \by (t,\btheta') = \by (t,\btheta) \qquad \implies \qquad \theta_i' = \theta_i,
    \end{equation*}
    \end{linenomath*}
    where $\btheta' = [\theta'_1, \ldots, \theta'_i, \ldots, \theta'_p]^\top$ and $\btheta = [\theta_1, \ldots, \theta_i, \ldots, \theta_p]^\top$.
\end{definition}
\textit{Global} structural identifiability, as in Definition~\ref{def:globalID},  implies that a parameter $\theta_i$ can be uniquely identified from data.
If a parameter is globally structurally identifiable, then there is a single unique value of that parameter that gives each observed output value with the same initial conditions~\cite{Hong2020global}.
Alternatively, a parameter may be \textit{locally} structurally identifiable if the condition in Definition~\ref{def:globalID} only holds in a local neighborhood, $\mathcal{V}(\theta_i)$, of parameter space around $\theta_i$, e.g., for $\theta_i \in \mathcal{V}(\theta_i)$~\cite{Wieland2021structural, Anstett-Collin2020-qn, Hong2020global}.
This condition implies that a finite number of values of a locally identifiable parameter can give the same output values~\cite{Hong2020global}.
Lastly, if a parameter is nonidentifiable, then infinitely many values of that parameter can give the same model output.
While models with locally structurally identifiable and nonidentifiable parameters are still valid over the entire parameter space, the existence of such parameters can confound successful parameter estimation.
Many computational methods have been developed to assess structural identifiability~\cite{Wieland2021structural, Anstett-Collin2020-qn, Hong2020global, Villaverde2019-tq} for ODE-based models.
In this work, we use the differential algebra and power series-based approach presented in~\cite{Hong2020global, Hong2019sian}, because the method specifically assesses global identifiability.
The SIAN (Structural Identifiability ANalyser) software~\cite{Hong2019sian} provides a numerical implementation of the approach proposed in~\cite{Hong2020global}.
The SIAN algorithm for assessing global structural identifiability uses a combination of symbolic and probabilistic computation~\cite{Hong2020global, Hong2019sian}.
First, SIAN uses Taylor expansions of the model equations to obtain a polynomial representation of the system.
Second, the algorithm truncates the polynomial system to produce a minimal system containing all parameter identifiability information.
Third, SIAN solves the identifiability problem for a single parameter set that is randomly selected to guarantee correctness up to a user-specified probability level $p$ (see Theorem 5 in~\cite{Hong2020global}).
Fourth, the algorithm uses the results in the third step to separate the parameters into globally identifiable, locally identifiable, and nonidentifiable sets.
SIAN is implemented in Maple (Maplesoft, Waterloo, ON) and Julia (The Julia Project~\cite{bezanson2017julia}).
We refer the reader to~\cite{Hong2020global} for additional mathematical details on SIAN.
    
We use the Julia implementation of the SIAN algorithm with the default probability of correctness,  $p = 0.99$ (available at \url{https://github.com/alexeyovchinnikov/SIAN-Julia}).
Furthermore, we set the additional \verb!p_mod! parameter to $2^{29} - 3$ to enable the algorithm to run faster~\cite{SIAN-GH}.
Any parameters that are not globally structurally identifiable are fixed to nominal values informed by the available literature following identifiability analysis.
While these parameters may convey meaningful biological information, nonidentifiability implies that it is mathematically impossible to identify them from the available data.
Next, global sensitivity analysis is used to further reduce the set of identifiable parameters.

\subsection{Variance-based global sensitivity analysis} \label{methods:gsa}

Parametric sensitivity analysis quantifies the contributions of the model parameters to variations in the model output~\cite{smith2013uncertainty, Saltelli2007-ta}.
Specifically, global sensitivity analysis aims to quantify the effects of the model parameters on the output quantity of interest over the entire parameter space~\cite{smith2013uncertainty, Saltelli2007-ta} and is well suited for studying parameters in nonlinear models~\cite{Saltelli2007-ta, Kent2013learn}.
This work applies Sobol's method~\cite{Sobol2001gsa} for variance-based sensitivity analysis because it provides a quantitative output to rank parameters and leverages the prior distributions defined for the parameters.
Sobol sensitivity analysis decomposes the variance of model outputs based on contributions from individual parameters and interactions between parameters~\cite{Saltelli2007-ta, Sobol2001gsa}.
The total variance of the output quantity $f(\btheta)$ is
\begin{linenomath*}
\begin{equation} \label{eq:total-var}
    D = \int_{\real^p} f^2(\btheta){\rm d}\btheta - f_0^2,
\end{equation}
\end{linenomath*}
where $f_0 \coloneqq \int_{\real^p} f(\btheta) {\rm d}\btheta$ is the mean of the output.
The following definition for the analysis of variance (ANOVA) representation provides an expansion for the output variance in a high-dimensional representation (HDMR), also known as a Sobol representation~\cite{smith2013uncertainty, Sobol2001gsa}. 
%
\begin{definition}[Analysis of variance (ANOVA) representation~\cite{smith2013uncertainty, Sobol2001gsa, Saltelli2007-ta}] \label{def:anova}
    The ANOVA expansion states that the output function $f(\btheta)$, for $\btheta \in \real^p$ defined as $\btheta =[\theta_1$, $\theta_2, \ \dots$, $\theta_p]\top$, can be represented as
        \begin{linenomath*}
        \begin{equation*} 
            f(\btheta) = \sum_i f_i(\theta_i) + \sum_{i < j} f_{i,j}(\theta_i, \theta_j) + \ldots + f_{1,2,\ldots,p}(\theta_1, \theta_2, \ldots, \theta_p),
        \end{equation*}
        \end{linenomath*}
    where the zero-, first-, and second-order terms are defined recursively as
    \begin{linenomath*}
    \begin{align*}
        f_0 &= \int_{\real^p} f(\btheta) {\rm d}\btheta\\
        f_i(\theta_i) &= \int_{\real^p} f(\btheta) {\rm d} \theta_{\sim i} - f_0 \\
        f_{i,j}(\theta_i, \theta_j)  &= \int_{\real^p} f(\btheta) {\rm d} \theta_{\sim \{i,j\}} - f_i(\theta_i) - f_j(\theta_j) - f_0 ,
    \end{align*}
    \end{linenomath*}
    respectively.
    The recursion is extended further for increasing numbers of parameters to compute higher-order terms.
    This definition assumes that the contribution terms are orthogonal (see Def~1~in~\cite{Sobol2001gsa}), and the notation $\sim i$ refers to the set excluding index $i$, for example: ${\rm d}\theta_{\sim i} = \{{\rm d}\theta_{1}$, $\ldots$, ${\rm d}\theta_{i-1}$, ${\rm d}\theta_{i+1}$, $\ldots$, ${\rm d}\theta_{p} \}$.
\end{definition}
The ANOVA representation (see Def~\ref{def:anova}) expands the total variance, Eq~\eqref{eq:total-var}, as
\begin{linenomath*}
\begin{equation} \label{eq:total-var-decomp}
    D = \sum_{i=1}^p D_i + \sum_{1 \leq i < j \leq p} D_{i,j},
\end{equation}
\end{linenomath*}
where the variances $D_i$ and $D_{i,j}$ are
\begin{linenomath*}
\begin{equation*}
    D_i = \int_{\real^p} f_i^2(\theta_i){\rm d}\theta_i \quad \text{and} \quad
    D_{i,j} = \int \int_{\real^p} f_{i,j}^2(\theta_i, \theta_j){\rm d}\theta_i {\rm d}\theta_j.
\end{equation*}
\end{linenomath*}
Note that it is possible to compute higher-order variances by increasing the dimension of the integral and following the recursion in Def~\ref{def:anova}, however we limit our discussion to second-order or lower variances for brevity. 
The first and second-order Sobol sensitivity indices are then defined using the variance terms in Eq~\eqref{eq:total-var-decomp} as
\begin{linenomath*}
\begin{equation*}
   S_i = \frac{D_i}{D} \quad {\rm and} \quad S_{i,j} = \frac{D_{i,j}}{D}.
\end{equation*}
\end{linenomath*}
The first-order sensitivity index $S_i$ quantifies the fraction of the total variance attributed to parameter $\theta_i$, and the second-order sensitivity index $S_{i,j}$ quantifies this for the interactions between $\theta_i$ and $\theta_j.$
Lastly, the \textit{total-order sensitivity} is 
\begin{linenomath*}
\begin{equation*} 
   S_{T_i} = S_i + \sum_{j=1}^p S_{i,j},
\end{equation*}
\end{linenomath*}
which quantifies all contributions from parameter $\theta_i$ on the output variance.

This work uses the UQLab toolbox~\cite{marelli2014uqlab, UQdoc_20_106} to perform Sobol sensitivity analysis in Matlab (MathWorks, Natick, MA) and the \verb!DifferentialEquations.jl! package~\cite{rackauckas2017differentialequations} for analysis in Julia.
Both softwares use Monte Carlo estimators to compute Sobol sensitivity indices from parameter samples (see~\cite{smith2013uncertainty, Saltelli2007-ta} and references therein for details).
In Matlab, unless otherwise specified, first and total sensitivity indices are computed using Sobol pseudo-random sampling (e.g. \verb!SOpts.Sobol.Sampling = `sobol'!) and the default estimator (\verb!SOpts.Sobol.Sampling = `janon'!), see~\cite{UQdoc_20_106} for details.
Additionally, the number of samples, \verb!SOpts.Sobol.SampleSize!, is set specifically for each problem in Section~\ref{sec:results}.
In Julia, we perform similar Sobol sampling with \verb!QuasiMonteCarlo.jl! (\url{https://github.com/SciML/QuasiMonteCarlo.jl)} with the \verb!SobolSample()! sampler, and the default estimator for the sensitivity indices, e.g. \verb!Ei_estimator! set to \verb!:jansen1999!.

Sobol sensitivity analysis computes sensitivity indices for scalar output quantities rather than for an entire trajectory.
Thus, we define quantities of interest (QoI) for each problem that capture the biologically relevant information in a trajectory, such as the steady state value of a certain biochemical species.
Using the computed sensitivity indices for the given QoI, we rank the parameters and aim to find the most influential parameters by selecting those with the greatest sensitivity indices.
The critical challenge is selecting a cutoff point that separates the influential parameters from the complete set.
To do so, we chose a threshold corresponding to a pronounced decline in the sensitivity index value or a logical value of that index, for example, $S_i \geq 0.1$.
All parameters with sensitivity indices above the cutoff value are left free for estimation, and the remaining parameters are fixed to their nominal values.
The following section recasts the parameter estimation problem (Problem~\ref{prob:problem-setting}) in the Bayesian framework to estimate the free, influential parameters.

\subsection{Bayesian parameter estimation} \label{methods:bayes}

Bayesian parameter estimation solves Problem~\ref{prob:problem-setting} by considering the model parameters, $\btheta$, as random variables that do not have a pre-specified parametric distribution.
Note that parametric distributions have probability density functions that are fully described by a closed-form parametric equation, such as a normal distribution that is characterized by its mean and variance.
The approach characterizes the posterior probability distribution for the parameters $p(\btheta | \cY_n)$ conditioned on a given dataset $\cY_n$ and provides the \textit{best guess} probability distribution for $\btheta$ \textit{given the data}.
We can use Bayes' rule to express the posterior distribution as
\begin{linenomath*}
\begin{equation} \label{eq:bayes-rule}
    p(\btheta | \cY_n) \propto p(\btheta) \cL( \btheta; \cY_n),
\end{equation}
\end{linenomath*}
where $p(\btheta)$ is known as the prior distribution, and $\cL(\btheta; \cY_n)$ is the likelihood function (see Fig~\subpanelref{fig:uqOverview}{B2} for a visual representation of Bayes' rule).

Intuitively, Bayes' rule updates our \textit{best guess} about the distribution of the model parameters as new data is being incorporated.
The prior distribution $p(\btheta)$ represents the \textit{best guess} before any data are collected and encodes any assumptions on the parameters.
For instance, the prior may convey the physiological ranges for parameter values or may weigh known values more heavily (see Section~\ref{methods:prior}).
The likelihood function $\cL(\btheta; \cY_n) = p(\cY_n | \btheta)$ updates our \textit{belief state} by measuring the misfit between the data and model predictions with a specific parameter set.
Parameter sets that are more likely to occur will produce model predictions that better match the data and thus have higher likelihood probabilities.
For example, although the prior in Fig~\subpanelref{fig:uqOverview}{B2}
places more probability on smaller values of $\theta$, the likelihood in Fig~\subpanelref{fig:uqOverview}{B2} places more probability mass towards larger values.
It is important to note that evaluation of the likelihood function requires model simulations.
For example, assuming Gaussian measurement noise with zero mean, a possible likelihood function is
\begin{linenomath*}
\begin{equation} \label{eq:guass-like}
    \cL(\btheta; \cY_n) = \frac{1}{(2 \pi)^{\frac{m}{2}} |\bGamma|^{\frac{1}{2}}} \prod_{k=1}^{n}\exp{\left( -\frac{1}{2}\left|\left|\by_k - \bH \hat{\bx}_k\right|\right|^2_{\bGamma}\right)},
\end{equation}
\end{linenomath*}
where $\hbx_k$ is the predicted state at time $t_k$ with the parameters $\btheta$, $|\cdot|$ denotes the matrix determinant, and the $\bC$-weighted norm is defined as $||\ba||^2_{\bC} \coloneqq \ba^\top \bC^{-1} \ba$, where $\bC$ is a symmetric positive definite matrix.
Lastly, the posterior distribution conveys the \textit{best guess} after collecting and incorporating the data $\cY_n$ into the statistical model and can be further refined if more data are included.
In Fig~\subpanelref{fig:uqOverview}{B2}, the posterior illustrates how the likelihood function re-weights the prior to update our \textit{belief state}.  
    
A fundamental difficulty in Bayesian parameter estimation is that Bayes' rule only enables evaluating the posterior distribution at specific points in parameter space.
This is in contrast to, for example, the formula for a Gaussian distribution (with mean $\mu$ and standard deviation $\sigma$)
\begin{linenomath*}
\begin{equation*} 
    p(x) = \frac{1}{\sqrt{2\pi \sigma^2}} \exp\left(\frac{-(x - \mu)^2}{\sigma^2} \right)
\end{equation*}
\end{linenomath*}
that can be analytically evaluated at all values of $x$.
Therefore, parameter samples are drawn from the posterior help to characterize the distribution over the entire parameter space.
Markov chain Monte Carlo (MCMC) algorithms enable sampling from arbitrary distributions, such as the posterior distribution (see Section~\ref{methods:MCMC} for details).
Before performing Bayesian estimation the next section introduces the constrained interval unscented Kalman filter Markov chain Monte Carlo (CIUKF-MCMC) algorithm that accounts for uncertainty in the model and the data.

\subsection{Constrained interval unscented Kalman filter Markov chain Monte Carlo (CIUKF-MCMC)} \label{methods:CIUKF-MCMC}

Complete uncertainty quantification of dynamical models in systems biology must account for uncertainty in the model form, parameters and noisy data.
In~\cite{Galioto2020-gq}, Galioto and Gorodetsky suggest adding a process noise term to Eq~\eqref{eq:state-eq} to account for model form uncertainty in the system.
Following this suggestion, the model in Eqs~(\ref{eq:state-eq}-\ref{eq:obs-eq}) is recast as a discrete time stochastic process 
\begin{linenomath*}
    \begin{alignat}{2}
        \bx_k &= \psi \left(\bx_{k-1}; \ \btheta_f \right) + \bxi_k, \qquad && \bxi_k \sim \cN\left(\mathbf{0}, \bSigma(\btheta_{\Sigma})\right) \label{eq:discrete_state} \\
        \by_k &= \bH \bx_k + \boldeta_k, \qquad && \boldeta_k \sim \cN\left(\mathbf{0}, \bGamma(\btheta_{\Gamma})\right), \label{eq:discrete-obs}
    \end{alignat}
\end{linenomath*}
where $k$ is the discrete time index for $t_k$, and $\psi(\cdot; \cdot)$ is the discrete state propagator that evolves the state from time $t_{k-1}$ to time $t_k$. Additionally, $\bxi_k$ and $\boldeta_k$ are Gaussian process and measurement noise (stochastic noise processes) with covariances $\bSigma(\btheta_\Sigma)$ and $\bGamma(\btheta_\Gamma)$, respectively.
The discrete state propagator $\psi(\cdot; \cdot)$ in Eq~\eqref{eq:discrete_state} is the discrete operator that evolves the state according to the differential equation in Eq~\eqref{eq:state-eq}, and, for example, could be a forward-Euler finite-difference approximation.
This work deploys the Matlab function \verb!ode15s()! to construct a state propagator that guarantees the necessary stability to handle systems biology models.

Bayesian estimation of the model parameters, $\btheta=[\btheta_f, \ \btheta_\Sigma, \ \btheta_\Gamma]\top$, of the extended system in Eqs~\eqref{eq:discrete_state}--\eqref{eq:discrete-obs} accounts for uncertainty in the data, model, and parameters.
The introduction of process noise increases the dimension of the parameters to estimate by requiring estimates for $\btheta_\Sigma$.
Further, the addition of stochastic process noise means the state variables are random variables; Bayes' rule for this system becomes
\begin{linenomath*}
\begin{equation} \label{eq:bayesWithStates}
        p(\btheta, \cX_n | \cY_n) \propto p(\btheta) \cL(\btheta; \cY_n,\cX_n),
\end{equation}
\end{linenomath*}
to account for the additional uncertainty in the states.
The key step of the UKF-MCMC algorithm is the marginalization of uncertainty in the states out of Eq~\eqref{eq:bayesWithStates} to enable estimation of the parameter posterior distribution~\cite{Galioto2020-gq}.

The UKF-MCMC algorithm begins by constructing an expression for the joint likelihood of the states and the parameters, $\cL\left(\btheta; \cY_n, \cX_n \right)$.
Two probability distributions implied by the stochastic system in Eqs~(\ref{eq:discrete_state}-\ref{eq:discrete-obs}) are needed to define an expression for the joint likelihood.
First, the probability of the current state $\bx_{k}$ given the past state $\bx_{k-1}$ is
\begin{linenomath*}
\begin{equation} \label{eq:prop-prob}
	p\left(\bx_k | \bx_{k-1}, \btheta_f, \bSigma(\btheta_\Sigma) \right) = \frac{\exp \left(-\frac{1}{2}\left|\left|\bx_k - \psi(\bx_{k-1}, \btheta_f)\right|\right|^2_{\bSigma}\right)}{(2 \pi)^{\frac{d}{2}}\left|\bSigma(\btheta_\Sigma)\right|^{\frac{1}{2}}},
\end{equation}
\end{linenomath*}
where the norm $\left|\left|\bx_k - \psi(\bx_{k-1}, \btheta_f)\right|\right|^2_{\bSigma}$ quantifies the misfit between the past state and the predicted current state.
Next, the probability of a measurement $\by_k$ given $\bx_k$ is
\begin{linenomath*}
\begin{equation}\label{eq:meas-prob}
	p\left(\by_k |\bx_{k}, \bGamma(\btheta_\Gamma) \right) = \frac{\exp \left(-\frac{1}{2}\left|\left|\by_k - \bH \bx_k\right|\right|^2_{\bGamma}\right)}{(2 \pi)^{\frac{m}{2}}\left|\bGamma(\btheta_\Gamma)\right|^{\frac{1}{2}}}, 
\end{equation}
\end{linenomath*}
where the norm $\left|\left|\by_k - \bH \bx_k\right|\right|^2_{\bGamma}$ quantifies the residual between the measurement and the true states.
By combining Eq~\eqref{eq:prop-prob} and Eq~\eqref{eq:meas-prob} the joint likelihood is
\begin{linenomath*}
\begin{equation}
	\cL\left(\btheta; \cY_n, \cX_n \right) =
	\prod_{k=1}^{n}\left[
	\frac{\exp \left(-\frac{1}{2}\left|\left|\bx_k - \psi(\bx_{k-1}, \btheta_f)\right|\right|^2_{\bSigma}\right)}{(2 \pi)^{\frac{d}{2}}\left|\bSigma(\btheta_\Sigma)\right|^{\frac{1}{2}}}
	\times
	\frac{\exp \left(-\frac{1}{2}\left|\left|\by_k - \bH \bx_k\right|\right|^2_{\bGamma}\right)}{(2 \pi)^{\frac{m}{2}}\left|\bGamma(\btheta_\Gamma)\right|^{\frac{1}{2}}}
	\right]. \label{eq:joint}
\end{equation}
\end{linenomath*}
Marginalizing out the uncertain states by integration yields the likelihood for the uncertain parameters
\begin{linenomath*}
\begin{equation} \label{eq:likelihood-marginalization}
    \cL\left(\btheta; \cY_n \right) = \int_{\real^d_{\geq 0}} \cdots \int_{\real^d_{\geq 0}}\cL\left(\btheta; \cY_n, \cX_n \right) {\rm d}\bx_1 \ldots {\rm d}\bx_n.
\end{equation}
\end{linenomath*}
However, there is no obvious computationally tractable approach to integrate over a set of uncertain states directly.
Theorem~\ref{thm:marginal-likelihood}, stated below, provides a recursive algorithm to marginalize the states out of the likelihood, e.g., to perform the integration in Eq~\eqref{eq:likelihood-marginalization}.
Although Theorem~\ref{thm:marginal-likelihood} assumes that the initial condition is uncertain (and it is therefore estimated), we do not use that estimate in this work, as we start with a known initial condition $\bx_0$.
\begin{theorem}[Marginal likelihood (Theorem 1 of~\cite{Galioto2020-gq} and 12.1 of~\cite{sarkka2013bayesian})] \label{thm:marginal-likelihood}
    Let $\cY_k$ denote the set of all observations up to time $t_k$ as defined in Section~\ref{methods:partial-obs}. Let the initial condition be uncertain with distribution $p\left(\bx_0 | \btheta \right)$. Then the marginal likelihood is defined recursively in three stages:
    \vspace{10pt}
    
    \noindent for $k = 1, 2, \ldots $
    \ben
        \item Predict the new state from previous data
        \begin{linenomath*}
    	\begin{equation*}
    		p\left(\bx_{k+1} \big| \btheta, \cY_{k} \right) = \int_{\real^d_{\geq 0}} p\left(\bx_k \big | \btheta, \cY_k\right) \frac{\exp \left(-\frac{1}{2}\left|\left|\bx_{k+1} - \psi(\bx_{k}, \btheta_f)\right|\right|^2_{\bSigma}\right)}{(2 \pi)^{\frac{d}{2}}\left|\bSigma(\btheta_\Sigma)\right|^{\frac{1}{2}}} {\rm d}\bx_k,
    	\end{equation*}
    	\end{linenomath*}
    	\item update the prediction with the current data
    	\begin{linenomath*}
    	\begin{equation*}
    		p\left(\bx_{k+1} \big | \btheta, \cY_{k+1} \right) = 	p\left(\bx_{k+1} \big | \btheta, \cY_{k} \right)  \frac{\exp \left(-\frac{1}{2}\left|\left|\by_{k+1} - \bH \bx_{k+1}\right|\right|^2_{\bGamma}\right)}{(2 \pi)^{\frac{m}{2}}\left|\bGamma(\btheta_\Gamma)\right|^{\frac{1}{2}}},
    	\end{equation*}
    	\end{linenomath*}
    	\item and marginalize out uncertainty in the states
    	\begin{linenomath*}
    	\begin{equation*}
    		\cL_{k+1}\left( \btheta \big | \cY_{k+1} \right)= \int_{\real^d_{\geq 0}} p\left(\bx_{k+1} \big | \btheta, \cY_k \right)  \frac{\exp \left(-\frac{1}{2}\left|\left|\by_{k+1} - \bH \bx_{k+1}\right|\right|^2_{\bGamma}\right)}{(2 \pi)^{\frac{m}{2}}\left|\bGamma(\btheta_\Gamma)\right|^{\frac{1}{2}}} {\rm d}\bx_{k+1}.
    	\end{equation*}
    	\end{linenomath*}
    \een
\end{theorem}  

The recursion defined in Theorem~\ref{thm:marginal-likelihood} closely resembles a Bayesian filter~\cite{sarkka2013bayesian}; thus, it is evaluated with Kalman filtering algorithms~\cite{Galioto2020-gq}.
For linear models, the standard linear Gaussian Kalman filter can be used to evaluate the recursion (see Algorithm 2 in~\cite{Galioto2020-gq}).
However, exact solutions to the recursion are not possible if the model or measurement processes are nonlinear.
Therefore, approximations such as extended Kalman filters (EKF), unscented Kalman filters (UKF), or ensemble Kalman filters (EnKF)~\cite{Galioto2020-gq, sarkka2013bayesian} must be employed.
The original implementations of the UKF-MCMC algorithm use the UKF~\cite{Julier2004unscented} for this approximation because the UKF is generally stable and can handle nonlinear models and measurement processes~\cite{Galioto2020-gq, Erazo2017-ls}.
However, the UKF is not suitable for systems biology models because it ignores constraints on the state variables, such as the nonnegativity of chemical concentrations.
Section~\ref{methods:CIUKF} describes the constrained interval unscented Kalman filter (CIUKF)~\cite{Teixeira2008-vy} implemented in this work to enforce state constraints during filtering.
Thus, we refer to the UKF-MCMC from~\cite{Galioto2020-gq} that uses the constrained interval unscented Kalman filter~\cite{Teixeira2008-vy} as CIUKF-MCMC.
 
\subsection{Constrained interval unscented Kalman filter (CIUKF)} \label{methods:CIUKF}

We implement the constrained interval Unscented Kalman filter (CIUKF)~\cite{Teixeira2008-vy, Vachhani2006-bo} algorithm to enforce all state constraints in CIUKF-MCMC.
This algorithm assumes that the state is subject to an interval constraint $\bx_{\textsf{LB}} \leq \bx \leq \bx_{\textsf{UB}}$.
Since we only seek to enforce nonnegativity in systems biology, the interval constraint is $\mathbf{0} \leq \bx \leq \boldsymbol{\infty}$.
We choose the CIUKF over alternative state constrained Kalman filters~\cite{Teixeira2008-vy, Simon2010-fz} because it enforces constraints in both the predict and update steps of the algorithm and retains the same structure as the standard UKF~\cite{sarkka2013bayesian}.
We outline the steps of the CIUKF below.
  
The CIUKF algorithm predicts the state $\bx_{k+1}$ from all preceding measurement data $\cY_n$.
Following the structure of the linear Kalman filter, CIUKF is a recursive algorithm that iterates over all data and performs prediction and update steps at each time point~\cite{sarkka2013bayesian}.
For simplicity, we outline a single iteration of the CIUKF that moves the state forward in time from $\bx_k$ to $\bx_{k+1}$.
Let $\bC^{xx}_k$ be the  state covariance matrix at time $t_k$, $\bSigma(\btheta_\Sigma)$ be the process noise covariance matrix, $\bGamma(\btheta_\Gamma)$ be the measurement noise covariance matrix, and $\btheta_f$ be the model parameters.
There are two steps to the CIUKF, prediction, and update.

First, the prediction step uses the the interval constrained unscented transform~\cite{Vachhani2006-bo, Teixeira2008-vy} (the state-constrained equivalent to the unscented transform~\cite{julier1997new, Julier2004unscented}) to predict the state and its covariance matrix at the next time point after propagation by the nonlinear model, e.g., Eq~\eqref{eq:discrete_state}.
The interval constrained unscented transform constructs a set of sigma points that capture the covariance $\bC^{xx}_k$ at time $t_k$.
Each sigma point is propagated in time by the nonlinear model to approximate the new state and its covariance at the next time $t_{k+1}$.
The set of $2d + 1$ sigma points, $\cX$, is given by
\begin{linenomath*}
\begin{subequations} \label{eq:pred-sigma}
\begin{align}
    \cX^{(0)}_k &= \bx_k \\
    \cX^{(i)}_k &= \bx_k + \xi_i [\sqrt{\bC^{xx}_k}]_i \\
    \cX^{(i+d)}_k &= \bx_k - \xi_{i+d} [\sqrt{\bC^{xx}_k}]_{i},
\end{align}
\end{subequations}
\end{linenomath*}
where $\xi_i$ is the $i$th sigma point coefficient, $[\bA]_i$ is the $i$th column of $\bA$, $\sqrt{\bA}$ is the matrix square root of $\bA$, and $i = 1, \ \ldots \, d$.
The sigma point coefficients $\xi_i$ control the distances of the sigma points around the initial state $\bx_k$ and are chosen to ensure that no sigma points violate the state constraints. 
They are
\begin{linenomath*}
\begin{align}
    \xi_i &= \text{min}([\bXi]_i) \label{eq:icut-w} \\ 
    \bXi(i,j) &\coloneqq \begin{cases} 
    \sqrt{d+\lambda}& \text{if} \ \bS(i,j) = 0 \\ 
    \text{min}\left(\sqrt{d+\lambda}, \frac{\bx_{\textsf{LB}} - \bx_k}{\bS(i,j)} \right)& \text{if} \ \bS(i,j) < 0 \\ 
    \text{min}\left(\sqrt{d+\lambda}, \frac{\bx_{\textsf{UB}} - \bx_k}{\bS(i,j)} \right)& \text{if} \ \bS(i,j) > 0
    \end{cases} \label{eq:icut-Wij} \\ 
    \bS &\coloneqq \begin{bmatrix}\sqrt{\bC^{xx}_k} & -\sqrt{\bC^{xx}_k} \end{bmatrix}, \label{eq:icut-S}
\end{align}
\end{linenomath*}
where $\lambda$ is a parameter of the algorithm.
Alternatively, in the standard unscented transform, the coefficients are all equal to $\sqrt{d + \lambda}$~\cite{ Julier2004unscented, julier1997new}.
Next, a set of weights, $w_i$, are assigned to each sigma point as
\begin{linenomath*}
\begin{subequations} \label{eq:pred-sigma-weights}
\begin{align}
    w_0 &= b \\
    w_i &= a \xi_i + b \\
    a &\coloneqq \frac{2 \lambda - 1}{2(d + \lambda)\left(\textstyle\sum\nolimits_{i=1}^d \xi_i - (2d + 1)\sqrt{d+\lambda}\right)} \\
    b &\coloneqq \frac{1}{2(d+\lambda)} - \frac{2 \lambda - 1}{2\sqrt{d + \lambda}\left(\textstyle\sum\nolimits_{i=1}^d \xi_i - (2d + 1)\sqrt{d+\lambda}\right)},
\end{align}
\end{subequations}
\end{linenomath*}  
where $\xi_i$ are as defined in Eqs~(\ref{eq:icut-w}-\ref{eq:icut-S}).
Importantly the sum of the weights equals one, $\sum_{i=0}^{2d} w_i = 1$.
The prediction step then uses the nonlinear model, Eq~\eqref{eq:discrete_state}, to propagate each sigma point forward in time 
\begin{linenomath*}
\begin{equation} \label{eq:nonlinear-prop}
    \hat{\cX}^{(i)}_{k} = \psi\left(\cX^{(i)}_{k}, \btheta_f \right)
 \end{equation}
\end{linenomath*}
for $i = 0, \ldots, 2d$.
The prediction mean $\bx^{*}_{k}$ and covariance $\bC^{*xx}_{k}$ are then respectively computed as
\begin{linenomath*}
\begin{align}
    \bx^*_{k} &= \sum_{i=0}^{2n} w_i \hat{\cX}^{(i)}_{k} \label{eq:pred-mean}\\
    \bC^{xx^*}_{k} &= \sum_{i=0}^{2n} w_i \left[\hat{\cX}^{(i)}_{k} - \bx^*_{k} \right]\left[\hat{\cX}^{(i)}_{k} - \bx^*_{k} \right]^\top + \bSigma(\btheta_\Sigma). \label{eq:pred-cov}
\end{align}
\end{linenomath*}
Equations~\eqref{eq:pred-sigma}-\eqref{eq:pred-cov} describe the constrained interval unscented transform that approximates the mean and covariance of the state after propagation by the nonlinear model.
The prediction mean, Eq~\eqref{eq:pred-mean}, and covariance, Eq~\eqref{eq:pred-cov}, provide the predicted state and its covariance, respectively, that are then updated using the available data $\by_{k}$.

The update step begins by constructing a new set of sigma points centered around $\bx^*_{k}$, where
\begin{linenomath*}
\begin{subequations} \label{eq:update-sigma}
\begin{align}
    \cX^{*(0)}_k &= \bx^*_k \\
    \cX^{*(i)}_k &= \bx^*_k + \sqrt{n+\lambda} \left[\sqrt{\bC^{xx^*}_{k}}\right]_i \\
    \cX^{*(i+d)}_k &= \bx^*_k -  \sqrt{n+\lambda} \left[\sqrt{\bC^{xx^*}_{k}}\right]_{i}.
\end{align}
\end{subequations}
\end{linenomath*}
Additionally, a new set of weights are
\begin{linenomath*}
\begin{alignat*}{3}
    w^{(m)}_0 &= \frac{\lambda}{d + \lambda}, & \quad & w^{(m)}_i &= \frac{1}{2(d + \lambda)}, \\
    w^{(c)}_0 &= \frac{\lambda}{d + \lambda} + (1 - \alpha^2 + \beta), & & w^{(c)}_i &=  \frac{1}{2(d + \lambda)},
\end{alignat*}
\end{linenomath*}
where $\{w_i^{(m)}\}$ are used to compute the mean and $\{w_i^{(c)}\}$ are used to compute the covariance matrix.
These weights are indeed equal for all sigma points and are equivalent to those used in the standard UKF.
Next, the measurement function is applied to each sigma point, yielding a set of predicted measurements
\begin{linenomath*}
\begin{equation*}
    \cY^{*(i)}_k = h\left(\cX^{*(i)}_k, \btheta_h\right),
\end{equation*}
\end{linenomath*}
where the measurement function $h(\cdot; \cdot)$ is possibly nonlinear with parameters $\btheta_h$.
Then the mean and covariance matrices of the predicted measurements are computed with the weighted sums,
\begin{linenomath*}
\begin{align*}
    \by^*_{k} &= \sum_{i=0}^{2n} w^{(m)}_i \cY^{*(i)}_{k} \\
    \bC^{yy^*}_{k} &= \sum_{i=0}^{2n} w^{(c)}_i \left[\cY^{*(i)}_{k} - \by^*_{k} \right]\left[\cY^{*(i)}_{k} - \by^*_{k} \right]^\top + \bGamma \\
    \bC^{xy^*}_{k} &= \sum_{i=0}^{2n} w^{(c)}_i \left[\cX^{*(i)}_{k} - \bx^*_{k} \right]\left[\cY^{*(i)}_{k} - \by^*_{k} \right]^\top,
\end{align*}
\end{linenomath*}
and the Kalman gain is
\begin{linenomath*}
\begin{equation*}
    \bK_k = \bC^{xy^*}_{k} \left(\bC^{yy^*}_{k}\right)^{-1}.
\end{equation*}
\end{linenomath*}
%
Lastly, the updated state $\bx_{k+1}$ is found by solving the following constrained nonlinear optimization problem,
\begin{linenomath*}
\begin{subequations} \label{eq:ciukf-update}
    \begin{alignat}{2}
    \bx_{k+1} = \ &\!\argmin_{\bx}        & \ & f(\bx) \label{eq:ciukf-update-prob} \\
    &\text{subject to} &      & \bx_{\textsf{LB}} \leq \bx \leq \bx_{\textsf{UB}}, \label{eq:ciukf-constraint}
    \end{alignat}
\end{subequations}
\end{linenomath*}
where the objective function is
\begin{linenomath*}
\begin{equation} \label{eq:ciukf-update-obj}
    f(\bx) = \left[\by_k - h(\bx, \btheta_h)\right] \bGamma^{-1} \left[\by_k - h(\bx, \btheta_h)\right]^\top + \left[\bx_k - \bx^*_k\right] (\bC_k^{xx^*})^{-1}\left[\bx_k - \bx^*_k\right]^\top. 
\end{equation}
\end{linenomath*}
This optimization problem can be solved in Matlab using the \verb!fmincon()! optimizer.
Additionally, the updated covariance matrix is given by
\begin{linenomath*}
\begin{equation*}
    \bC^{xx}_{k+1} = \bC^{xx^*}_k - \bK_k \bC_k^{yy^*} \bK_k^\top.
\end{equation*}
\end{linenomath*}
In offline state estimation problems, such as CIUKF-MCMC, this filter is iterated over all available data, e.g. from time $t_0$ to time $t_n$ if $n$ data points are available~\cite{Galioto2020-gq, Erazo2017-ls}.

In practice, the CIUKF algorithm is substantially more compute-intensive than the standard UKF~\cite{Teixeira2008-vy} because the CIUKF update step involves solving a constrained nonlinear optimization problem, e.g., Eq~\eqref{eq:ciukf-update}.
However, the objective function in Eq~\eqref{eq:ciukf-update-obj} can be simplified given the linear measurement assumptions made in Section~\ref{methods:partial-obs}.
The simplified objective function becomes
\begin{linenomath*}
\begin{equation} \label{eq:ciukf-update-linearobj}
    f(\bx) = \left[\by_k - \bH\bx \right] \bGamma^{-1} \left[\by_k - \bH \bx\right]^\top + \left[\bx_k - \bx^*_k\right] (\bC_k^{xx^*})^{-1}\left[\bx_k - \bx^*_k\right]^\top.
\end{equation}
\end{linenomath*}
Expanding this and recognizing that minimizing $f(x) = y(x) + b$ is equivalent to minimizing $f(x) = y(x)$, gives
\begin{linenomath*}
\begin{equation} \label{eq:ciukf-update-quadobj}
    f^*(\bx) = \bx^\top \left[\bH^\top \bGamma^{-1} \bH + (\bC_k^{xx^*})^{-1} \right]\bx - 2\left[\by_k^\top \bGamma^{-1}\bH +  \bx^*_k \bC_k^{xx^*} \right]\bx,
\end{equation}
\end{linenomath*}
which is equivalent objective function to Eq~\eqref{eq:ciukf-update-linearobj} and is a quadratic form.
Thus, the constrained optimization problem in Eq~\eqref{eq:ciukf-update} becomes a quadratic program when using Eq~\eqref{eq:ciukf-update-quadobj} as the objective function.
We use the \verb!quadprog()! function in Matlab to solve the quadratic program with the \verb!`Algorithm'! option set to \verb!`trust-region-reflective'!.
We keep all other settings for the solver as defaults.
As expected, solving the quadratic program is substantially more efficient than solving the general nonlinear problem.

Throughout this work, we set $\lambda = 1$, $\alpha=1\times 10^{-3}$, and $\beta=1$ for the CIUKF.
Furthermore the Cholesky decomposition, $\bA = \bL \bL^\top$, is used to compute the matrix square roots in Eq~\eqref{eq:pred-sigma} and Eq~\eqref{eq:update-sigma}, because covariance matrices are always positive definite.
Additionally, we compute all covariance matrices as $\bP^* = \frac{1}{2}(\bP + \bP^\top) + \epsilon \bI$, where $\epsilon=1\times 10^{-10}$, to ensure all they remain symmetric positive definite.
The next section discusses how to choose prior distributions for Bayesian estimation. 

\subsection{Prior distribution} \label{methods:prior}

The prior distribution, $p(\btheta)$ encodes our \textit{belief state} about parameters before collecting data and performing Bayesian estimation~\cite{smith2013uncertainty, gelman2013bayesian}.
The form of the prior distribution allows us to incorporate varying levels of prior knowledge into our models.
If the values of a parameter are known, informative priors can be used to shift to the possible values for that parameter towards the known values~\cite{gelman2013bayesian}; for example, a log-normal prior distribution can be used to center the prior around experimental measurements of a parameter~\cite{Tsigkinopoulou2018informative}.
However, the Bayesian inference literature often warns that informative priors should only be used in combination with good information on the parameter values~\cite{gelman2013bayesian} as it can take a large amount of data to overcome a ``bad" prior.
Alternatively, noninformative or weakly-informative priors reflect a lack of good prior knowledge about the parameters.
For example, if we only know a parameter's physiological ranges, we could construct a uniform prior that states there is an equal probability of any parameter value within this range.
Noninformative or weakly-informative priors rely on the data to provide information on the parameters, so the choice of such priors is often safer when prior knowledge of the parameters is limited~\cite{gelman2013bayesian}.
    
In applying the CIUKF-MCMC algorithm, this work constructs prior distributions for two sets of parameters, the biological model parameters, and the noise covariance parameters.
We choose to use uniform priors for the biological model parameters, $\btheta_f$, to replicate the typical modeling setting where only the possible ranges for model parameters are known.
Supplemental Tables in~\ref{sup:tables} list the upper and lower bounds of all biological model parameters. 
Furthermore, we follow the choices in~\cite{Galioto2020-gq} and use right-half-normal priors for the measurement and process noise covariance parameters, $\btheta_\Sigma$ and $\btheta_\Gamma$, respectively (this is further motivated in section 7.1 of~\cite{Gelman2006priors}).
The choice of covariance and upper bound of these priors was found to significantly affect the convergence of MCMC sampling.
Thus, the prior distributions for the noise covariance parameters were scaled to match the respective state variable.
For example, the prior for a measurement noise covariance $\theta_{\Sigma^i}$ that corresponds to measurements $\{y_1^i, y_2^i, \ldots, y_n^i\}$ would be a right-half-normal distribution with mean zero and standard deviation equal to a fraction of the standard deviation $\sigma_{y^i}$ of the data.
The prior is then $\theta_{\Sigma_i} \sim \text{Right-Half-Normal}(0, b \sigma_{x_i})$ truncated to $[0, \ \sigma_{x_i}]\top$, where $b < 1$ and $\sigma_{x_i}$ is the standard deviation of the state $x_i$.
We chose $b = \frac{1}{3}$ unless otherwise specified.
These choices are motivated by the observation that measurement noise and process noise covariances will always be smaller than or equal to the covariance of the available data if there is any meaningful information in the data.
The next section introduces MCMC sampling to characterize posterior distributions that use CIUKF to approximate the likelihood function.

\subsection{Markov chain Monte Carlo sampling} \label{methods:MCMC}

Markov chain Monte Carlo (MCMC) algorithms enable sampling from arbitrary probability distributions~\cite{smith2013uncertainty, gelman1996efficient, mcbook, sokal1997monte}.
The key idea of MCMC is to construct a Markov chain of samples $\btheta^1, \ \btheta^2,\ldots, \btheta^N$ whose distribution converges to the target distribution, $\pi (\btheta)$, that we wish to sample~\cite{gelman2013bayesian, mcbook}.
We apply MCMC sampling to Bayesian parameter estimation by constructing a Markov chain where the target distribution is the posterior distribution, that is $\pi(\btheta) = p\left(\btheta|\cY_n\right)$.
In this work, we use two MCMC sampling algorithms, delayed rejection adaptive Metropolis (DRAM)~\cite{Haario2006dram} and affine invariant ensemble sampler (AIES)~\cite{Goodman2010aies}.
These samplers build upon the classical Metropolis-Hastings algorithm~\cite{metropolis1953equation, hastings1970monte} that we introduce in Section~\ref{methods:MH}.
We outline DRAM in Section~\ref{methods:DRAM} and AIES Section~\ref{methods:AIES}. 
Lastly, we discuss convergence assessment with the integrated autocorrelation time~\cite{Goodman2010aies} in Section~\ref{methods:iact}.
We focus these discussions on the practical aspects of MCMC and refer the reader to~\cite{smith2013uncertainty, gelman2013bayesian, mcbook} for additional theoretical details.
    
\subsubsection{Metropolis-Hastings} \label{methods:MH}

The Metropolis-Hastings (MH) algorithm~\cite{metropolis1953equation, hastings1970monte} constructs a Markov chain whose probability distribution is guaranteed to converge to the target distribution and forms the foundation for a large family of MCMC samplers~\cite{gelman2013bayesian, mcbook}.
The MH algorithm consists of two steps, a proposal and an accept-reject step, repeated to draw the set of samples $\btheta^1, \ \btheta^2,\ldots, \btheta^N$.
The Markov chain starts with an initial sample $\btheta^0$ that the user chooses.
We outline one iteration of the MH algorithm moving from step $i$ to step $i+1$.
    
The first step in Metropolis-Hastings, called the proposal step, is to propose a new sample $\btheta^*$.
We draw $\btheta^*$ from the proposal distribution $q(\btheta^*|\btheta^i)$.
The proposal distribution is specific to each MCMC algorithm, but, for example, could be a normal distribution centered around the previous sample such that $\btheta^* \sim \cN\left(\btheta^i, \sigma^2 \bI \right)$ where $\sigma$ is specified (this is random walk Metropolis (RWM)~\cite{mcbook}).
    
Next, the accept-reject step decides if the proposal is accepted, set $\btheta^{i+1} = \btheta^*$, or rejected,  set $\btheta^{i+1} = \btheta^{i}$.
In the MH algorithm, this decision is probabilistic, e.g., the proposal is accepted or rejected with acceptance probability $\alpha(\btheta^*| \btheta^{i})$.
The acceptance probability is
\begin{linenomath*}
\begin{equation} \label{eq:MH-accept}
    \alpha(\btheta^*| \btheta^{i}) = \text{min}\left(1, \frac{\pi(\btheta^*)}{\pi(\btheta^{i})} \frac{q(\btheta^i|\btheta^*)}{q(\btheta^*|\btheta^i)} \right),
\end{equation}
\end{linenomath*}
which guarantees that the stationary distribution of the samples (the distribution that the samples converge to in the infinite sample limit) equals the target distribution~\cite{gelman2013bayesian, mcbook}.
These steps, propose and accept-reject, are repeated until the distribution of the set of samples has \textit{converged} to its stationary distribution.
Although convergence is guaranteed in the infinite sample limit~\cite{ gelman2013bayesian, mcbook}, assessing convergence in practice is nontrivial.
We chose to use an approach from~\cite{Goodman2010aies} that uses the integrated autocorrelation time to asses convergence as outlined in Section~\ref{methods:iact} (see \cite{smith2013uncertainty, gelman2013bayesian, mcbook} for additional approaches).

\subsubsection{Delayed rejection adaptive Metropolis (DRAM)} \label{methods:DRAM}

The delayed rejection adaptive Metropolis (DRAM) algorithm is based on the random walk Metropolis algorithm and combines delayed rejection with adaptive Metropolis~\cite{Haario2006dram}.
DRAM closely follows the Metropolis-Hastings algorithm but specifically uses a Gaussian proposal distribution such that $\btheta^* \sim \cN\left(\btheta^i, \bC \right)$, where $\bC$ is the covariance matrix.
Delayed rejection (DR)~\cite{Haario2006dram, tierney1994markov} and adaptive Metropolis (AM)~\cite{Haario2006dram, haario2001adaptive} are two modifications to the random walk Metropolis algorithm that help to improve the convergence rate of the Markov chain.
    
First, delayed rejection adds an additional round of Metropolis-Hastings (propose and accept-reject) if the first proposal is rejected.
That is, if $\btheta^*$ is rejected a new proposal, $\btheta^{*2} \sim q_2(\btheta^{*2}|\btheta^*, \btheta^i)$ is drawn, where $q_2(\cdot)$ is the new proposal distribution.
The new proposal distribution is also Gaussian, however the covariance is scaled to a fraction of the original proposal covariance, e.g. $q_2(\btheta^{*2}|\btheta^*, \btheta^i) = \cN\left(\btheta^i, \gamma \bC \right)$, where $\gamma < 1$ and is a tuning parameter of the algorithm~\cite{Haario2006dram} .
Thus the second proposal is closer to the previous point and is more likely to be accepted.
DRAM evaluates the new proposal with a Metropolis-Hastings accept-reject step with the acceptance probability
\begin{linenomath*}
\begin{align*}
    \alpha_2(\btheta^{*2}|\btheta^*, \btheta^{i}) &= 
    \text{min}\left(1, \frac{\pi(\btheta^{*2})}{\pi(\btheta^{i})} \frac{q(\btheta^*|\btheta^{*2})}{q(\btheta^*|\btheta^i)}
    \frac{\left(1-\alpha(\btheta^*| \btheta^{*2})\right)}{\left(1-\alpha(\btheta^*| \btheta^{i})\right)}
    \right),
\end{align*}
\end{linenomath*}
where $\alpha(\cdot| \cdot')$ is as defined in Eq~\eqref{eq:MH-accept}.
Although most implementations impose a single DR step~\cite{smith2013uncertainty}, delayed rejection can be repeated more than once, where the proposal distribution and acceptance probability are modified accordingly at additional each round.

Adaptive Metropolis acts separately from delayed rejection and aims to move the proposal distribution closer to the target distribution~\cite{Haario2006dram},  by replacing the covariance matrix of the proposal, $\bC$, with the covariance matrix of the samples.
In practice, adaptation begins after a set number of $i_0$ samples have been drawn and updates the covariance matrix at every step as
\begin{linenomath*}
\begin{equation*} 
    \bC_i = \begin{cases} 
        \bC_0 & i \leq i_0 \\
        s_p\text{Cov}[\btheta_1, \ldots, \btheta_i] + s_p\epsilon\bI & i > i_0,
    \end{cases} 
\end{equation*}
\end{linenomath*}
where $\bC_0$ is the initial covariance matrix, $s_p$ is an algorithm tuning parameter that is often set to $s_p = 2.38^2/p$~\cite{Haario2006dram, gelman1996efficient} where $p$ is the dimension of $\btheta$, and $\epsilon$ is a small positive number.
    
We implement the DRAM algorithm with a single round of delayed rejection following~\cite{Galioto2020-gq} with $\gamma  = 0.01$, $i_0 = 200$, and $\epsilon=1\times10^{-10}$.
Furthermore, the initial covariance matrix $\bC_0$ is tuned to accept between $20$--$40 \%$ of proposals.
We refer the reader to~\cite{smith2013uncertainty, Haario2006dram} for further details on DRAM and tuning the initial covariance matrix.
Additionally, each Markov chain is initialized to the MAP point by using \verb!fmincon()! (default options except: \verb!`UseParallel'! set to \verb!true! and \verb!`MaxFunctionEvaluations'! set to $10,000$) in Matlab to minimize the negative log-posterior (equivalent to maximizing the posterior) as in~\cite{Galioto2020-gq}.

\subsubsection{Affine invariant ensemble sampler (AIES)} \label{methods:AIES}

While random walk Metropolis-based algorithms such as DRAM can adequately sample complex posterior distributions, these methods will show very slow convergence when the target distribution is highly anisotropic~\cite{Goodman2010aies}.
The posterior distribution with CIUKF-MCMC in systems biology are anisotropic because the scales of model parameters can vary by several orders of magnitude, and the noise covariance parameters often have different scaling than the model parameters.
Fortunately, AIES provides an algorithm to sample such anisotropic distributions~\cite{Goodman2010aies} effectively.
The motivation for affine invariance is that anisotropic distributions can be transformed to isotropic distributions with an affine transformation.
Thus an algorithm that is invariant to such transformations will effectively sample an isotropic distribution when sampling an anisotropic distribution~\cite{Goodman2010aies}.
    
The AIES algorithm differs from DRAM and random walk Metropolis because it leverages an ensemble of Markov chains rather than a single chain.
Each chain in the ensemble of $N_e$ Markov chains is called a walker, and we denote the set of walkers at step $i$ with $\{\btheta_1^i,\ldots,\btheta_{Ne}^i\}$ where the subscript is the walker index and the superscript is the step-index.
Note that the number of walkers must be larger than the dimension of $\btheta$ that is $N_e > p$, for $\btheta \in \real^p$.
At the end of $N$ steps, a set of $N_e$ Markov chains with $N$ samples is obtained.
The first chain in the ensemble is, for example, $\btheta_1^1, \btheta_1^2, \ldots, \btheta_1^N$.
Note that the total ensemble will have $N \cdot N_e$ samples.
Each step of the AIES algorithm involves a proposal and a Metropolis-Hastings update for each chain in the ensemble.
One iteration of these steps to move from $i$ to $i+1$ for a single walker, $\btheta_n^i$, is outlined below and these steps are repeated to update the entire ensemble.

First, a proposal for the current walker $\btheta_n^i$ is chosen using the stretch move~\cite{Goodman2010aies} that ensures affine invariance of the sampler.
The stretch move proposes a new point that lies along the line 
\begin{linenomath*}
\begin{equation*} 
    \btheta_n^* = \btheta_k^i + Z \cdot (\btheta_n^i - \btheta_k^i) 
\end{equation*}
\end{linenomath*}
that connects the current walker $\btheta_n^i$ and another walker in $\btheta_k^i$ randomly chosen from the ensemble.
Here, $Z$ is a random variable that is sampled following
\begin{linenomath*}
\begin{equation*} 
    g(z) = \begin{cases}
    \frac{1}{\sqrt{z}}& \text{if} \ \frac{1}{a} \leq z \leq a \\
    0& \text{else},
    \end{cases}
\end{equation*}
\end{linenomath*}
where $a > 1$ is an algorithm tuning parameter that the user must specify.

Second, the proposal is accepted or rejected using a Metropolis-Hastings-like accept-reject step.
The acceptance probability is
\begin{linenomath*}
\begin{equation*} 
    \alpha(\btheta_n^* | \btheta_n^i) = \text{min}\left(1,  Z^{p-1}\frac{\pi(\btheta_n^*)}{\pi(\btheta_n^i)}\right),
\end{equation*}
\end{linenomath*}
where $Z$ is as defined above, and $p$ is the dimension of $\btheta$.
This formulation of the acceptance probability guarantees convergence to the target distribution~\cite{Goodman2010aies}.
    
We use the Matlab implementation of the AIES algorithm~\cite{UQdoc_20_113} from the UQLab toolbox~\cite{marelli2014uqlab}.
Unless otherwise specified, the ensemble size is $N_e = 150$ because we observed improved sampling with a large ensemble.
To accelerate sampling, the likelihood is evaluated for each ensemble member in parallel using a parallel for loop (e.g., \verb!parfor! in Matlab) with at most 24 parallel threads.
Additionally, default value of $a = 2$ is used for the stretch move tuning parameter.
Lastly, each Markov chain in the ensemble is initialized to a random point drawn uniformly over the support of the prior.

\subsubsection{Markov chain burn-in} \label{methods:burn-in}
In MCMC, Markov chains (or ensembles of chains) often display an initial transient, called \textit{burn-in}, before converging to their stationary distributions~\cite{smith2013uncertainty, gelman2013bayesian, sokal1997monte}.
Importantly, these samples during burn-in are not distributed according to the stationary distribution and should therefore be excluded from the final set of samples.
Common practice in the MCMC literature is to simply discard these initial samples to remove the effects of burn-in~\cite{smith2013uncertainty, gelman2013bayesian}.
The choice of the burn-in length is often nontrivial and is best informed by an analysis of the Markov chain~\cite{sokal1997monte}.
Unless otherwise specified the integrated autocorrelation time (described below) dictates the number of samples to discard as burn-in in this work.
Specifically, we compute the integrated autocorrelation time after collecting many samples and set the burn-in length to 5--10 times the computed value.

\subsubsection{Convergence assessment with the integrated autocorrelation time} \label{methods:iact}

A key challenge in Markov chain Monte Carlo sampling is determining the appropriate number of samples $N$ to collect.
In this work, we use the integrated autocorrelation time~\cite{Goodman2010aies, sokal1997monte} to determine when the Markov chain has approximately converged to its stationary distribution.
We outline the theory and motivation behind the integrated autocorrelation time for a single Markov chain and refer the reader to~\cite{Goodman2010aies} for a discussion of ensemble methods. 
The use of the integrated autocorrelation time is motivated by the typical use of MCMC sampling to compute an expectation
\begin{linenomath*}
\begin{equation*} 
    \bbE[\bx] = \int \bx \pi(\bx) {\rm d}\bx.
\end{equation*}
\end{linenomath*}
Given a Markov chain of length $N$, we can estimate the expectation with the Monte Carlo estimator
\begin{linenomath*}
\begin{equation*} 
    \hat{\mu} =  \frac{1}{N} \sum_{i=1}^{N} \bx^i.
\end{equation*}
\end{linenomath*}
In general it is common to consider the variance of $\hat{\mu}$, $\text{var}(\hat{\mu})$, as the estimation error~\cite{Goodman2010aies}.
This variance is given by 
\begin{linenomath*}
\begin{equation*} 
    \text{var}(\hat{\mu}) = \frac{\text{var}(\bx)}{N / \tau_s},
\end{equation*}
\end{linenomath*}
where the integrated autocorrelation time $\tau_s$ is given by 
\begin{linenomath*}
\begin{equation*} 
    \tau_s = \sum_{T=-\infty}^\infty \frac{C_s(T)}{C_s(0)}.
\end{equation*}
\end{linenomath*}
Here, the autocovariance function $C_s(T)$ with lag $T \in \nat$ is given by 
\begin{linenomath*}
\begin{equation*} 
    C_s(T) = \lim_{t^* \to \infty} \text{cov} [ \bx^{t^* + T}, \ \bx^T],
\end{equation*}
\end{linenomath*}
where $t^* \in \nat$.
Thus, the Monte Carlo estimation error is proportional to integrated autocorrelation time for a fixed chain length.
The integrated autocorrelation time can be interpreted as the time it takes for the samples in a Markov chain to become uncorrelated~\cite{sokal1997monte}.
Additionally, the effective number of samples, $N_{\text{eff}}$, can be defined using the integrated autocorrelation time as, $N_{\text{eff}} = N / \tau_s$.

In this work, we use the integrated autocorrelation time for two purposes.
First, the computation of the integrated autocorrelation time is used to choose the correct burn-in length.
We compute the integrated autocorrelation time after collecting many samples and then discard between 5--10 times $\tau_s$ as burn-in.
Second, after discarding the initial samples, the integrated autocorrelation time helps to determine if enough samples have been collected, e.g., $N_{\text{eff}}$ is large.
Should the effective sample size be small, the MCMC sampler is run longer to collect more samples.
We compute the integrated autocorrelation time using a Matlab function associated with~\cite{Wolff2004error} (available at \url{https://www.physik.hu-berlin.de/de/com/UWerr_fft.m}) with default algorithm parameters.
We compute $\tau_s$ for each parameter and take the maximum of these values.
For an ensemble, we compute the mean $\tau_s$ for the walkers of each parameter and take the maximum value of the means.
After completing MCMC and evaluating the chains, we leverage the posterior samples to quantify uncertainty in model predictions.

\subsection{Ensemble simulation and output uncertainty analysis} \label{methods:ensemble-model}

Markov chain Monte Carlo sampling provides a set of samples, $\{\hat{\btheta}_1, \ldots \hat{\btheta}_n\}$, that converge in distribution to the posterior distribution (e.g. Fig~\subpanelref{fig:uqOverview}{B3}).
One is often interested in how uncertainty in the parameter estimates, which is conveyed by the posterior distribution, propagates to uncertainty in the model predictions.
Fortunately, an ensemble of simulations (see Fig~\subpanelref{fig:uqOverview}{B1}) distributed according to the posterior can be run using the posterior samples.
This approach to uncertainty propagation is known as sampling-based uncertainty propagation~\cite{smith2013uncertainty} and is feasible because the simulation of dynamical systems biology models is computationally efficient.
We refer the reader to~\cite{Mortlock2021-lk, Miskovic2011uncertainty} for examples of sampling-based uncertainty propagation in systems biology.

Each simulation is run by solving the differential equation with the \verb!ode15s()! integrator in Matlab with a unique sample from $\{\hat{\btheta}_1, \ldots \hat{\btheta}_n\}$ for the parameters.
We use default tolerances for the integrator and supply the Jacobian matrix (specify the \verb!`Jacobian'! option) to improve computations.
We compute the Jacobian by hand and evaluate it with the same parameter as the differential equation.
    
Following ensemble simulation, statistical analyses of the ensemble of predicted trajectories and any relevant quantities of interest (QoI) can be performed.
For example, Fig~\subpanelref{fig:uqOverview}{B2} highlights the uncertainty in the trajectory with $95 \%$ credible intervals to show the region where $95 \%$ of the trajectories fall.
Additionally, we compute the statistics of relevant QoIs, such as the steady state values or limit cycle period, from the ensemble.
The next section discusses how to choose single point estimates that best represents the estimated parameters.

\subsection{Parameter point estimates from posterior samples} \label{methods:point-est}

\begin{figure}[ht!]
    \centering
    \includegraphics[width=256pt]{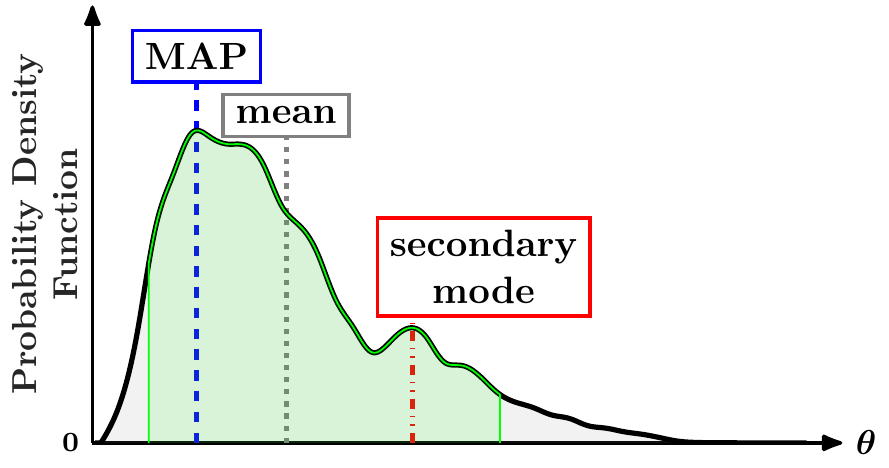}
    \caption{Examples of point estimates depicted on an arbitrary probability density function (black line).
    The MAP (maximum \textit{a posteriori}) point is located at the most probable point (blue dashed line).
    The long tail of the distribution shifts the mean (gray dotted line) away from the MAP point.
    Secondary modes (red dashed-dotted line) can effect the quality of a point estimate.
    Additionally, the green shaded region highlights the $95 \%$ credible interval, the region between the $2.5$th and $97.5$th percentiles, that is used to capture the uncertainty in an estimate.}
    \label{fig:point-estimate}
\end{figure}

One often wants to compute a point estimate for each of the parameters in addition to characterizing the entire posterior distribution (black line in Fig~\ref{fig:point-estimate}).
Common choices for point estimates in Bayesian statistics include the mean, median, or mode of the posterior distribution~\cite{gelman2013bayesian}.
Fig~\ref{fig:point-estimate} highlights the mean and mode (denoted MAP), along with a secondary mode that may confound choosing a point estimator.
The mode of the posterior distribution is a strong choice because it provides the most probable set of parameters and is often called the maximum \textit{a posteriori} (MAP) point.
However, the mean or median may provide better point estimates when the posterior is multimodal (there are multiple modes).
    
These point estimates are computed from posterior samples acquired via MCMC sampling.
While sample statistics, such as the mean and median, are computed directly from the samples, computing the MAP point requires estimating the posterior of the probability density function and finding the maximal point.
Direct computation of the sample mode will not yield the MAP point because the parameters are continuous random variables, so no two-parameter samples are expected to be identical.
One approach to estimate the MAP point involves computing a histogram of samples and using the center of the bin with the most associated probability as the MAP.
However, we found that the histogram of posterior samples is often noisy and can be sensitive to the bin size, so that the MAP estimate may be erroneous.
We chose to take an alternative approach that fits a kernel density estimator~\cite{bowman1997applied} to approximate the posterior distribution and subsequently computes that MAP point. 
The kernel density estimator provides a non-parametric approximation of the posterior distribution and, intuitively, smooths the posterior histogram.
We use the \verb!ksdensity()! function for kernel density estimation in Matlab.
The \verb!`support'! option is set to be the region bounded by the prior distribution and the \verb!`BoundaryCorrection'! option is set to use the \verb!`reflection'! method to account for these bounds. 
Lastly, we use the default values for the \verb!`Bandwidth'! unless otherwise specified.
All other options are kept to the defaults as defined in the Matlab documentation.
The MAP point is the point with maximum probability in the \verb!ksdensity()! output.
The next section moves from the details of CIUKF-MCMC and Bayesian estimation to outline how synthetic data is generated for the examples in this paper.

\subsection{Synthetic data generation for numerical experiments} \label{method:synthetic-data}

The synthetic data in this work aims to replicate noisy data found in biological experiments.
We generate noisy synthetic data by drawing samples from deterministic model simulations and simulate measurement noise by adding independently and identically distributed (iid) perturbations to each sample.
First, a nominal set of biological model parameters and an initial condition are chosen for data generation.
These values become the \textit{ground truth} for the estimation problem, and are informed by the available literature when possible.
Next, a numerical solution to the system provides \textit{true} trajectories of the state variables.
Unless otherwise specified the \verb!ode15s()! integrator in Matlab is used with default tolerances, and the Jacobian matrix is supplied (specify the \verb!`Jacobian'! option with the analytical Jacobian Matrix) to improve computations.
We then apply the measurement function and sub-sample the \textit{true} trajectory to simulate sparse sampling.
Lastly, we corrupt the data by adding a realization of an iid noise stochastic process to each sample.
To meet the assumptions of the CIUKF-MCMC algorithm (see Section~\ref{methods:CIUKF-MCMC}) we use mean-zero, normally distributed perturbations with the diagonal covariance matrix, $\bGamma$.
We chose the entries of $\bGamma$ to be proportional to the variances of the respective state variables, e.g.
\begin{linenomath*}
\begin{equation*}
    \bGamma = b \begin{bmatrix} 
        \text{var}(x^1_0, \ldots,x^1_n) & & 0 \\ 
        & \ddots & \\ 
        0 & & \text{var}(x^d_0, \ldots,x^d_n)
    \end{bmatrix},
\end{equation*}
\end{linenomath*}
where $b$ is a positive constant typically chosen to be less than one that controls the noise level.
The last section discusses how to compute the amplitude and period of a limit cycle oscillation.

\subsection{Limit cycle analysis} \label{methods:limit-cycle}

Limit cycle amplitude and period are used to characterize limit cycle oscillations for global sensitivity and output uncertainty analysis.
These quantities are relevant in intracellular signaling because the strength and timing of signals, amplitude and period, respectively, are thought to encode different inputs~\cite{Shankaran2010-xi}.
The limit cycle amplitude, $y_{\rm lca}$, quantifies the difference between the maximum and minimum values of the oscillations and is defined as
\begin{linenomath*}
\begin{equation} \label{eq:LCA}
    y_{\rm lca} \coloneqq x_{\rm max} - x_{\rm min},
\end{equation}
\end{linenomath*}
where $x_{\rm min}$ and $x_{\rm max}$ denote the minimum and maximum values of the state $x$ over a single complete oscillation.
Further, the limit cycle period, $y_{\rm period}$, is the time to complete an oscillation and is defined as
\begin{linenomath*}
\begin{equation} \label{eq:period}
    y_{\rm period} \coloneqq t^* \iff x(t) = x(t + t^*)\ \quad \text{for all} \ t.
\end{equation}
\end{linenomath*}
To compute these quantities we find trajectories that show limit cycle oscillations and then extract the two quantities of interest.
This approach leverages the \verb!findpeaks()! function that returns the locations of the local maxima (peaks) in a trajectory for these computations in Matlab.
The \verb!findpeaks()! function can also find the local minima by applying it to the negative of the trajectory.

The first task in computing the limit cycle features is to detect actual limit cycle oscillations.
A trajectory is discarded if it reaches a fixed point (steady state) when no peaks are detected (\verb!findpeaks()! returns an empty set).
Next, the difference between the heights of the identified peaks is used to discard trajectories that show decaying oscillations to a fixed point.
A threshold on one half of this difference, called for the \verb!helpPeakThreshold!, is set to $17.0$ unless otherwise specified; a trajectory is discarded if its difference in peak values exceeds this threshold.
Any remaining trajectories will show limit cycles or will contain numerical artifacts which are falsely detected as limit cycles.

Lastly, the limit cycle amplitude and period are computed.
The limit cycle amplitude is the mean difference between each pair of detected peaks and minima that correspond to one oscillation of the limit cycle.
The limit cycle period is then computed as the mean time between two peaks and the time between two minima.
Trajectories with numerical artifacts are eliminated by discarding those with a limit cycle amplitude that is smaller than the \verb!LCAminThresh!, with a range of limit cycle values greater than the \verb!Decaythresh! or those that return no limit cycle period (e.g., the empty matrix on Matlab).
Unless otherwise specified, we set the \verb!LCAminThresh! to $1.0$ and the \verb!Decaythresh! to $5.0$.
These computations assume that the period of the limit cycle is stable and that no frequency modulation occurs.
We note that a small number of oscillating trajectories can be misclassified as fixed points due to numerical errors in the simulated trajectory.
We specifically observed this when there was large peak-to-peak variability.
However, misclassification is rare, so we do not attempt manual relabelling of these trajectories.


\section{Results}\label{sec:results} 

We applied the Bayesian parameter estimation framework to three different models that represent signal transduction cascades of increasing biological and mathematical complexity.
Section~\ref{results:two-comp} uses the first model, a simple kinetic scheme, to describe a series of computational experiments that illustrate the effects of measurement noise and data sparsity on estimation uncertainty (Fig~\ref{fig:two-comp}).
Next, we tested our framework on two models that are more representative of the nonlinearities and overparameterization observed in systems biology models.
Section~\ref{results:mapk} analyzes a representative model of the mitogen-activated protein kinase (MAPK) pathway~\cite{Nguyen2015dyvipac} that exhibits multistability depending on the choice of model parameters,  and Supplemental Section~\ref{sup:sup-text} uses this model to illustrate the necessity of structural identifiability and global sensitivity analyses.
Finally, Section~\ref{results:kinase-phosphatase} analyzes a simplified model of synaptic plasticity~\cite{Pi2008coupled} that illustrates the effects of higher parameter uncertainty even in phenomenological biological representations.

\subsection{Measurement noise and sparsity increase estimation uncertainty in a simple model of signal transduction}\label{results:two-comp}


\begin{figure}[htp!]
    \centering
    \includegraphics[width=0.985\textwidth]{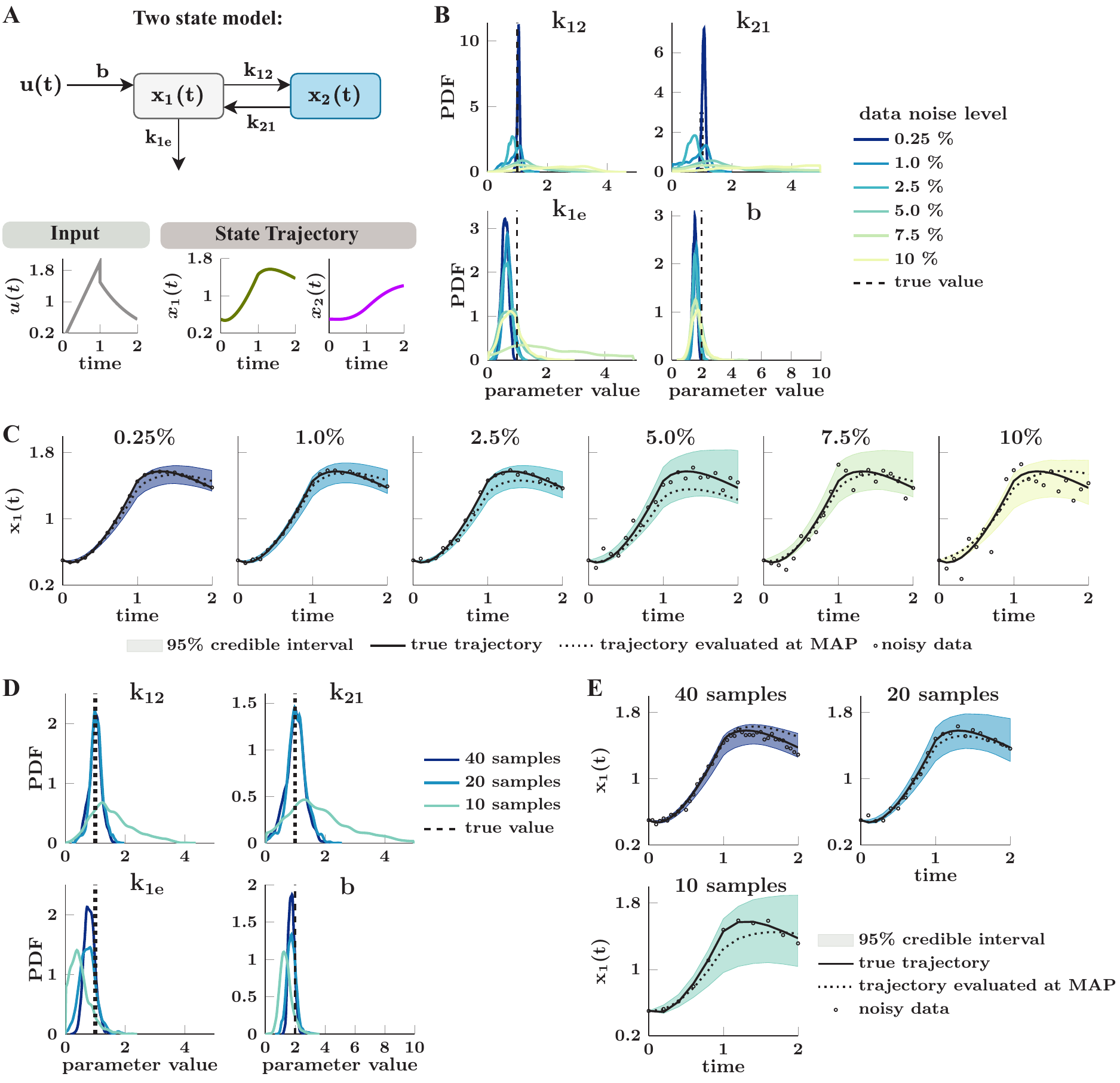}
    \caption{
        Parameter estimation for a simple two-state model.
        (\textbf{A}) \textit{Top row:} Network diagram of the two-state model with states, $x_1(t)$, $x_2(t)$, input function $u(t)$, and four unknown parameters, $\btheta_f = [k_{1e}, \ k_{12}, \ k_{21}, \ b]^\top$.
        \textit{Bottom row:} Trajectories of the input function $u(t)$ and corresponding state trajectories.
        The input has at least one non-zero derivative to ensure that all model parameters are globally structurally identifiable following~\cite{Villaverde2019-zc}.
        (\textbf{B}) Marginal posterior distributions of the model parameters show increasing uncertainty in the parameter estimates (e.g. widening and flattening) with increasing levels of additive normally distributed measurement noise with mean zero.
        We control the noise level by setting the noise covariances to the specified percentage of the standard deviation of each state variable.
        The dashed black vertical lines indicate each parameter's nominal (true) value.
        Marginal posteriors are visualized by fitting a kernel density estimator to $20{,}000$ MCMC samples obtained using CIUKF-MCMC with the delayed rejection adaptive Metropolis (DRAM) MCMC algorithm after discarding the first $10{,}000$ samples as burn-in.
        (\textbf{C}) Posterior distributions of the trajectory of $x_1(t)$ reflect increasing parameter estimation uncertainty in panel B.
        The true trajectory (solid black line) shows the dynamics with the nominal parameters, dashed black lines show that trajectory with the most probable set of parameters (MAP point), and the empty circles show the noisy data at the specified noise level.
        The $95 \%$ credible interval shows the region between the $2.5$th and $97.5$th percentiles that contains $95 \%$ of the $5{,}000$ trajectories.
        (\textbf{D}) Marginal posterior distributions of the model parameters show increasing uncertainty (widening and flattening) with increasing data sparsity (fewer samples).
        We simulate data sparsity by sampling the simulation from $0 \leq t \leq 2$ with three time steps, $\Delta t = 0.05$ (40 experimental samples), $\Delta t = 0.1$ (20 experimental samples) and $\Delta t = 0.2$ (10 experimental samples).
        Marginal posteriors are fit to $20{,}000$ MCMC samples obtained as in panel B.
        (\textbf{E}) Posterior distributions of the trajectory of $x_1(t)$ reflect increasing parameter estimation uncertainty seen in panel (\textbf{D}).
    }
    \label{fig:two-comp}
\end{figure}

The first model we consider is a relatively simple two-state model from \cite{Villaverde2019-zc} shown in Fig~\subpanelref{fig:two-comp}{A}.
The governing equations for this model are
\begin{linenomath*}
    \begin{subequations} \begin{align}\label{eq:two-comp}
        \frac{{\rm d} x_1(t)}{{\rm d}t} &= -(k_{1e} + k_{12})\cdot x_1(t) + k_{21}\cdot x_2(t) + b \cdot u(t)\\
        \frac{{\rm d} x_2(t)}{{\rm d}t} &= k_{12}\cdot x_1(t) - k_{21}\cdot x_2(t),
    \end{align} \end{subequations}
\end{linenomath*}
where the states are $\bx(t) = [x_1(t), \ x_2(t)]^\top$, the biological model parameters are $\btheta_f = [k_{1e}, \ k_{12}, \ k_{21}, \ b]^\top$, and $u(t)$ is the input function.
The input function (illustrated in Fig~\subpanelref{fig:two-comp}{A})
\begin{linenomath*}
\begin{equation*}
    u(t) = \begin{cases} t + 0.5& \text{if} \ 0 \leq t \leq 1 \\ 1.5e^{1-t}& \text{if} \ t > 1 \end{cases},
\end{equation*}
\end{linenomath*}
ensures that all four biological model parameters are structurally identifiable because the input function has at least one nonzero derivative~\cite{Villaverde2019-zc}.
Supplemental Table~\ref{tab:twostate-params} lists the nominal parameter values and the initial condition was $\bx_0 = [0.5$, $0.5]^\top$.
Sensitivity analysis was not performed on this model because the state variables are linearly dependent on the parameters.

Synthetic data with full state measurements (see Section~\ref{method:synthetic-data}) was used to perform two parameter estimation experiments---one with increasingly noisy data and another with increasingly sparse data---to investigate how measurement noise and data sparsity affect parameter and output uncertainty.
The noise levels of each dataset were controlled by taking a fraction of the maximum values of the true trajectory for the corresponding state variables.
In each estimation experiment, we chose the prior distributions as outlined in Section~\ref{methods:prior} and used CIUKF-MCMC (Section~\ref{methods:CIUKF-MCMC}) with DRAM (Section~\ref{methods:DRAM}) to draw $30{,}000$ posterior samples conditioned on this noisy data.
To ensure sample counts were constant across noise and sparsity levels, we chose a constant burn-in length of $10{,}000$ samples to discard from every Markov chain (see Section~\ref{methods:burn-in} for details).
Supplemental Fig~\ref{supfig:two-comp-noise-MCMC} shows the Markov chains for the measurement noise experiments, and Supplemental Fig~\ref{supfig:two-comp-samples-MCMC} shows those for data sparsity experiments.
The following remark highlights an important distinction between data points and MCMC samples.
\begin{remark}
Data points are different than MCMC samples.
The experiments in this work produce at most 40 (simulated) data points, i.e., noisy measurements of the states, for parameter estimation.
However, MCMC algorithms draw 10,000s-1,000,000s of sample parameter sets to characterize the posterior distribution, which requires evaluating the likelihood and, therefore, simulating the model.
\end{remark}

A first hypothesis we tested was that noisy data increases uncertainty because measurement noise limits the ability to constrain the dynamics of the state variables.
To test this hypothesis, we performed Bayesian parameter estimation from synthetic data with increasing measurement noise (circle marks in Fig~\subpanelref{fig:two-comp}{C}).
We observed that the marginal posterior distributions in Fig~\subpanelref{fig:two-comp}{B} (kernel density estimator fit to the posterior samples as in Section~\ref{methods:point-est}) widen and flatten with increasing noise levels, indicating increased uncertainty.
However, the most probable value for each parameter, the MAP point, lies close to the nominal parameter values (dashed lines in Fig~\subpanelref{fig:two-comp}{B}) for every noise level, suggesting that the data provide information about the parameter irrespective of the noise level. 
Additionally, Fig~\subpanelref{fig:two-comp}{C} shows that the width of the $95 \%$ credible interval for the dynamics of $x_1(t)$ grew as the noise increased from the lowest level ($2.5 \%$) to $5.0 \%$ and then remained similarly wide at the highest values (see Supplemental Fig~\subpanelref{supfig:two-comp-x2dyn}{A} for the respective trajectories of $x_2(t)$).
While the uncertainty bound did not widen above the $5.0 \%$ noise level, the shape of the trajectories began to shift further from the truth (dotted line in Fig~\subpanelref{fig:two-comp}{C}), indicating the estimates began to take on a bias.
These experiments validated our hypothesis that even in a simple dynamical system measurement noise increases estimation uncertainty of kinetic parameters.

Next, we hypothesized that data sparsity (fewer data points) would increase the uncertainty in parameter estimates.
To test this, we fixed the measurement noise to the $2.5 \%$ level (see Fig~\ref{fig:two-comp}) and varied the number of measurements (e.g., the sampling rate) that were included in the data used for estimation.
We tested three sparsity levels with $40$ experimental samples ($\Delta t = 0.05$), $20$ experimental samples ($\Delta t = 0.1$) and $10$ experimental samples ($\Delta t = 0.2$) over the simulation time $0 \leq t \leq 2$.
Fig~\subpanelref{fig:two-comp}{D} highlights the widening of the estimated marginal posterior distributions for each model parameter as we decreased the number of data points (increased sparsity).
Additionally, Fig~\subpanelref{fig:two-comp}{E} shows that the increased parameter estimation uncertainty translates to increased uncertainty (wider $95 \%$ credible interval) in the trajectory of $x_1(t)$ (see Supplemental Fig~\subpanelref{supfig:two-comp-x2dyn}{B} for the trajectories of $x_2(t)$).
In both of these experiments, the proposed uncertainty quantification framework qualitatively and quantitatively confirmed that increasing the noise or sparsity level increases estimation uncertainty.

\subsection{Parameter estimation for a model of the MAPK cascade}\label{results:mapk}

\begin{figure}[hb!]
    \centering
    \includegraphics[width=\textwidth]{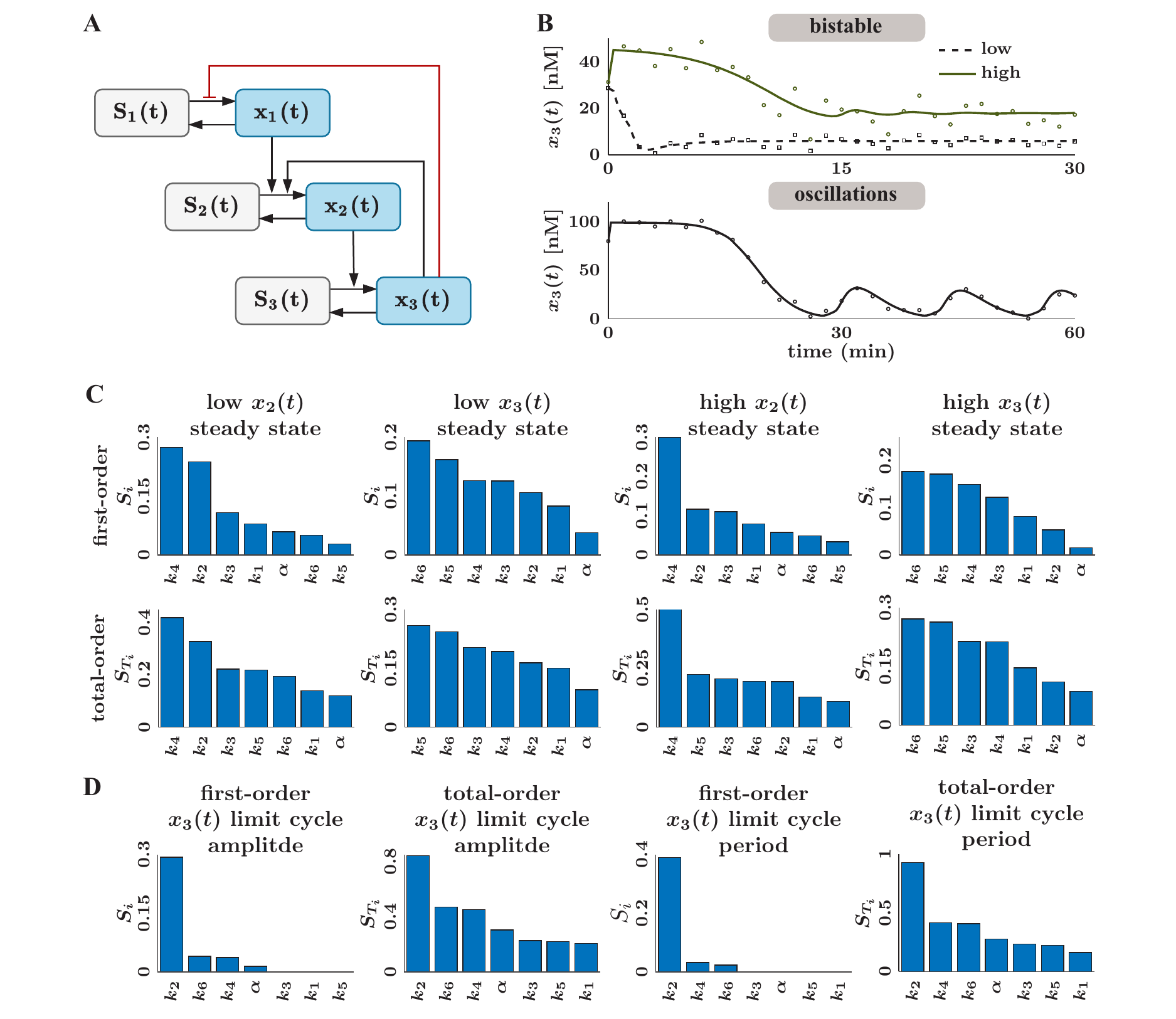}
    \caption{
        Parameter estimation for a simplified MAPK cascade that exhibits multistability.
        (\textbf{A}) Network diagram of the model of the core MAPK signaling cascade.
        The red line indicates inhibition; the black lines indicate activation.
        (\textbf{B}) Trajectories of $x_3(t)$ with the sets of nominal parameters that produce bistability (top) and limit cycle oscillations (bottom).
        The two dynamical regimes correspond to two different sets of nominal parameter values.
        The low (black dashed line) and high (solid green line) steady states are reached by manipulating the initial condition $\bx_0$.
        The initial condition for the high steady state is $\bx_{0, \rm high} = [0.1245, \ 2.4870,\ 31.2623]^\top$ and that for the low steady state is $\bx_{0, \rm low} = [0.0015,\ 3.6678, \ 28.7307]^\top$.
        (\textbf{C} and \textbf{D}) Sobol sensitivity for the MAPK model parameters.
        All parameters except the total concentrations, $S_{1t}$, $S_{2t}$ and $S_{3t}$, exponents, $n_1$ and $n_2$, and locally identifiable $K_1$ and $K_2$, are varied uniformly over the identified ranges (see Supplemental Table~\ref{tab:mapk-params}). We use $5{,}000$ and $15{,}000$ samples for the bistable and limit cycle regimes, respectively.
        (\textbf{C}) Sensitivity indices for bistable behavior dynamics.
        We use the steady state value of $x_2(t)$ and $x_3(t)$ for both the high and low steady states as quantities of interest.
        By selecting the two most sensitive parameters for the four quantities of interest, we reduce the set of free parameters to $\btheta_f = [k_2, \ k_4, \ k_5, \ k_6]^\top$.
        (\textbf{D}) Sobol sensitivity indices for a set of free parameters that contribute to limit cycle behavior.
        We show the first-order sensitivity indices $S_i$ and the total-order indices $S_{T_i}$ for the limit cycle amplitude and period of $x_3(t)$.
        We reduce the number of free parameters by selecting those with $S_i > 10^{-3}$ across both output quantities, that is, $\btheta_f = [k_2, \ k_6, \ k_4, \ \alpha]^\top$.
    }
    \label{fig:mapk-intro}
\end{figure}

We chose a simplified model of the highly conserved mitogen-activated protein kinase (MAPK; also known as the MEK/ERK cascade) signaling pathway~\cite{shaul2007mek} as a second test case for our parameter estimation framework.
This pathway is known to exhibit bifurcations in its dynamical behavior~\cite{Shankaran2010-xi, Vera2010-mf}; the system can reach a stable steady state or exhibit limit cycle oscillations.
We focused on a phenomenological model of the MAPK pathway from~\cite{Nguyen2015dyvipac} (see diagram in Fig~\subpanelref{fig:mapk-intro}{A}) that includes the mixed feedback (negative and positive feedback) necessary to predict the range of dynamical behavior observed in experiments.
This model has three states, $x_1(t), \ x_2(t), \ x_3(t)$ that correspond to phosphorylated RAF, MEK, and MAPK/ERK, respectively~\cite{Kholodenko2000negative, Kholodenko2010ballet} and 14 model parameters.
The differential equations are 
\begin{linenomath*}
\begin{subequations} \begin{align}\label{eq:mapk}
    \frac{{\rm d} x_1(t)}{{\rm d}t} &= k_1\cdot \left(S_{1t} - x_1(t)\right)\cdot \left[\frac{K_1^{n_1}}{K_1^{n_1} + x_3(t)^{n_1}}\right] - k_2\cdot x_1(t) \\
    \frac{{\rm d} x_2(t)}{{\rm d}t} &= k_3\cdot \left(S_{2t} - x_2(t)\right)\cdot x_1(t) \cdot \left[1 + \frac{\alpha \cdot x_3(t)^{n_2}}{K_2^{n_2} + x_3(t)^{n_2}}\right] - k_4 \cdot x_2(t) \\
    \frac{{\rm d} x_3(t)}{{\rm d}t} &= k_5\cdot \left(S_{3t} - x_3(t)\right) \cdot x_2(t) - k_6 \cdot x_3(t),
\end{align} \end{subequations}
\end{linenomath*}
with the biological model parameters $\btheta_f = [k_1$, $k_2$, $k_3$, $k_4$, $k_5$, $k_6$, $K_1$, $K_2$ $S_{1t}$, $S_{2t}$, $S_{3t}$, $\alpha$, $n_1$, $n_2]^\top$.
A previous analysis of the model in~\cite{Nguyen2015dyvipac} found that it can predict three regimes of dynamical behavior that depend on the model parameters; these regimes are limit cycle oscillations, bistability and mixed multistability.
Here we focus on using CIUKF-MCMC to estimate model parameters that produce two of the three dynamical regimes--bistability and limit cycle oscillations (see Fig~\subpanelref{fig:mapk-intro}{B} for example trajectories).
Supplemental Tables~\ref{tab:mapk-params}~and~\ref{tab:mapk-IC} list the nominal parameter values and initial conditions (as defined in~\cite{Nguyen2015dyvipac}) used to produce each of these dynamics.


First, we performed identifiability and sensitivity analysis to find the subsets of relevant parameters to estimate each dynamical regime.
The SIAN software~\cite{Hong2019sian} (see Section~\ref{methods:strucID}) showed that 12 of the 14 biological model parameters are structurally identifiable from measurements of all three state variables; however, $K_1$ and $K_2$ are only locally structurally identifiable.
SIAN cannot assess parameters that appear in an exponent~\cite{Hong2020global, Hong2019sian}, so we fixed $n_1$ and $n_2$ to their nominal values listed in Supplemental Table~\ref{tab:mapk-params}.
To avoid global versus local identifiability complications, we fixed $K_1$ and $K_2$ to their nominal values.
Additionally, we omitted the total concentration parameters, $S_{1t}$, $S_{2t}$, and $S_{3t}$, from further analysis because we assume they would be specified according to the cell type that corresponds to the available data. 
As a result, we narrowed the free parameters down to a set of 9 parameters, from a set of 14 originally.


\begin{figure}[hb!]
    \centering
    \includegraphics[width=\textwidth]{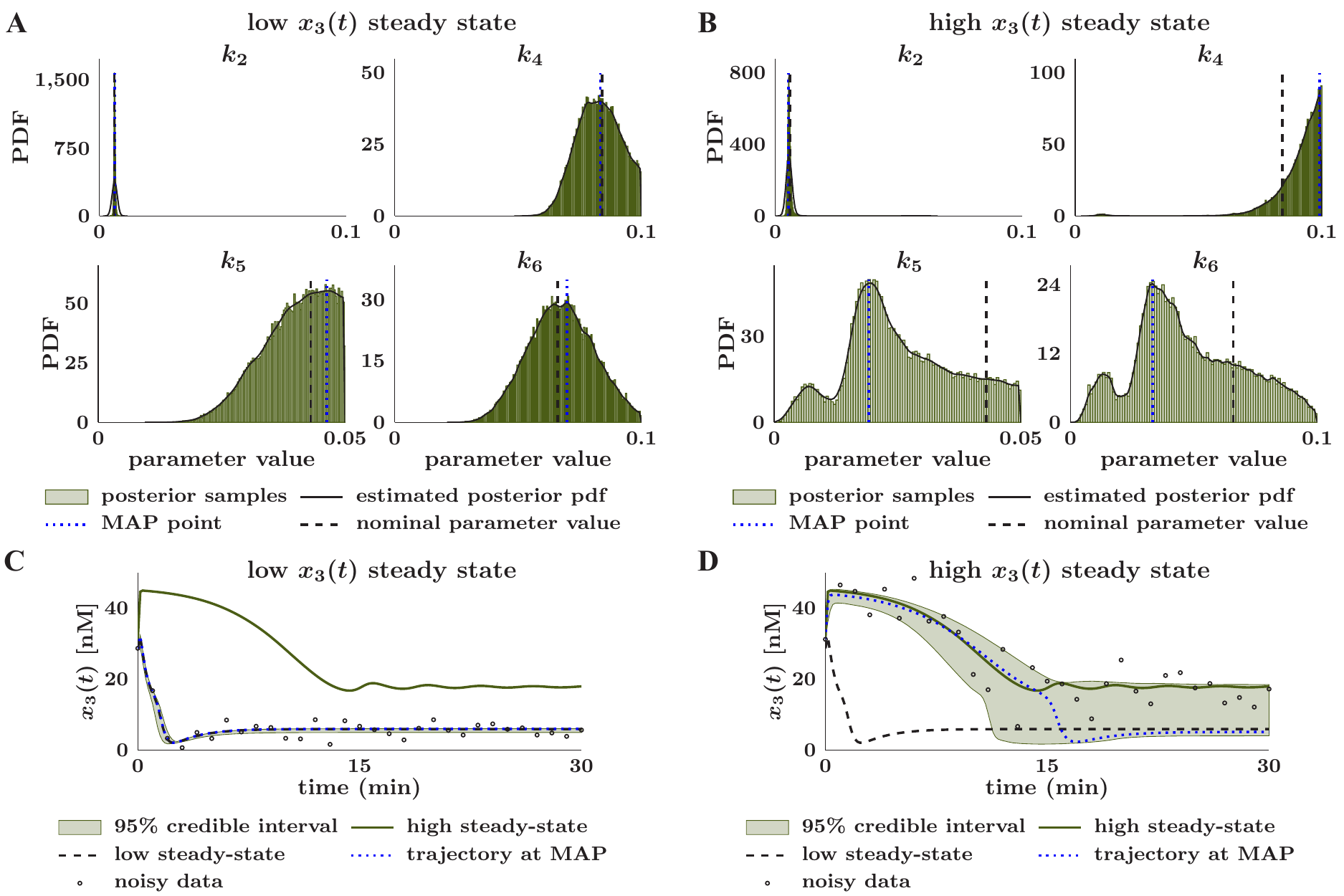}
    \caption{
        Varying levels of uncertainty in the parameters associated with the MAPK model impact steady state prediction.
        (\textbf{A}) Marginal posterior distributions of the model parameters for parameter estimation from noisy data of the low steady state.
        Posterior distributions are visualized by fitting a kernel density estimator to $325{,}200$ ($150$ walkers with $2{,}660$ steps each) MCMC samples obtained using CIUKF-MCMC with the affine invariant ensemble sampler (AIES) for MCMC after discarding the first $840$ samples per walker as burn-in.
        (\textbf{B}) Marginal posterior distributions of the model parameters for parameter estimation from noisy data of the high steady state reveal larger uncertainty in the model parameters when compared to the low steady state.
        We visualize distributions by fitting a kernel density estimator to $347{,}700$ ($150$ walkers with $2{,}644$ steps each) MCMC samples obtained using CIUKF-MCMC with the affine invariant ensemble sampler (AIES) for MCMC after discarding the first $856$ samples per walker as burn-in.
        (\textbf{C}) Posterior distribution of the trajectory of $x_3(t)$ with initial conditions that yield the low steady state highlights low uncertainty in the predicted dynamics.
        The true trajectory (dashed black line) shows the dynamics with the nominal parameters, the dotted blue line shows the trajectory evaluated at the MAP point, and the empty circles show the noisy data (covariance is $50 \%$ of the standard deviation of the true trajectory).
        The $95 \%$ credible interval shows the region between the $2.5$th and $97.5$th percentiles that contains $95 \%$ of $30{,}000$ posterior trajectories.
        (\textbf{D}) Posterior distribution of the trajectory of $x_3(t)$ with initial conditions that yield the high steady state highlights the ambiguity between which steady state is reached.
        All lines and computations are the same as in panel (\textbf{A}), except simulations were run using an initial condition that results in the high steady state.
    }
    \label{fig:mapk-bistable-results}
\end{figure}

Next, we used global sensitivity analysis as described in Section~\ref{methods:gsa} to further reduce the number of free parameters.
We choose the quantities of interest for sensitivity analysis for the bistable and oscillatory regimes separately.
The quantities of interest for the bistable regime are the steady state values of $x_2$ and $x_3$ (the values at $t=30$ min). 
Those for the oscillatory regime are the limit cycle amplitude and limit cycle period (computed following Section~\ref{methods:limit-cycle}).
{In the computations, we allowed the biological model parameters to vary uniformly over the ranges listed in Supplemental Table~\ref{tab:mapk-params} with $5{,}000$ samples for the bistable cases and $15{,}000$ samples for the limit cycle regime.}
Figures~\subpanelref{fig:mapk-intro}{C}~and~\subpanelref{fig:mapk-intro}{D} show the computed Sobol sensitivity indices for the parameters ranked by decreasing sensitivity index.
We selected the parameters with the \textit{greatest} sensitivity indices, that is, the parameters to which the output quantities of interest are \textit{most} sensitive.
For the bistable case, we selected the four most sensitive parameters, which were $k_2$, $k_4$, $k_5$, and $k_6$.
For the oscillatory case, we selected the parameters with a first-order sensitivity index $S_i$ greater than $10^{-3}$, which were $k_2$, $k_4$, $k_6$, and $\alpha$.
All remaining biological model parameters were fixed to the nominal values listed in Supplemental Table~\ref{tab:mapk-params}.
Sensitivity analysis highlighted that $k_2$, $k_4$, and $k_6$ are important in predicting both dynamical regimes.
Meanwhile, parameters such as $k_5$ and $\alpha$ are only important for the bistable and oscillatory regimes, respectively.

After identifiability and sensitivity analyses, we applied the CIUKF-MCMC method to estimate model parameters that predict the correct steady state in the bistable case.
To simulate noisy experimental data, we generated two synthetic datasets (see Section~\ref{method:synthetic-data}), sampled from the high steady state and the low steady state (circle and square markers in Fig~\subpanelref{fig:mapk-intro}{B}).
Each dataset had $30$ full-state measurements evenly spaced over $0 \leq t \leq 30 \ (\text{min})$ with measurement noise covariances set to $2.5 \%$ of the variances of the true trajectories.
The prior assumptions were specified according to Section~\ref{methods:prior} for all model parameters with the bounds listed in Supplemental Table~\ref{tab:mapk-params}.
Using the CIUKF-MCMC algorithm with AIES (Section~\ref{methods:AIES}), we ran an ensemble of $150$ Markov chains with $3{,}500$ samples per chain ($525{,}000$ total samples) for each of the datasets (as shown in Supplemental Fig~\subpanelref{supfig:mapk-bistable-mcmc}{A-B}).
The maximum integrated autocorrelation times (Section~\ref{methods:iact}) were
$120.06$ for the low steady state and $169.03$ for the high steady state, leading us to discard $840$ samples for the low steady state and $1{,}183$ samples for the high steady state as burn-in.

The estimated marginal posterior distributions (solid black line; Fig~\subpanelref{fig:mapk-bistable-results}{A} and Fig~\subpanelref{fig:mapk-bistable-results}{B} for the low and high steady states, respectively) indicated varying levels of uncertainty between the model parameters and across the two steady states.
For example, in both the low and high steady states, the marginal posterior for $k_2$ has most of its probability mass centered around the nominal value (dashed black lines), while that for $k_6$ has probability mass spread over a broader range of the prior support (range of the prior bounds).
Additionally, the MAP points (dotted blue line) for the low steady state closely correspond to the nominal values for all model parameters, whereas there is a significant discrepancy between the MAP and the nominal values for $k_4$, $k_5$ and $k_6$ in the high steady state case.

An ensemble of $30{,}000$ simulations with randomly selected posterior samples (see Section~\ref{methods:ensemble-model}) represented the posterior distributions of the dynamics for both steady states.
For the low steady state, the trajectory evaluated at the MAP point (dotted blue line in Fig~\subpanelref{fig:mapk-bistable-results}{C}) closely matches the true trajectory (dashed black line) and the $95 \%$ credible interval (green shaded region) tightly constrains these trajectories.
However, the trajectory at the MAP point for the high steady state (Fig~\subpanelref{fig:mapk-bistable-results}{D}) reaches the low steady state rather than the high steady state (solid green line).
Furthermore, the $95 \%$ credible interval for the high steady state closely follows the initial transient ($0 \leq t \leq 10 \ (\text{min})$), but it covers both steady states by the end of the simulation, e.g., for $10 \leq t \ (\text{min})$.
The considerable uncertainty and bias in the estimated dynamics of the high steady state are unsurprising, given the uncertainty observed in the marginal posterior distributions.
This comprehensive uncertainty analysis of the bistable MAPK dynamics showed that the presence of multiple steady states makes parameter estimation harder for the same set of parameters and governing equations. 
In particular, the estimation uncertainty is much lower when data from the low steady state is supplied for estimation than when data from the high steady state is used.


\begin{figure}[hb!]
    \centering
    \includegraphics[width=\textwidth]{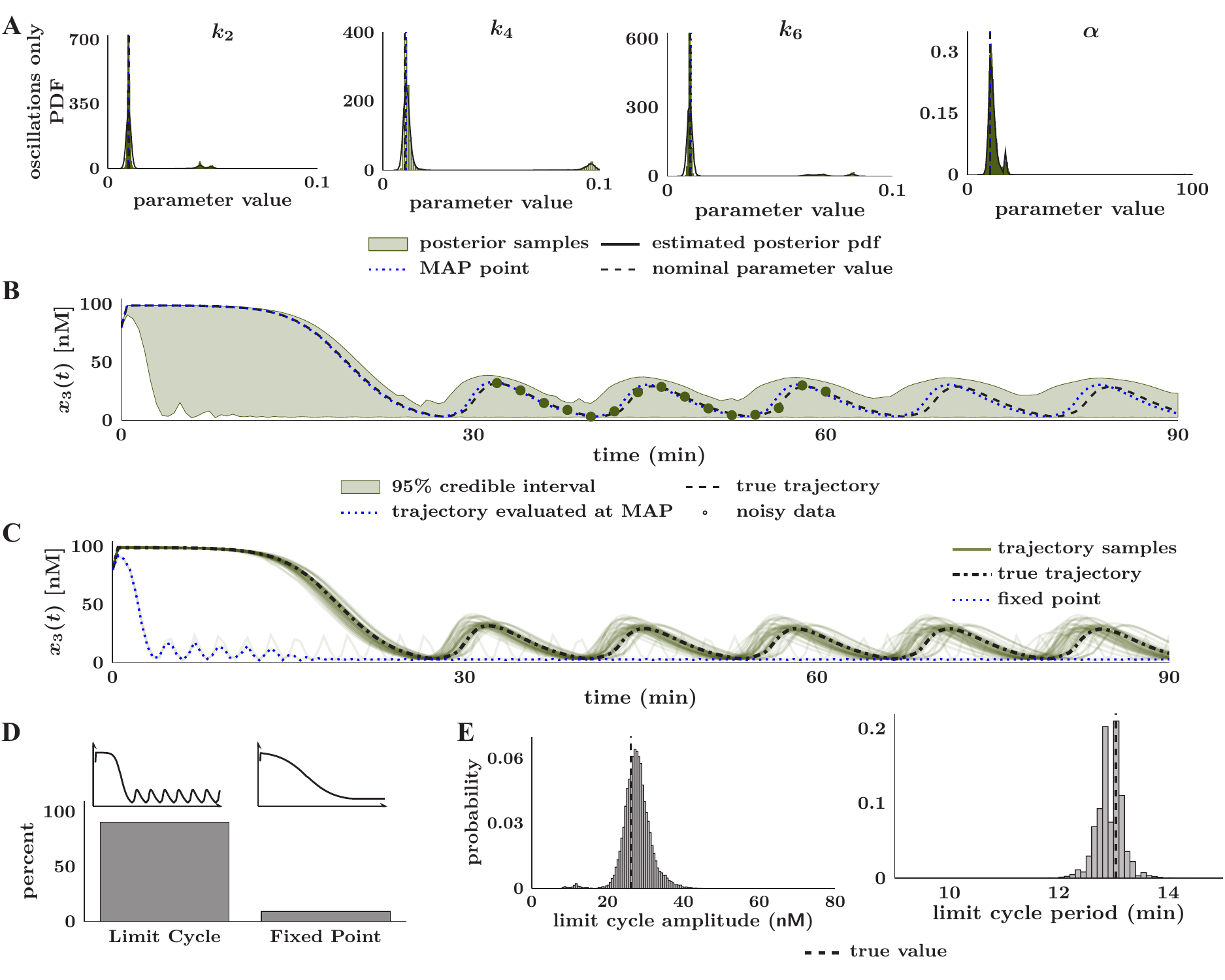}
    \caption{
        Parameter estimation results for the MAPK model in the limit cycle regime with data only sampled from the oscillations. 
        (\textbf{A}) Marginal posterior distributions of the model parameters. 
        Distributions are visualized by fitting a kernel density estimator to $1{,}305{,}720$ ($30$ walkers with $43{,}524$ steps each) MCMC samples obtained using CIUKF-MCMC with the affine invariant ensemble sampler (AIES) after discarding the first $7{,}447$ samples per walker as burn-in. 
        (\textbf{B}) Posterior distribution of the trajectory of $x_3(t)$ in the limit cycle regime. The true trajectory (dashed black line) shows the dynamics with the nominal parameters, the dotted blue line shows the trajectory evaluated at the MAP point, and the points show the noisy data (covariance is $1 \%$ of the variance of the true trajectory). The $95 \%$ credible interval shows the region between the $2.5$th and $97.5$th percentiles that contains $95 \%$ of $30{,}000$ posterior trajectories.
        (\textbf{C}) Sample posterior trajectories ($50$ out of $30{,}000$ total) reveal that most trajectories closely match the true limit cycles.
        Additionally, several trajectories that reach a fixed point are shown.
        (\textbf{D}) Quantification of the fraction of the $30{,}000$ sample trajectories that produce limit cycles oscillations, $90.6 \%$ ($27{,}182$ samples), or reach a fixed point, $9.4 \%$ ($2{,}818$ samples).
        (\textbf{E}) Histograms quantify the variability in limit cycle amplitude and period for the $27{,}182$ trajectories that show limit cycle oscillations.
        We define the limit cycle amplitude as the peak-to-peak difference for one oscillation, and the period is the time to complete an oscillation.
        The vertical black lines show these quantities for the true trajectory.
    }
    \label{fig:mapk-osc-results}
\end{figure}

\begin{figure}[htp!]
    \centering
    \includegraphics[width=0.93\textwidth]{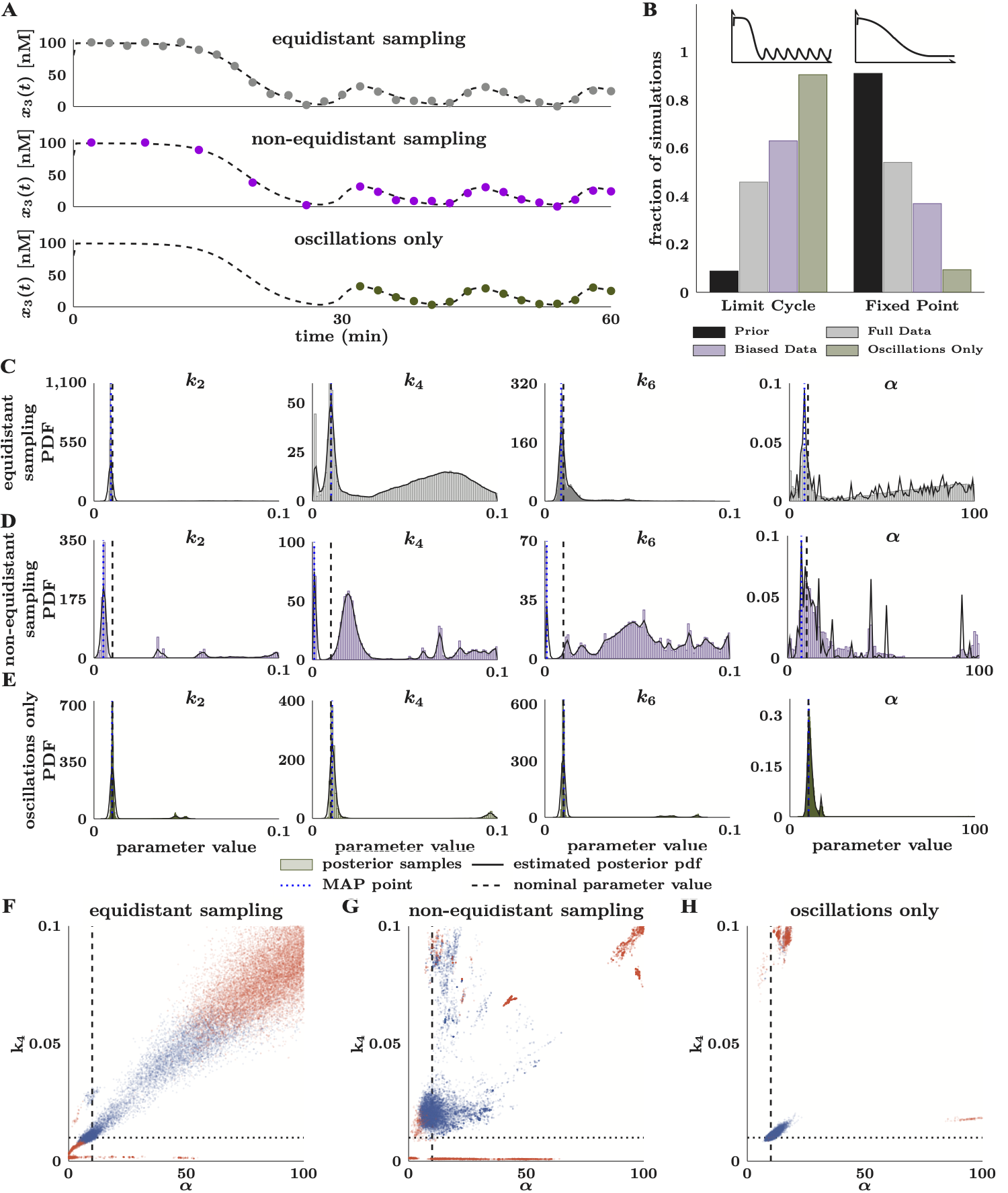}
    \caption{
        Parameter estimation results for the MAPK model in the limit cycle regime with varied sampling strategies.
        (\textbf{A}) The \textit{equidistant sampling} data includes $30$ samples taken every $2$ minutes over $0 < t \leq 60$ (min).
        The \textit{non-equidistant sampling} data includes $20$ total samples with two sampling rates; there are $5$ samples taken every $5$ minute for the first 30 minutes and $15$ samples taken every $2$ minutes for 30 additional minutes.
        The \textit{oscillations only} data set only includes samples from the oscillations with $15$ samples taken every two minutes over the interval $t \in (30, 60]$.
        (\textbf{B})~The fraction of the $30{,}000$ simulations that yield limit cycle or fixed point trajectories, with parameter samples from the prior, and the posterior distribution associated with each data set.
        (\textbf{C}--\textbf{E}) Marginal posterior distributions of the model parameters. Distributions are visualized by fitting a kernel density estimator to $2{,}524{,}800$ samples for the equidistant sampling data, $763{,}080$ for the non-equidistant sampling data, and $1{,}305{,}720$ for the oscillations only data.
        (\textbf{F}--\textbf{H}) Two-dimensional scatter plots reveal relationships between $k_4$ and $\alpha$ that are necessary to produce limit oscillations.
        Simulations with blue points produce limit cycle oscillations, and those with red points produce fixed points.
        Darker regions indicate a higher probability of observing the corresponding parameter values.
    }
    \label{fig:mapk-osc-dataCompare}
\end{figure}

Next, we used CIUKF-MCMC to estimate posterior distributions for the reduced set of model parameters that predict limit cycle oscillations in $x_3(t)$.
Synthetic data with $15$ samples evenly spaced over $30 \leq t \leq 60 \ (\text{min})$ simulated noisy measurements from the oscillating trajectory at a noise level of $1 \%$ of the variance of the true trajectory (circle markers in Fig~\subpanelref{fig:mapk-osc-dataCompare}{A}).
We refer to this data as the \textit{oscillations only} data, because it only includes samples from the limit cycle and not from the initial transient. 
Using CIUKF-MCMC with AIES, we ran $30$ Markov chains (shown in Supplemental Fig~\ref{supfig:mapk-osc-only-mcmc}) with $51{,}000$ steps per chain to sample the parameter posterior distributions.
We discarded $7{,}477$ samples per chain, seven times the integrated autocorrelation time of $1{,}068$, to account for burn-in.
The marginal posterior distributions estimated from the remaining posterior samples (Fig~\subpanelref{fig:mapk-osc-results}{A}) tightly concentrate the probability around the nominal parameter values.
This high level of certainty in the parameter values leads to a posterior distribution of $x_3(t)$ (represented with an ensemble of $30{,}000$ simulations in Figure~\subpanelref{fig:mapk-osc-results}{B}) that closely follows the true limit cycle oscillations (see Supplemental Fig~\subpanelref{supfig:mapk-x1x2-post}{C} for those of $x_1(t)$ and $x_2(t)$).
Additionally, the trajectory evaluated at the MAP point (dotted blue line) follows the true trajectory (dashed black line) and shows very similar oscillations.

Closer examination of a subset of $50$ out of the $30{,}000$ posterior trajectories (green lines in Fig~\subpanelref{fig:mapk-osc-results}{C}) revealed that the $95 \%$ credible interval includes many limit cycles that are similar to the true trajectory (dashed black lines), a smaller number of limit cycles that do not match the true oscillations, and yet an even smaller number of trajectories that reach fixed points (for example, the dotted blue line in Fig~\subpanelref{fig:mapk-osc-results}{C}).
We further leveraged the posterior samples to quantify the variability in the predicted limit cycles.
First, we characterized all of the trajectories in the ensemble (Fig~\subpanelref{fig:mapk-osc-results}{D}) and found that $90.6 \%$ ($27{,}182$ samples) correctly produce limit cycle oscillations while $9.6 \%$ ($2{,}818$ samples) reach a fixed point.
Quantification of the limit cycle amplitude and period for each of the $27{,}182$ oscillating trajectories showed that despite the small uncertainties in model parameters, we still observe variability in the characteristics of predicted limit cycles.
We find that the most probable limit cycle amplitude and period (see histograms in Fig~\subpanelref{fig:mapk-osc-results}{E}) are close to the true values ($26.12$ for the amplitude and $13.04$ for the period; indicated by vertical dashed black lines).
However, we still observe variability in these quantities despite the high level of certainty in the parameter estimates.
Specifically, we find that the limit cycle amplitude often deviates by up to $30 \%$ (10 nM) from the true amplitude.
While the period deviates by a smaller amount, up to $15 \%$ (2 min), we find that its distribution is skewed to periods that are shorter than the true period.
Based on these findings, we conclude that small uncertainties in the model parameters result in limit cycles with varied amplitudes and periods, but do not strongly affect our ability to predict oscillatory dynamics.

Next, we investigated whether the nature of the data supplied for estimation affects our ability to predict limit cycle oscillations.
To investigate this, we repeated parameter estimation with two additional noisy limit cycle data sets (shown in Fig~\subpanelref{fig:mapk-osc-dataCompare}{A} with the noise level set to $1 \%$ of the variance of the true trajectory).
We refer to these new data as the \textit{equidistant sampling} data and the \textit{non-equidistant sampling} data.
The equidistant sampling data includes $20$ samples taken every two minutes over $0 < t \leq 60$ minutes and covers the initial transient and several periods of oscillations (see Fig~\subpanelref{fig:mapk-osc-dataCompare}{A}). 
The non-equidistant sampling data also covers the first 60 minutes, however it only has five samples taken every five minutes for the initial transient (the first 30 minutes) and the same $15$ samples taken every two minutes for the remaining $30$ minutes ($20$ samples in total).
We compared this to the results for the oscillations only data (Fig~\ref{fig:mapk-osc-results}) to investigate how samples from the initial transient affects estimation.
We performed MCMC with $150$ walkers for $22{,}500$ steps for the equidistant sampling data (Supplemental Fig~\ref{supfig:mapk-osc-equidistant-mcmc}) and $30$ AIES walkers for $36{,}500$ steps for the non-equidistant sampling data (Supplemental Fig~\ref{supfig:mapk-osc-nonequidistant-mcmc}).
To account for the MCMC burn in, we discarded the first $5{,}669$ samples per chain for the equidistant data and $11{,}065$ samples for the non-equidistant data.
Additionally, we simulated ensembles of $30{,}000$ trajectories with random posterior samples for the two new data sets; Supplemental Fig~\ref{supfig:mapk-newData-dyn-post} shows the resulting posterior distributions of the dynamics.


Comparison of the fraction of the $30{,}000$ simulated trajectories that show oscillations clearly indicates that the input data affects our ability to predict limit cycles (see Fig~\subpanelref{fig:mapk-osc-dataCompare}{B}).
As a baseline, we compare the three data sets to simulations with samples from the prior distribution, which we think of as the setting where no data are provided.
Figure~\subpanelref{fig:mapk-osc-dataCompare}{B} shows that all cases with data predict a higher faction of limit cycle trajectories than the prior samples, where only $ 17.6 \%$ of simulations oscillated.
We found that only $45.9 \%$ ($13{,}794$ samples) of the simulations corresponding to the equidistant sampling data show sustained oscillations.
However as the number of samples from the initial transient (data from $0$ to $30$ minutes) is decreased, the fraction of oscillating simulations increased; $63.1 \%$ ($18.921$ samples) and $90.6 \%$ ($27{,}182$ samples) of the simulations corresponding to the non-equidistant sampling and the oscillation only data, respectively, showed limit cycle oscillations.

We additionally found that uncertainty in the nature of the dynamics is closely tied to the uncertainty in the model parameters.
Figure~\subpanelref{fig:mapk-osc-dataCompare}{C-E} shows the marginal posterior distributions corresponding to the equidistant sampling, non-equidistant sampling, and oscillations only data, respectively.
Specifically, we observed that we can always predict $k_2$ with high certainty, but the amount of uncertainty in $k_4$, $k_6$ and $\alpha$ varies between the three data sets. 
For example, for the equidistant sampling data, the marginal distribution for $k_4$ (Fig~\subpanelref{fig:mapk-osc-dataCompare}{C}) has three distinct peaks of high probability (modes); one concentrated at very small values, another centered near the nominal value and a wider mode at higher values.
The $k_4$ marginal distribution for the non-equidistant sampling data in Fig~\subpanelref{fig:mapk-osc-dataCompare}{D} places less probability at the two modes far from the nominal value, but the mode near the nominal value is shifted towards higher values of $k_4$.
Lastly, for the oscillations only data, we see only one prominent mode for $k_4$ that is very close to the nominal value (Fig~\subpanelref{fig:mapk-osc-dataCompare}{E}).
Similar trends are seen for the $\alpha$ marginal distributions.
The posterior distribution of dynamics in Supplemental Fig~\ref{supfig:mapk-newData-dyn-post} and the distributions of the limit cycle characteristics in Supplemental Fig~\ref{supfig:mapk-newData-limit-cycle} show that the increased uncertainty in these model parameters increases variability in the predicted dynamics.
In cases with high uncertainty in the model parameters, the $95 \%$ credible intervals for $x_3(t)$, for example in Supplemental Fig~\subpanelref{supfig:mapk-newData-dyn-post}{A}, do not tightly constrain the dynamics, and the limit cycle characteristics vary more (Supplemental Fig~\subpanelref{supfig:mapk-newData-limit-cycle}{A}).

Lastly, we used the results of our Bayesian analysis to discover relationships between parameter pairs and the nature of the corresponding predicted dynamics.
Based on observations that increased uncertainty in both $k_4$ and $\alpha$ results in higher fractions of incorrectly predicted fixed point trajectories and prior knowledge of how the model parameters interact to control the interplay of positive and negative feedback~\cite{Nguyen2015dyvipac}, we hypothesized that we might observe correlations between pairs of model parameters and the predicted dynamics.
To test this, we plotted two-dimensional scatter plots of all pairs of parameters for each data set in Supplemental Fig~\ref{supfig:mapk-oscOnly-scatter}, Supplemental Fig~\ref{supfig:mapk-equidistant-scatter}, and Supplemental Fig~\ref{supfig:mapk-non-equidistant-scatter}, and for $k_4$ and $\alpha$ in Fig~\subpanelref{fig:mapk-osc-dataCompare}{F--H}.
The points are colored blue if the corresponding trajectory produced a limit cycle and red if the trajectory reached a fixed point.
We observed the strongest correlation between $k_4$ and $\alpha$, where, for the equidistant sampling data, (see Fig~\subpanelref{fig:mapk-osc-dataCompare}{F}), the values of both parameters needed to be above the smallest values, but not too large, to produce oscillations.
The plots for the non-equidistant sampling data (Fig~\subpanelref{fig:mapk-osc-dataCompare}{G}) and oscillations only data (Fig~\subpanelref{fig:mapk-osc-dataCompare}{H}) further highlight the relationship between $k_4$ and $\alpha$.
Based on these results, we conclude that $k_4$ and $\alpha$ are correlated such that their values must be within specific ranges to produce limit cycles.
We note that we were able to discover this correlation through a Bayesian analysis; while, it is visually most apparent for the equidistant sampling data due to the high estimation uncertainty, it is a property of the model itself and not of the data.

In summary, comprehensive UQ for the MAPK model highlighted how the existence of multistability introduces additional uncertainties into parameter estimation.
Specifically, sensitivity analysis identified two parameters of the MAPK model, $k_5$ and $\alpha$, that were only important for the bistable dynamics and the oscillatory dynamics, respectively.
In both cases, Bayesian parameter estimation was able to predict the correct type of dynamics but showed remaining uncertainty in the specific characteristics of the dynamics.
By varying the data supplied for estimation, we found that the nature of the data set provided to CIUKF-MCMC affects our ability to predict limit cycles.
Interestingly, we found that more data does not always improve estimation and that having less data that is more focused on the limit cycles reduced estimation uncertainties and gave better posterior parameter distributions.
Additionally, we discovered that the values of both $k_4$ and $\alpha$ need to be within a certain range to predict oscillatory MAPK dynamics.
Overall, the proposed framework for comprehensive UQ found parameters that were important to each dynamical regime and directly quantified how uncertainties in these parameters contributes to uncertainty in the dynamics.

\subsection{Parameter estimation in a phenomenological model of long-term potentiation/depression} \label{results:kinase-phosphatase}

A phenomenological model of coupled kinase and phosphatase switches whose activities affect the level of membrane-bound AMPAR (alpha-amino-3-hydroxy-5-methyl-4-isoxazolepropionic acid receptor) as a reporter of synaptic plasticity was proposed in~\cite{Pi2008coupled} to capture the key events in synaptic plasticity.
This kinase-phosphatase model has three states $pK(t)$, $P(t)$, and $A(t)$ that correspond to active forms of kinase (CaMKII in~\cite{Pi2008coupled}), phosphatase (PP2A in~\cite{Pi2008coupled}), and membrane-bound AMPAR, respectively.
The differential equations are
\begin{linenomath*}
    \begin{subequations} \begin{align}\label{eq:kinase-phosphatase}
	    \frac{{\rm d} pK(t)}{{\rm d}t} &=
	    \begin{aligned}[t]
			& k_1 \cdot pK(t) \cdot \frac{K_{\text{tot}} - pK(t)}{K_{m_1} + \left(K_{\text{tot}} - pK(t)\right)} - k_2 \cdot \left(P(t) + P_0\right) \cdot \frac{pK(t)}{K_{m_2} + pK(t)} \\
			&\quad + k_3 \cdot K_0 + k_4 \cdot \left(K_{\text{tot}} - pK(t)\right) \cdot \frac{Ca^{2+}(t)^4}{K_{m_3}^4 + Ca^{2+}(t)^4} 
		\end{aligned} \\
	    \frac{{\rm d} P(t)}{{\rm d}t} &= 
	    \begin{aligned}[t]
	        & k_5 \cdot P(t) \frac{P_{\text{tot}} - P(t)}{K_{m_4} + \left(P_{\text{tot}} - P(t)\right)} - k_6 \cdot \left(pK(t) + K_0 \right) \cdot \frac{P(t) }{K_{m_5} + P(t)} \\
		    &\quad + k_7 \cdot P_0 - k_8 \cdot \left(P_{\text{tot}} - P(t)\right) \cdot \frac{Ca^{2+}(t)^3}{K_{m_3}^3 + Ca^{2+}(t)^3}
	    \end{aligned} \\
	 \frac{{\rm d} A(t)}{{\rm d}t} &= \left(c_1 \cdot pK(t) + c_3\right) \cdot \left(A_{\text{tot}} - A(t)\right) - \left(c_2  \cdot P(t) + c_4 \right) \cdot A(t),
    \end{align} \end{subequations}
\end{linenomath*}
where the 24 biological model parameters are $\btheta_f = [k_1$, $k_2$, $k_3$, $k_4$, $k_5$, $k_6$, $k_7$, $k_8$, $K_{m_1}$, $K_{m_2}$, $K_{m_3}$, $K_{m_4}$, $K_{m_5}$, $K_0$, $P_0$, $K_{\text{tot}}$, $P_{\text{tot}}$, $A_{\text{tot}}$, $c_1$, $c_2$, $c_3$, $c_4]\top$.
The nominal values and physiological ranges for these parameters are listed in Supplemental Table~\ref{tab:kinphos-params}.

\begin{figure}[ht!]
    \centering
    \includegraphics[width=\textwidth]{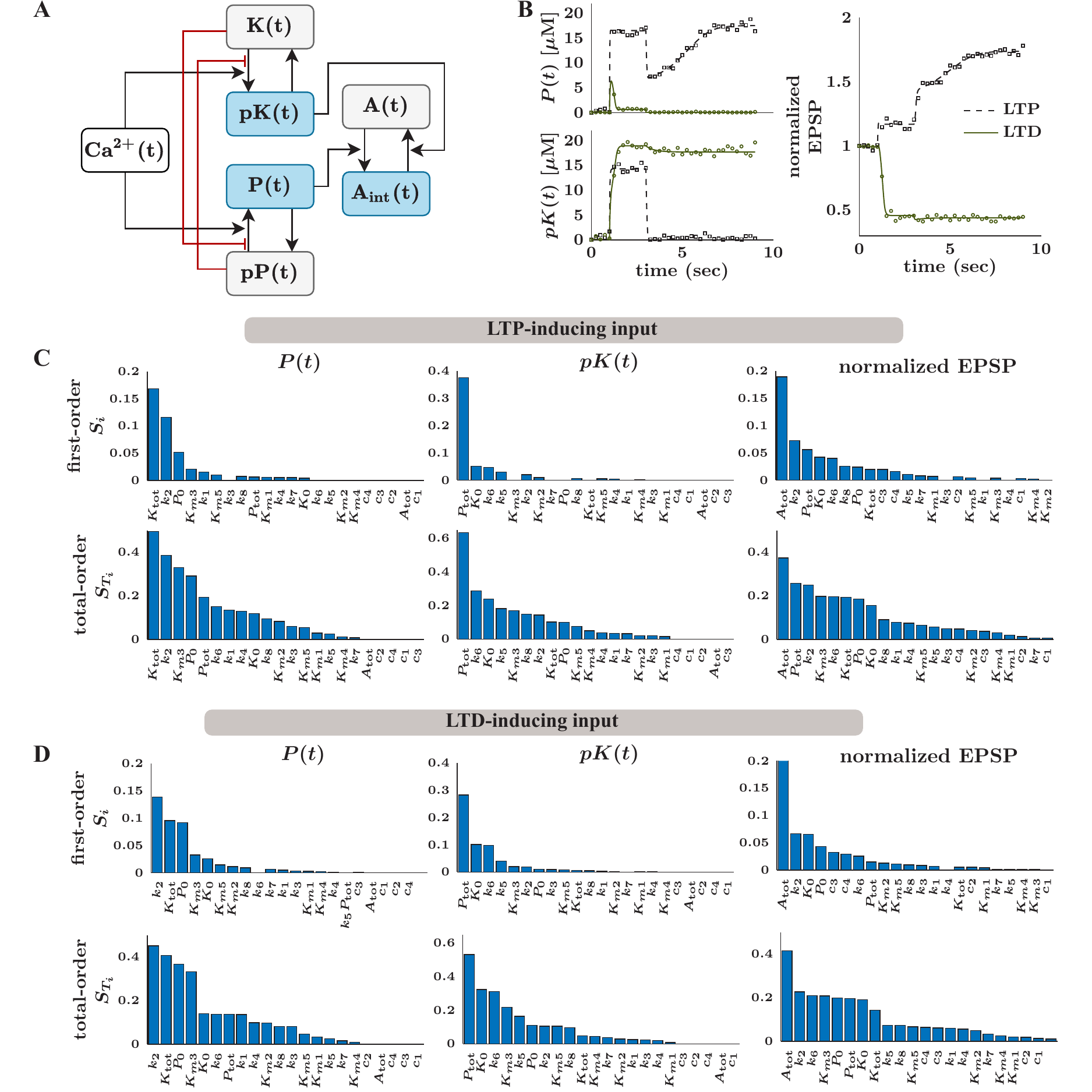}
    \caption{
        Parameter estimation for a coupled kinase-phosphatase switch for long-term potentiation and long-term depression in neurons as a function of calcium input.
        (\textbf{A}) Network diagram of the simplified coupled kinase-phosphatase signaling model where calcium $Ca^{2+}(t)$ acts as the input.
        (\textbf{B}) Trajectories of the three state variables in response to long-term potentiation (LTP; pulse of $Ca^{2+}(t) \equiv  4.0 \ [\mu\text{M}]$ from $2 \leq t \leq 3 \ (\text{sec})$) and long-term depression (LTD; pulse of $Ca^{2+}(t) \equiv 2.2 \ [\mu\text{M}]$ from $2 \leq t \leq 3 (\text{sec})$) inducing calcium inputs.
        The calcium level is set to a baseline of $Ca^{2+}(t) \equiv 0.1 \ [\mu\text{M}]$ before and after stimulus.
        We compute normalized EPSP by normalizing $A(t)$ to its initial condition as described in~\cite{Pi2008coupled}.
        The synthetic noisy data for the LTP and LTD cases are indicated by the black square and green circle marks, respectively, with the noise covariance equal to $1 \%$ of the variance of the data.
        (\textbf{C} and \textbf{D}) Sobol sensitivity indices for all free model parameters in response to LTP-inducing and LTD-inducing inputs, respectively. 
        The quantities of interest are the steady state values of each state variable.
        We show both the first-order sensitivity indices $S_i$ and the total-order indices $S_{T_i}$.
        We select a reduced set of free parameters by choosing the parameters whose first-order sensitivity index is greater than $0.05$, e.g., $S_i > 0.05$.
        This gives us the same set of free parameters, $\btheta_f = [k_2$, $k_6$, $K_0$, $P_0$, $K_{\text{tot}}$, $P_{\text{tot}}$, $A_{\text{tot}}]\top$, for both the LTP and LTD cases.
       Remaining model parameters are fixed to the nominal values in Supplemental Table~\ref{tab:kinphos-params}.
    }
    \label{fig:kinpho-intro}
\end{figure}

This model predicts tristability (three steady states) in the level of excitatory postsynaptic potential (EPSP) as a function of the calcium $Ca^{2+}(t)$ input.
The normalized EPSP is the membrane-bound AMPAR $A(t)$ level, normalized to the initial condition, e.g., $\text{normalized EPSP} = A(t) / A(t=0)$, as defined in~\cite{Pi2008coupled}.
Fig~\subpanelref{fig:kinpho-intro}{B} shows simulations of the three expected responses with the nominal parameter values from~\cite{Pi2008coupled}.
The initial condition $\bx_0 = [0.0228$, $0.0017$, $0.4294]\top$ used for all simulations was determined by allowing the system to reach steady state with the baseline $Ca^{2+}(t)$ level of $Ca^{2+}(t) \equiv 0.1 \ [\mu\text{M}]\top$.
The three steady states are the initial baseline, the higher long-term potentiation state (LTP; trajectory depicted in dashed black Fig~\subpanelref{fig:kinpho-intro}{B}), and the lower long-term depression state (LTD;  trajectory depicted in solid green Fig~\subpanelref{fig:kinpho-intro}{B}).
The LTP state is obtained by applying a constant stimulus of $Ca^{2+}(t) \equiv 4.0 \ [\mu\text{M}]$  from $1 \leq t \leq 3 \ (\text{sec})$, while the LTD state is reached by applying a constant stimulus of $Ca^{2+}(t) \equiv 2.2 \ [\mu\text{M}]$ in the same time interval; $Ca^{2+}(t)$ is set to the baseline level before ($t < 1$ (sec)) and after ($ t > 3$ (sec)) the stimulus is applied.
We investigated how well the proposed uncertainty quantification framework could estimate the model parameters for LTP and LTD from synthetic data of an LTP-inducing calcium input.


Following the proposed framework, the parameter space is reduced by performing identifiability and sensitivity analysis.
First, identifiability analysis showed that all model parameters are globally structurally identifiable from full-state measurements.
Next, global sensitivity analysis of the steady states in response to an LTP-inducing and an LTD-inducing input ranked the 22 globally identifiable model parameters. 
All free parameters were uniformly varied over two orders of magnitude centered around the nominal values, $\theta^*_i$, for each parameter, e.g. $\theta_i \sim \cU(0.1\cdot\theta^*_i, \ 10\cdot\theta^*_i)$.
In order to maintain conservation of mass, the lower bounds of the total concentration parameters, $K_{\text{tot}}$, $P_{\text{tot}}$, $A_{\text{tot}}$, were chosen to be greater than or equal to the initial condition.
Figure~\subpanelref{fig:kinpho-intro}{C} shows the computed Sobol sensitivity indices for the LTP-inducing input and Fig~\subpanelref{fig:kinpho-intro}{D} shows those for the LTD-inducing input.
The sensitivity analyses point to the same group of seven model parameters for both the LTP-inducing and LTD-inducing inputs.
These are $\btheta_f = [k_2$, $k_6$, $K_0$, $P_0$, $K_{\text{tot}}$, $P_{\text{tot}}$, $A_{\text{tot}}]\top$, whose first-order indices were greater than $0.05$, e.g., $S_i > 0.05$.
We chose to estimate these seven parameters and fix the remaining model parameters to the nominal values listed in Supplemental Table~\ref{tab:kinphos-params}.

Next, we used the CIUKF-MCMC algorithm with AIES to estimate the posterior distribution for the reduced set of parameters.
The parameters were estimated from noisy synthetic data with $36$ full-state measurements of the LTP response.
The data are spread uniformly over the domain $0 \leq t \leq 9 \ (\text{sec})$ at a noise level of $1 \%$ of the variance of the true trajectory for the respective states.
The maximum integrated autocorrelation time of the ensemble $150$ Markov chains with $8{,}000$ steps per chain was $621.58$, leading us to discard $4{,}351$ samples as burn-in.
Traces of the ensemble of Markov chains for all parameters are shown in Supplemental Fig~\ref{supfig:kinpho-mcmc}.
Figure~\subpanelref{fig:kinpho-results}{A} shows the estimated marginal posterior distributions for the free model parameters, $k_2$, $k_6$, $P_0$, $K_0$, $P_{\text{tot}}$, $K_{\text{tot}}$, $A_{\text{tot}}$.
We observed different levels of uncertainty across the estimated parameters.
For instance, the marginal posterior for $A_{\text{tot}}$ indicated a very high level of certainty with almost all of the probability mass, and thus the MAP point, aligned to the nominal value.
However, the marginal posterior distributions for $P_0$ and $K_0$ show much more significant uncertainty because we observe posterior probability spread over the entire support of the prior.
Additionally, the marginal posterior distributions for $k_2$, $k_6$, $P_{\text{tot}}$, and $K_{\text{tot}}$ show a large mode around the MAP point that is shifted from the nominal value and a smaller mode at the nominal value.

Using the posterior samples, an ensemble of $30{,}000$ simulations (see Section~\ref{methods:ensemble-model}) represented the posterior distribution of the predicted dynamics in response to an LTP-inducing input, as shown in Fig~\subpanelref{fig:kinpho-results}{B}.
We observed that the $95 \%$ credible interval for the normalized EPSP covered both LTP (normalized EPSP $> 1$) and LTD (normalized EPSP $< 1$) responses even though the input was LTP-inducing.
However, the trajectory evaluated at the MAP point (dotted blue line) matched the true trajectory (dashed black line), indicating that most trajectories align with the expected LTP response.
Examination of the individual trajectories within the ensemble simulation (Fig~\subpanelref{fig:kinpho-results}{C}, top row) and the normalized EPSP steady state values (Fig~\subpanelref{fig:kinpho-results}{D}, top row) confirmed that there are both LTP (blue traces) and LTD (black traces) responses to the LTP-inducing input.
Specifically, we found that $76.59 \%$ ($22{,}972$ samples) of the responses correctly predict LTP, and $23.41 \%$ ($7{,}024$ samples) of response incorrectly predict LTD (Fig~\subpanelref{fig:kinpho-results}{E}).
Therefore, despite the high-quality data (many measurements and low noise), we still observed substantial uncertainty in the predicted normalized~EPSP.


\begin{figure}[htp!]
    \centering
    \includegraphics[width=\textwidth]{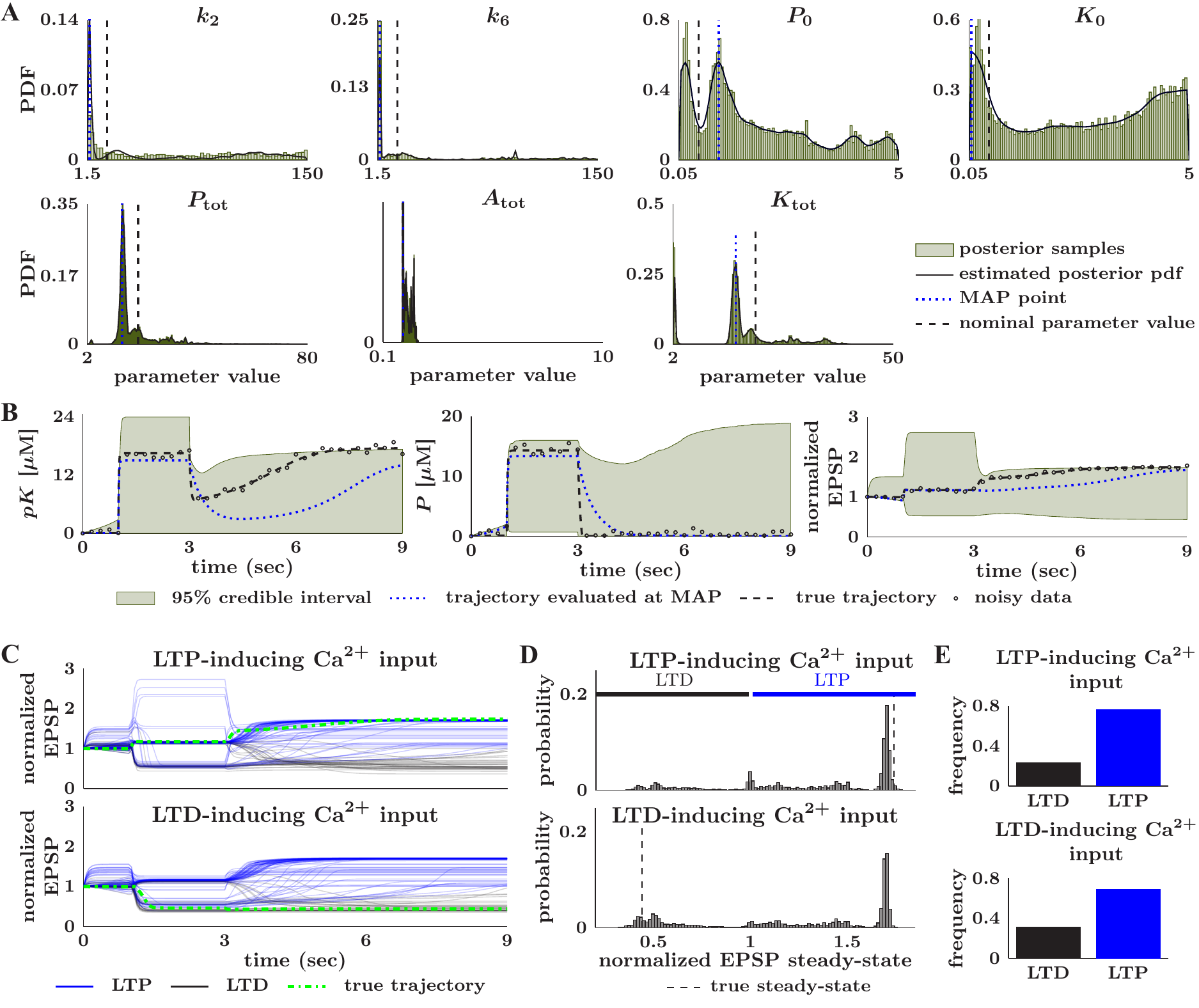}
    \caption{
        Comprehensive parameter estimation and uncertainty quantification reveal failures to predict the correct long-term model behavior.
        (\textbf{A}) We estimated marginal posterior distributions of the model parameters from noisy data with an LTP-inducing calcium input.
        Distributions are visualized by fitting a kernel density estimator to $547{,}500$ ($150$ walkers with $3{,}649$ steps each) MCMC samples obtained using CIUKF-MCMC with AIES for MCMC after discarding the first $4{,}351$ samples per walker as burn-in.
        (\textbf{B}) Posterior distribution of the trajectories of the state variables show LTP (normalized EPSP $> 1$) and LTD (normalized EPSP $< 1$) responses for an LTP inducing input. 
        The true trajectory (dashed black line) shows the dynamics with the nominal parameters, the dotted blue line shows the trajectory evaluated at the MAP point, and the points show the noisy data (covariance is $1 \%$ of the variance of the true trajectory).
        The $95 \%$ credible interval shows the region between the $2.5$th and $97.5$th percentiles that contains $95 \%$ of $30{,}000$ posterior trajectories.
        (\textbf{C}) Sample posterior trajectories ($100$ out of $30{,}000$ total) highlight the LTP (blue lines) and LTD (black lines) in response to the LTP inducing input (top) and the LTD inducing input (bottom).
        The dashed green lines show the true trajectories for the respective calcium inputs.
        (\textbf{D}) Histograms reveal the distribution of the long-term responses to the LTP-inducing input (top) and the LTD-inducing input (bottom).
        The dashed black lines show the true response for the respective calcium inputs.
        (\textbf{E}) Quantifying the percentage of the $30{,}000$ sample trajectories that produce LTD and LTP responses for each calcium input.
        The LTP-inducing input yields $76.59 \%$ ($22{,}972$ samples) of the responses in the LTP state and $23.41 \%$ ($7{,}024$ samples) in the LTD state. 
        The LTD inducing input yields $68.93 \%$ ($20{,}680$ samples) of the responses in the LTP state and $31.07 \%$ ($9{,}320$ samples) in the LTD state. 
    }
    \label{fig:kinpho-results}
\end{figure}

Lastly, we investigated if the posterior distribution estimated in the LTP regime can predict the response to an LTD-inducing input.
An initial hypothesis was that in response to an LTD-inducing input, most trajectories would predict LTD, but a significant subset would predict the incorrect response, LTP.
An additional ensemble of $30{,}000$ simulations, with an LTD-inducing input of $Ca^{2+}(t) \equiv 2.2 \ [\mu \text{M}]$ (pulse from $2 \leq t \leq 3 \ (\text{sec})$) was used to determine the posterior distribution of the dynamics.
Visualization of $100$ of the $30{,}000$ trajectories in Fig~\subpanelref{fig:kinpho-results}{C} (bottom row) again showed both LTP (blue) and LTD (black) responses; however, we unexpectedly observed more LTP than LTD in response to the LTD-inducing input.
The distribution of normalized EPSP responses (bottom row of Fig~\subpanelref{fig:kinpho-results}{D}) confirms that there are a large number of responses in the LTP regime with a minor mode around the correct response.
Quantification of these results highlights that only $31.07 \%$ ($9{,}320$ samples) of responses are in the LTD state (expected response) while $68.93 \%$ ($20{,}680$ samples) of the responses are in the LTP state (unexpected response).
In summary, this model formulation can correctly capture the same LTP behavior over a range of model parameters while losing the ability to predict LTD behavior.

Despite the LTP and LTD responses being sensitive to the same set of parameters (see Fig~\subpanelref{fig:kinpho-intro}{C-D}), the posterior distribution estimated from measurements of LTP places more probability on parameter sets that robustly predict LTP over those that correctly predict both responses.
From these results, we conclude that for this model, we need to learn the model parameters with a high degree of certainty in order to disambiguate the LTP versus the LTD response because sensitivity analyses revealed that the same set of parameters governs these two different outputs. 
This finding highlights that sensitivity analyses are not sufficient to distinguish parameter uncertainty for systems with multistability and a comprehensive framework as outlined here is necessary to shed light on such model complexities.


\section{Discussion} \label{sec:discussion}

In this work, we developed a framework (see Section~\ref{methods:framework}) for comprehensive uncertainty quantification of dynamical models in systems biology.
The proposed framework leverages identifiability and sensitivity analysis to reduce the parameter space (sections~\ref{methods:strucID} and~\ref{methods:gsa}) followed by Bayesian parameter estimation with CIUKF-MCMC (Section~\ref{methods:CIUKF-MCMC}).
We applied this framework to three systems biology models to demonstrate its applicability and highlight how a focus on uncertainty can transform modeling-based studies.
First, we performed two computational experiments on a simple two-state model that showed how noise and data sparsity contribute to estimation uncertainty.
Next, we applied our framework to two models, the MAPK and the synaptic plasticity models, which better resemble the models used to capture biological readouts.
Using these models, we highlight how comprehensive uncertainty quantification enables quantitative analysis of two biologically relevant dynamical behaviors, limit cycles and steady state responses.
We also found that good quality data cannot always overcome uncertainty due to the model structure.
These examples provide an essential starting point for applying our framework in practice and interpreting systems biology studies under uncertainty.

Our results highlight how a focus on uncertainty quantification can give new insights in modeling-based studies.
For example, in Section~\ref{results:mapk} we were able to learn a posterior distribution for the parameters that predict limit cycles with high probability.
The posterior distribution is our \textit{best guess} for the distribution of the model parameters after incorporating the available data into the statistical model.
Therefore, the posterior distribution for the dynamics (approximated via an ensemble simulation as in Section~\ref{methods:ensemble-model}) provides our \textit{best guess} for the dynamics, given everything we know about the model and the data.
For the MAPK limit cycle oscillations, this \textit{best guess} includes dynamics with a range of limit cycle properties.
Additional analysis of the parameters that produce limit cycle oscillations revealed correlations between $k_4$ and $\alpha$ and that their values must both be within a certain range to produce the expected dynamics.
Therefore, incorporating uncertainty quantification into modeling provides additional context and understanding of the system of interest.

We also observe that predictions do not always capture the correct bistability in the example of high MAPK steady state (Section~\ref{results:mapk}) and the LTP/LTD response (Section~\ref{results:kinase-phosphatase}).
In both cases, the $95 \%$  credible intervals of the ensemble of predicted dynamics cover both the higher and lower steady states (LTP and LTD in the synaptic plasticity example).
These results imply that the estimated posterior distributions for these models include parameter sets that no longer show bistability at the specified input (the initial condition for the MAPK model or the calcium level for the synaptic plasticity model).
Further analysis of these systems could test if these parameter sets lead to a complete loss of bistability or merely shift the bifurcation point, the value of the input that changes which steady state is reached.
Overall our results point to a complex interplay between model parameters and inputs that potentially confounds parameter estimation of multistable systems.

As highlighted in Sections~\ref{results:two-comp}~and~\ref{results:mapk}, we showed that estimation uncertainty is a result of the quality, quantity, and nature of the available experimental data.
As expected, we demonstrated, for the simple two state model (Fig~\ref{fig:two-comp},) that data with more noise and fewer samples leads to estimates with wider uncertainty intervals.
Furthermore, for the MAPK oscillations (Fig~\ref{fig:mapk-osc-dataCompare}), we found that the sampling strategy, either including data from the initial transient or focusing on sampling from the oscillations, drastically affects estimation uncertainty.
This finding contradicts the common intuition that \textit{more data is always better}, as we had the smallest estimation uncertainty in predictions made using data that had the fewest samples, the oscillations only data.
We suspect that these observations can be understood by considering that the likelihood function measures the mismatch between the data and estimated trajectory with an $l^2$-like norm, for example see Eq~\eqref{eq:meas-prob}.
When we include data from both the initial transient and oscillations, as in the equidistant and non-equidistant sampling cases in Section~\ref{results:mapk}, the likelihood probability is similar for all trajectories that capture the initial transient, regardless of whether they oscillate or reach a fixed point.
However, the oscillations only data focuses the likelihood on the long-time dynamics and places lower probability on trajectories that reach a fixed point.
While $l^2$-norm-based likelihood functions are often used in Bayesian methods, one can also design a likelihood function to explicitly determine if a trajectory displays the same features of the data, such as the feature-based approach in~\cite{Lunderman2020-nh}.
These findings imply that thoughtful experimental design can reduce unnecessary and potentially costly data collection and can increase certainty in parameter estimates and model predictions.

In the proposed framework, we stress that global identifiability and sensitivity analyses are essential to practical parameter estimation and uncertainty quantification.
By running variations of the proposed framework without these analyses in Supplemental Section~\ref{sup:sup-text}, we found that each analysis reduces estimation uncertainty and leads to an easier estimation problem.
In the complete framework, we isolate globally identifiable parameters and choose a cutoff value of the Sobol sensitivity index (see Section~\ref{methods:gsa}) to select the influential parameter subset.
Throughout this work, we choose a logical sensitivity index threshold that corresponds either to a sharp drop in or a logical value of the sensitivity index, such as the $S_i \geq 0.1$ criterion that we apply in several of the example problems.
However, we caution that selecting the subset of influential parameters is an intricate problem that requires users to decide the acceptable amount of information to discard from the uncertainty analysis~\cite{Wentworth2016-mu, Vittadello2022-yv}.
Alternative methods to the thresholding approach include those discussed in~\cite {Wentworth2016-mu}, group sensitivity analysis~\cite{Saltelli2007-ta} or lasso regression~\cite{Coleman2016-gn}.
While these methods aim to select a subset of influential parameters, they are similar to the thresholding approach in that each method hinges upon determining the acceptable number of degrees of freedom to exclude from the analysis.
Future research should develop methods for fully quantitative parameter subset selection methods.

While the computations remained tractable for the three example models in this paper, it is essential to consider how complete uncertainty quantification can increase the computational costs of parameter estimation.
In~\cite{Galioto2020-gq}, the authors showed that the original UKF-MCMC algorithm required more floating point operations than methods that do not account for all sources of uncertainty.
Similarly, we found (see Supplemental Section~\ref{sup:sup-text-timing}) the a single CIUKF likelihood evaluation takes $81$ times longer than an evaluation of a similar likelihood function that does not account for uncertainty in the model form.
A deeper look into the execution of the CIUKF likelihood function revealed that approximately $60 \%$ of the runtime was spent on calls to \verb!ode15s()!, which we use to discretize the differential equation model.
The added computational cost of CIUKF may be reduced in the future by developing optimized codes for discretizing the differential equation model and evaluating the likelihood function.

Bayesian methods, such as CIUKF-MCMC, enable complete uncertainty quantification; however, the need for MCMC sampling and the associated repeated likelihood evaluations can pose an increased computational expense.
In particular, as models include greater numbers of state variables and parameters, the added expense of simulation, the increased dimensionality, and the potentially complex geometry of the posterior distributions can challenge standard MCMC approaches.
Innovative methods from the broader UQ community, combined with the increasing accessibility of high-performance computing resources, enable Bayesian analysis of models with large numbers of state variables and parameters~\cite{smith2013uncertainty, Blei2017variational, bardsley2014randomize} . 
Additionally, while we showed that identifiability and sensitivity analyses could significantly reduce the dimensionality of the parameter space, this reduced space may remain relatively high dimensional for specific applications.
Many MCMC sampling algorithms that incorporate information about the geometry of the distribution or run parallel Markov chains can sample high-dimensional and possibly complex posterior distributions.
These methods include parallel MCMC samplers~\cite{Shockley2018-rb, Gupta2018-my}, Hamiltonian Monte Carlo~\cite{gelman2013bayesian} and dimension independent, likelihood-informed methods~\cite{Cui2016-pa}.
Although comprehensive uncertainty quantification comes with added computational costs, recent innovations in UQ methods can alleviate some of the added computational burdens of comprehensive analysis.

Throughout this work, we make several assumptions regarding the availability of the model equations and the statistical properties of associated errors.
First, we assume that the model equations are known prior to parameter estimation.
This assumption reflects standard modeling practices where models use biochemical theories that assume equations for the kinetics of biochemical reactions.
In using the CIUKF-MCMC we somewhat weaken this assumption because it introduces process noise to account for uncertainty in the model form~\cite{Galioto2020-gq}.
In reality, all models have some level of uncertainty because they rely on assumptions about the system. 
Therefore, accounting for model form uncertainty regularizes the dynamics to account for a mismatch between the predicted dynamics and the data~\cite{Galioto2020-gq}.
However, in cases where the model is not known \textit{a priori}, it may be necessary to estimate the model structure and parameters from the data simultaneously.
One approach to disambiguate a model structure is to learn the biochemical reaction network~\cite{Aijo2016-ow, Casadiego2017-ke} or the mathematical model directly from data~\cite{Schmidt2009-fp, Brunton2016-vf, Mangan2016-lj, Hoffmann2019-nz, Kaheman2020-pq}.
Additionally, it is possible to cast these problems in the Bayesian perspective to learn the model form and the associated uncertainties~\cite{Galioto2020-gq, Hirsh2021-dt}.

Furthermore, we make several assumptions when choosing statistical models for measurement and process noise and simulating biological measurement data.
First, we assume Gaussian measurement and process noises in applying the CIUKF-MCMC method. 
Standard Kalman filtering approaches revolve around assumptions of normally distributed noise processes~\cite{sarkka2013bayesian}.
However, we may be better able to describe noise in biological systems and experiments with alternate probability distributions (see, e.g.,~\cite{Ghasemi2011estimation}).
Assuming a non-Gaussian distribution for the measurement noise would require reformulating the likelihood function (it would alter the distribution in Eq~\eqref{eq:meas-prob}); however, incorporating alternative distributions for the process noise would require approximations for the marginal likelihood that go beyond Kalman filtering.
Next, this work considers linear measurement functions, e.g., Eq~\eqref{eq:linear-meas}, but CIUKF-MCMC is also well-suited to handle nonlinear measurement functions~\cite{Galioto2020-gq}. 
Lastly, we assume that we only have access to a single time series of measurements, e.g., one trial of an experiment; however, most experiments in biology perform several repeated trials.
A straightforward approach to incorporate all available data to inform the statistical model would use the mean of each time point and estimate parameters from the time series of means.
Additionally, one could estimate parameters from each time series separately and then analyze several posterior distributions that one could merge via meta-analysis or information fusion principles~\cite{gasca2012data, marin2010weighting, clemen1999combining}, or construct a statistical model that accounts for the multiple time series~\cite{gelman2013bayesian} simultaneously.

The framework introduced in this work can be applied and extended to enable comprehensive uncertainty quantification for most dynamical models encountered in systems biology.
Future research should focus on rigorous methods for parameter subset selection, applying MCMC samplers that are well suited for complex, high-dimensional distributions to systems biology, and incorporating more specific statistical models of biological data and noise.
In summary, research at the intersection of uncertainty quantification and systems biology modeling will strengthen parameter estimation and 
enable models that more accurately predict the dynamics observed in experimental measurements.

\section{Code availability statement}
The code necessary to reproduce this work is available on GitHub at \url{https://github.com/RangamaniLabUCSD/CIUKF-MCMC}.
We store all models in the systems biology markup language (SBML) for reproducibility and provide a function to translate these to MATLAB.
Additionally, we provide a tutorial in the GitHub as a starting point for future applications of the proposed uncertainty quantification framework.
In the tutorial, we also list Python and Julia-based software for parameter estimation and sensitivity analysis as alternatives to the MATLAB-based software we use in this study.

\section{Acknowledgements}
The authors would like to thank Miriam Bell, Sage Malingen, Mar\'{i}a Hern\'{a}ndez Mesa, Lingxia Qiao, Mayte Bonilla Quintana, and other members of the Rangamani and Kramer groups for critical discussions and feedback on the manuscript.
Additionaly the authors would like to thank Matthias Morzfeld, Alex Gorodetsky and Nick Galioto for their methodological insights and helpful discussions.
Nathaniel Linden acknowledges support from the National Institute of Biomedical Imaging and Bioengineering (NIBIB) of the National Institutes of Health (NIH) under award number T32EB9380 and a UCSD Sloan Scholar Fellowship.
Padmini Rangamani acknowledges support from Air Force Office of Scientific Research (AFOSR) Multidisciplinary University Research Initiative (MURI) grant FA9550-18-1-0051.


\newpage
\begin{appendix}
	\section{Supplemental Text} 
	
	\subsection{Structural identifiability and global sensitivity analyses are key to successful parameter estimation} \label{sup:sup-text}
	
	We propose an uncertainty quantification framework that applies structural identifiability and global sensitivity analysis to reduce the number of parameters to be estimated.
	This preprocessing is necessary as failure to reduce the parameter set to the identifiable and influential parameters led to a more difficult estimation problem and increased parameter uncertainty.
	To highlight these effects, we varied the preprocessing analysis that we used to reduce the set of free parameters for the MAPK model in the bistable regime.
	In particular, we tested two additional cases; one where we only used structural identifiability analysis to reduce the parameter space and another where we did not perform any preprocessing.
	Based on our previous experience, we predicted that parameter uncertainty would increase as we eliminated each parameter reduction preprocessing step. 
	
	\begin{figure}[ht!]
		\centering
		\includegraphics[width=\textwidth]{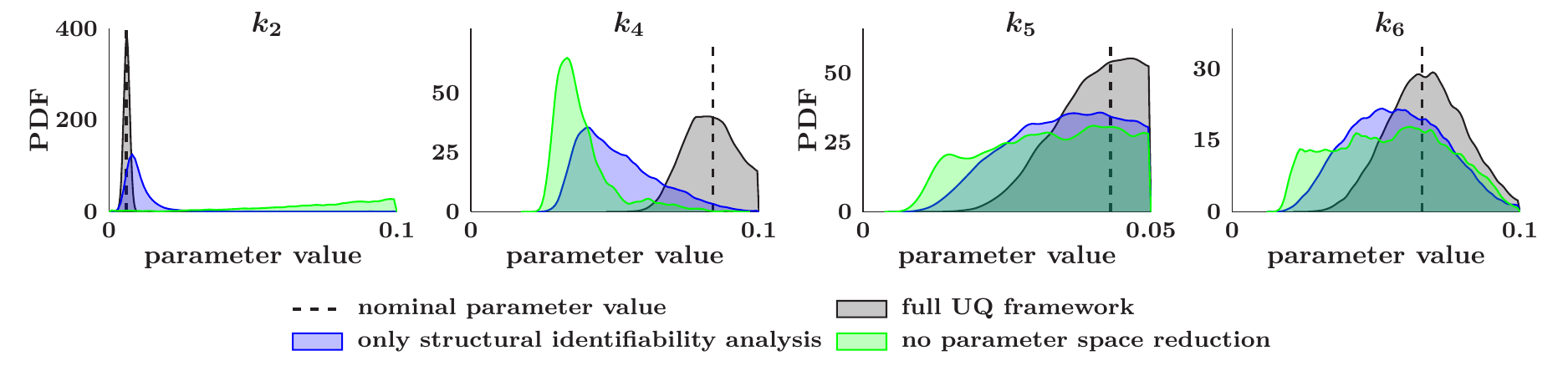}
		\caption{
			Marginal posterior distributions for $k_2$, $k_4$, $k_5$, and $k_6$ highlight the effects of excluding one or both of the structural identifiability and global sensitivity analyses.
			We visualize all distributions by fitting a kernel density estimator to MCMC samples.
			The grey-shaded distributions, labeled ``full UQ framework," are the posterior distributions for the estimated parameters after applying the entire proposed UQ framework to reduce the free parameters to $\btheta_f = [k_2, \ k_4, \ k_5, \ k_6]^\top$.
			The blue-shaded distributions, labeled ``only structural identifiability analysis," are the posterior distributions after using only structural identifiability analysis to reduce the parameter set to $\btheta_f^{\text{no GSA}} = [k_1, \ k_2, \ k_3 \ k_4, \ k_5, \ k_6, \ \alpha]^\top$.
			The green-shaded distributions, labeled ``no parameter space reduction," are the posterior distributions for the full set of free parameters $\btheta_f^{\text{no reduction}} = [k_1, \ k_2, \ k_3 \ k_4, \ k_5, \ k_6, \ K_1, \ K_2, \ \alpha]^\top$.
			We only show the distributions for the four parameters that are common to all three analyses.
			In each case, we used CIUKF-MCMC with AIES to draw $3{,}500$ MCMC samples with $150$ walkers.
			We discarded $840$, $1{,}342$, and $1{,}485$ samples per walker as burn-in for the full framework, structural identifiability only, and no reduction cases, respectively.
		}
		\label{supfig:mapk-bistable-paramReduction}
	\end{figure}
	
	Based on the marginal parameter distributions in Figure~\ref{supfig:mapk-bistable-paramReduction}, we conclude that structural identifiability and global sensitivity analysis are necessary for practical parameter estimation and uncertainty quantification.
	We estimated all parameters from the same data set generated from the MAPK low steady state in the bistable regime (results shown in Fig~\ref{fig:mapk-bistable-results}).
	The posterior distributions of the reduced set of free model parameters $\btheta_f = [k_2, \ k_4, \ k_5, \ k_6]^\top$ that we found by applying the complete proposed framework (black line with grey shading) have peaks well-aligned with the nominal parameter values.
	Repeating the estimation with the set of free parameters $\btheta_f^{\text{no GSA}} = [k_1, \ k_2, \ k_3 \ k_4, \ k_5, \ k_6, \ \alpha]^\top$ that we found by only using structural identifiability analysis for parameter reduction yields estimated posterior distributions (blue line and shading) with much greater uncertainties.
	Here, the peaks of the distributions for $k_2$ and $k_4$ do not align with the nominal values, and all four distributions are wider than in the previous case.
	Lastly, estimation of the full set of free model parameters (no preprocessing to reduce the parameter set) $\btheta_f^{\text{no reduction}} = [k_1, \ k_2, \ k_3 \ k_4, \ k_5, \ k_6, \ K_1, \ K_2, \ \alpha]^\top$ resulted in posterior distributions with even greater uncertainty which indicate lack of MCMC convergence.
	In this case, the posterior distributions place little probability on the nominal values of $k_2$ and $k_4$; thus, the predictions for these parameters would be incorrect. 
	We used AIES for MCMC with $3{,}500$ samples and $150$ walkers for all three cases.
	These standard MCMC settings highlight how reducing the parameter space with structural identifiability and global sensitivity analyses lead to a better-posed estimation problem.
	Overall, each preprocessing step improves MCMC convergence and reduces uncertainty in the estimated parameters.
	
	\subsection{Runtime analysis of the CIUKF-based likelihood function} \label{sup:sup-text-timing}
	
	UKF-MCMC, and thus CIUKF-MCMC, poses an increased computational cost compared to estimation methods that do not consider all sources of uncertainty~\cite{Galioto2020-gq}.
	To better understand how this increased computational burden affects parameter estimation, we used the \verb!timeit()! function in Matlab to time the execution of a CIUKF-based likelihood function and a Gaussian likelihood function Eq~\eqref{eq:guass-like} that ignores model form uncertainty and that is representative of standard approaches~\cite{smith2013uncertainty, Mortlock2021-lk, Ghasemi2011estimation, Wilkinson2007-rt}.
	Specifically, we ran \verb!timeit()! 20 times for each likelihood function evaluated for the MAPK model with the data from the low steady state, seen in Fig~\subpanelref{fig:mapk-bistable-results}{A}.
	We found that, on average, the CIUKF-based likelihood function takes $81$ times longer ($0.278$ seconds) than the Gaussian likelihood ($0.0034$ seconds).
	All timing was performed on a laptop with a 2.6Ghz 6-core Intel Core i7 (Intel Corporation, Santa Clara, CA) processor with 32 GB of RAM.
	
	To better understand what computations contribute to the observed runtimes of the CIUKF-based likelihood, we used the \verb!profiler! in Matlab to analyze a single call to the likelihood function.
	We found that the CIUKF-based likelihood execution spent approximately $60 \%$ of the time on calls to \verb!ode15s()!, which we used to discretize the ordinary differential equation model.
	Additionally, another $30 \%$ of the time was spent running \verb!quadprog()! during the update step of the constrained interval unscented Kalman filter.
	Based on these findings, we conclude that developing better-optimized codes for these steps will likely accelerate the CIUKF-based likelihood function and, thus, shorten MCMC runtimes.
	
	\section{Supplemental Table Captions} \label{sup:tables}
This section provides captions for the tables that included in the supporting information.


\begin{table}[ht!]
    \centering
    \caption{
        Two-compartment model~\cite{Villaverde2019-zc} model parameters and relevant ranges. 
        All listed values have units of one over time. 
    }
    \begin{tabular}{|c|c|c|}
        \hline
        \textbf{Parameter} & \textbf{Nominal Value} & \textbf{Range}  \\
        \hline
        $k_{1e}$ & $1$ & $[0, \ 5]$\\
        $k_{12}$ & $1$ & $[0, \ 5]$\\
        $k_{21}$ & $1$ & $[0, \ 5]$\\
        $b$      & $2$ & $[0, \ 10]$\\
        \hline
    \end{tabular}
    \label{tab:twostate-params}
\end{table}


\begin{table}[ht!]
    \centering
    \caption{
    MAPK model parameters and relevant ranges from~\cite{Nguyen2015dyvipac}.
    Note: For the oscillatory dynamics, the range for $k_5$ is $[1\times10^{-5} \ 0.05]$.
    }
    \begin{tabular}{|c|c|c|c|}
        \hline
        \textbf{Parameter} & \textbf{ Nominal Value: Limit-cycle} & \textbf{Nominal Value: Bistable} & \textbf{Range} \\ \hline
        $S_{1t}$ & $100 \ \textsf{nM}$ & $0.22 \ \textsf{nM}$ & $[0, \ 100]$\\  \hline
        $S_{2t}$ & $100 \ \textsf{nM}$ & $10\ \textsf{nM}$    & $[0, \ 100]$\\  \hline
        $S_{3t}$ & $100 \ \textsf{nM}$ & $53 \ \textsf{nM}$  & $[0, \ 100]$\\  \hline
        $k_{1}$  & $0.1 \ \textsf{nM}\cdot\textsf{s}^{-1}$   & $0.0012 \ \textsf{nM}\cdot\textsf{s}^{-1}$ & $[0, \ 0.05]$\\ \hline
        $k_{2}$  & $0.01 \ \textsf{nM}\cdot\textsf{s}^{-1}$  & $0.006 \ \textsf{nM}\cdot\textsf{s}^{-1}$  & $[0, \ 0.1]$\\  \hline
        $k_{3}$  & $0.01 \ \textsf{nM}\cdot\textsf{s}^{-1}$  & $0.049 \ \textsf{nM}\cdot\textsf{s}^{-1}$  & $[0, \ 0.05]$\\ \hline
        $k_{4}$  & $0.01 \ \textsf{nM}\cdot\textsf{s}^{-1}$  & $0.084 \ \textsf{nM}\cdot\textsf{s}^{-1}$  & $[0, \ 0.1]$\\  \hline
        $k_{5}$  & $0.01 \ \textsf{nM}\cdot\textsf{s}^{-1}$  & $0.043 \ \textsf{nM}\cdot\textsf{s}^{-1}$  & $[0, \ 0.05]$\\ \hline
        $k_{6}$  & $0.01 \ \textsf{nM}\cdot\textsf{s}^{-1}$  & $0.066 \ \textsf{nM}\cdot\textsf{s}^{-1}$  & $[0, \ 0.1]$\\  \hline
        $n_{1}$  & $10$    & $5$      & $[5, \ 10]$\\   \hline
        $K_{1}$  & $1 \ \textsf{nM}$     & $9.5 \ \textsf{nM}$    & $[0, \ 10]$\\   \hline
        $n_{2}$  & $15$    & $10$     & $[5, \ 10]$\\   \hline
        $K_{2}$  & $8 \ \textsf{nM}$     & $15 \ \textsf{nM}$     & $[0, \ 20]$\\   \hline
        $\alpha$ & $10$    & $95$     & $[0, \ 100]$ \\ \hline
    \end{tabular}
    \label{tab:mapk-params}
\end{table}

\begin{table}[ht!]
    \centering
    \caption{
    Initial conditions used for simulations of the MAPK model from~\cite{Nguyen2015dyvipac}. \textcolor{white}{Placeholder. Placeholder. Placeholder. Placeholder. }
    }
    \begin{tabular}{|c|c|c|c|}
        \hline
        \textbf{State-variable} & \textbf{Limit cycle} & \textbf{Bistable: low steady state} & \textbf{Bistable: high steady state} \\ \hline
        $x_1(0)$ & $10 \ \textsf{nM}$ & $0.0015 \ \textsf{nM}$ & $0.1245 \ \textsf{nM}$ \\  \hline
        $x_2(0)$ & $80 \ \textsf{nM}$ & $3.6678 \ \textsf{nM}$    & $2.4870 \ \textsf{nM}$\\  \hline
        $x_3(0)$ & $80 \ \textsf{nM}$ & $28.7307 \ \textsf{nM}$  & $31.2623 \ \textsf{nM}$\\ \hline
    \end{tabular}
    \label{tab:mapk-IC}
\end{table}


\begin{table}[ht!]
    \centering
    \caption{
        Long-term potentiation/depression model parameters from~\cite{Pi2008coupled} and ranges.
        The ranges are given by $[0.1\cdot\theta^*_i, \ 10\cdot\theta^*I]$, where $\theta^*_i$ is the nominal value.
        Note: we do not include ranges for $n_1$ and $n_2$ because these parameters are always set to the nominal values.
    }
    \begin{tabular}{|c|c|c|c|}
        \hline
        \textbf{Parameter} & \textbf{Nominal Value}  & \textbf{Range} \\ \hline
        $k_1$ & 2 \textsf{s}$^{-1}$ & $[0.2, \ 20]$ \\  \hline
        $k_2$ & 15 \textsf{s}$^{-1}$ & $[1.5, \ 150]$ \\  \hline
        $k_3$ & 1 \textsf{s}$^{-1}$ & $[0.1, \ 100]$ \\  \hline
        $k_4$ & 120 \textsf{s}$^{-1}$ & $[12, \ 1200]$ \\  \hline
        $k_5$ & 2 \textsf{s}$^{-1}$ & $[0.2, \ 20]$ \\  \hline
        $k_6$ & 15 \textsf{s}$^{-1}$ & $[1.5, \ 150]$ \\  \hline
        $k_7$ & 1 \textsf{s}$^{-1}$ & $[0.1, \ 10]$ \\  \hline
        $k_8$ & 80 \textsf{s}$^{-1}$ & $[8.0, \ 800]$ \\  \hline
        $c_1$ & 1 & $[0.1, \ 10]$ \\  \hline
        $c_2$ & 1 & $[0.1, \ 10]$ \\  \hline
        $c_3$ & 6 \textsf{s}$^{-1}$ & $[0.6, \ 60]$ \\  \hline
        $c_4$ & 8 \textsf{s}$^{-1}$ & $[0.8, \ 80]$ \\  \hline
        $K_{m1}$ & 10 $\mu$\textsf{M} & $[1.0, \ 100]$ \\  \hline
        $K_{m2}$ & 0.3 $\mu$\textsf{M} & $[0.03, \ 3]$ \\  \hline
        $K_{m3}$ & 4 $\mu$\textsf{M} & $[0.4, \ 40]$ \\  \hline
        $K_{m4}$ & 10 $\mu$\textsf{M} & $[1.0, \ 100]$ \\  \hline
        $K_{m5}$ & 1 $\mu$\textsf{M} & $[0.1, \ 10]$ \\  \hline
        $K_0$ & 0.5 $\mu$\textsf{M} & $[0.05, \ 5]$ \\  \hline
        $P_0$ & 0.5 $\mu$\textsf{M} & $[0.05, \ 5]$ \\  \hline
        $K_{\text{tot}}$ & 20 $\mu$\textsf{M} & $[2, \ 200]$ \\  \hline
        $P_{\text{tot}}$ & 20 $\mu$\textsf{M} & $[2, \ 200]$ \\  \hline
        $A_{\text{tot}}$ & 1 & $[0.1, \ 10]$ \\  \hline
        $n_1$ & 4 & -- \\ \hline
        $n_2$ & 3 & -- \\ \hline
    \end{tabular}
    \label{tab:kinphos-params}
\end{table}
\clearpage
\section{Supplemental Figure Captions }\label{sup:figures}
\begin{figure}[htp!]
    \centering
    \includegraphics[width=\textwidth]{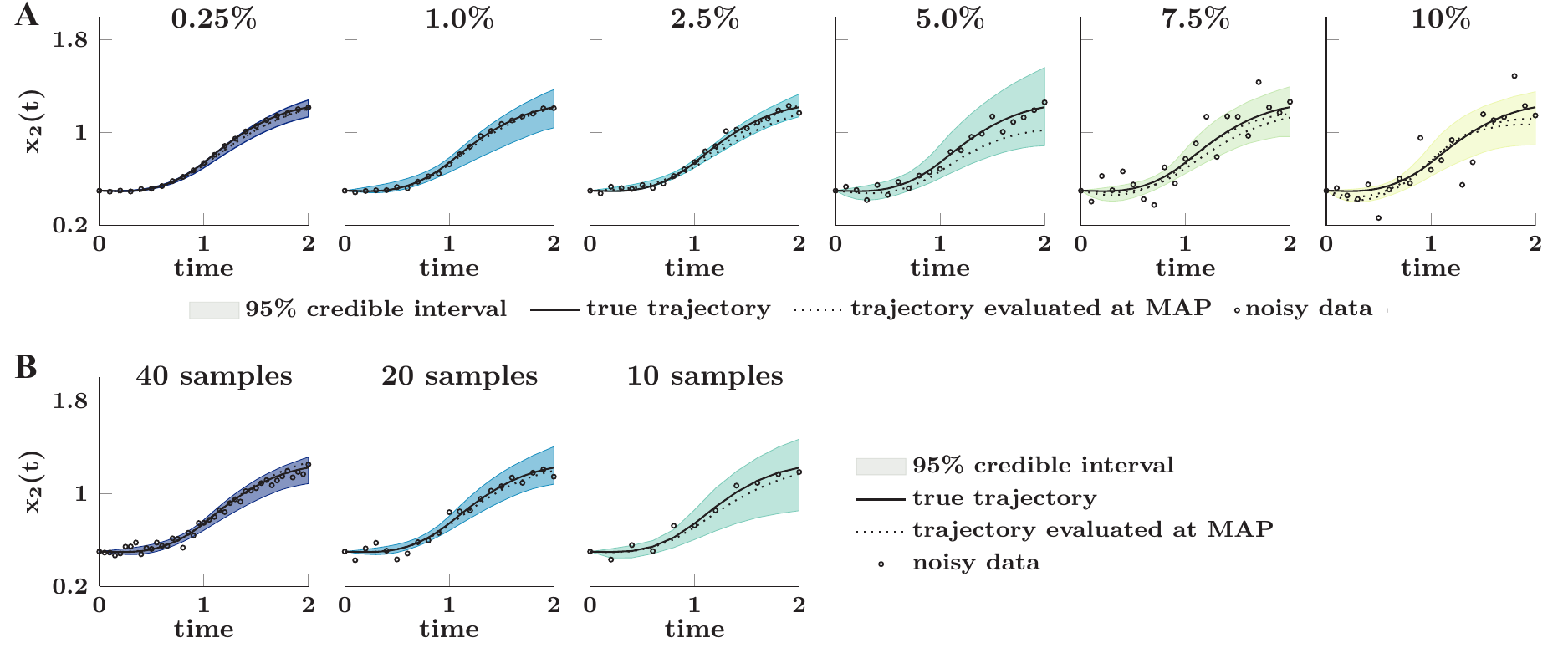}
    \caption{
        Posterior distributions of $x_2(t)$ in the two-state model experiments.
        (\textbf{A}) corresponds to the measurement noise experiment, Fig~\subpanelref{fig:two-comp}{C}.
        We observe that increasing the data noise level increased the uncertainty in the predicted dynamics.
        We control the noise level by setting the noise covariances to the specified percentage of the standard deviation of each state variable.
        The dashed black vertical lines indicate each parameter's nominal (true) value.
        (\textbf{B}) corresponds to the data sparsity experiment, Fig~\subpanelref{fig:two-comp}{E}.
        We observe that decreasing the number of experimental samples supplied for estimation increased estimation uncertainty.
        The noise level in the data was fixed to the $2.5 \%$ level shown above.
    }
    \label{supfig:two-comp-x2dyn}
\end{figure}

\begin{figure}[ht!]
    \centering
    \includegraphics[width=\textwidth]{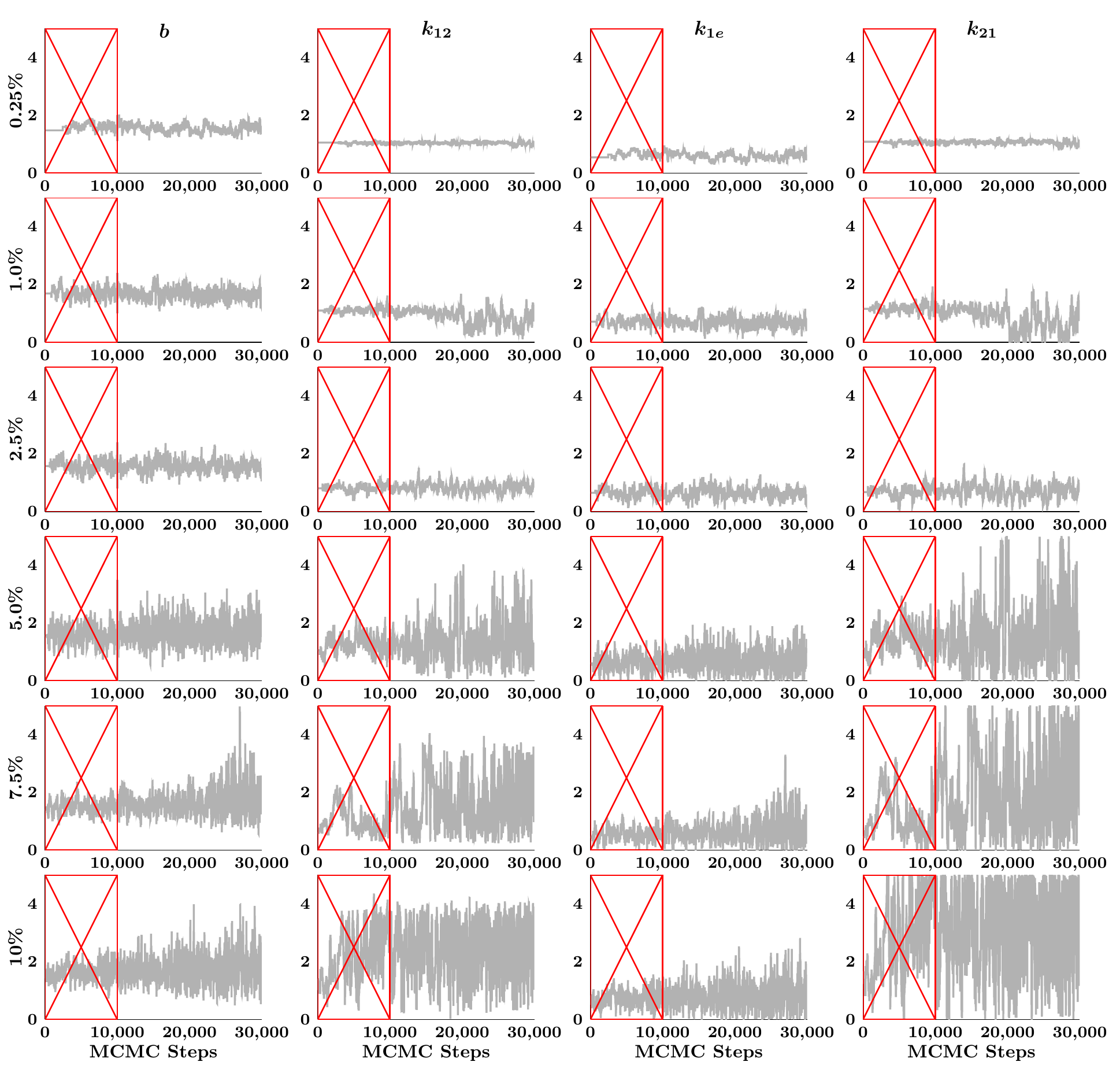}
    \caption{
        Markov chains for the two-state model parameters at increasing noise levels in the measurement noise experiment, Fig~\subpanelref{fig:two-comp}{B-C}.
        Each row corresponds to a noise level; noise increases down the figure.
        The red boxes indicate samples discarded as burn-in.
    }
    \label{supfig:two-comp-noise-MCMC}
\end{figure}

\begin{figure}[ht!]
    \centering
    \includegraphics[width=\textwidth]{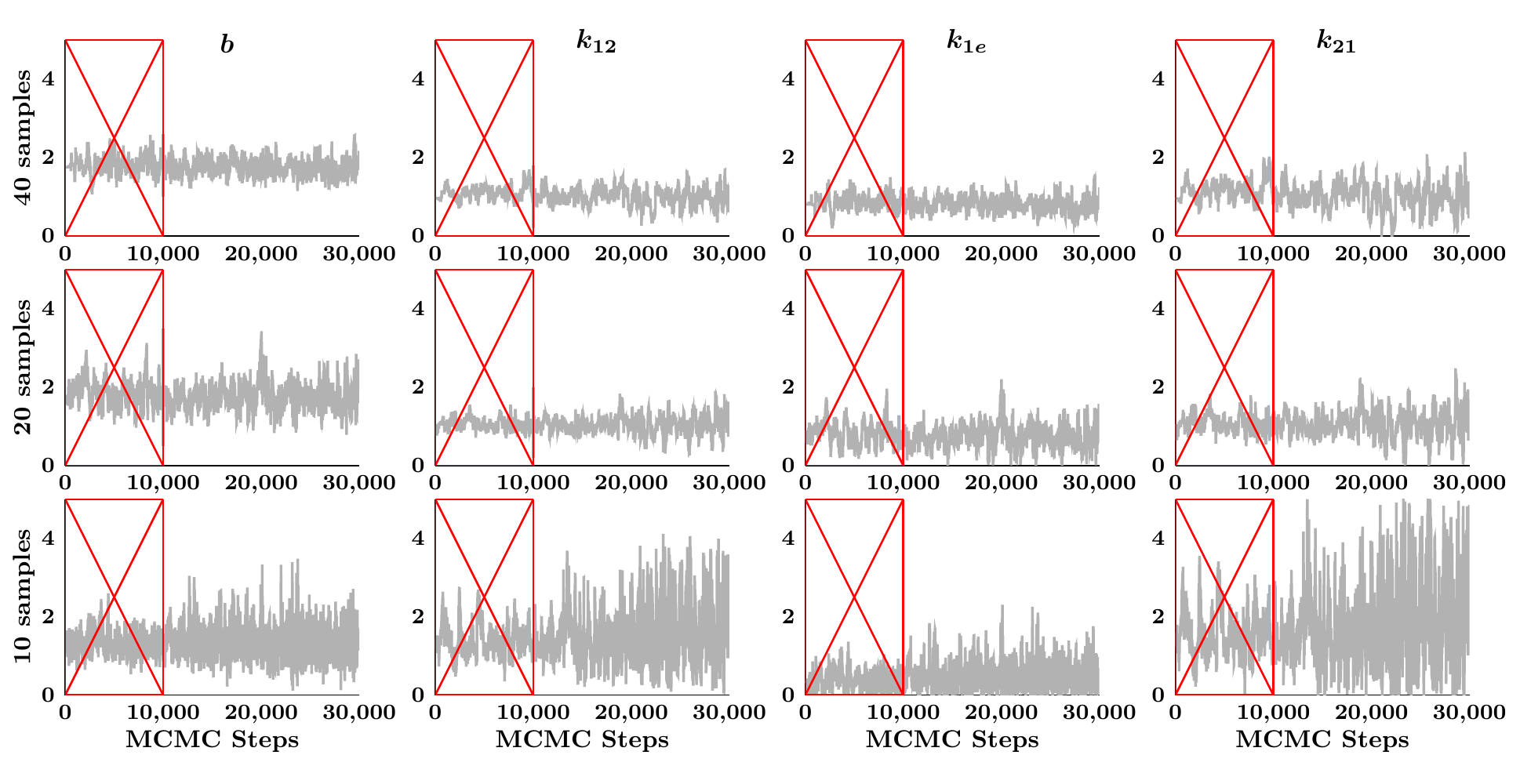}
    \caption{
        Markov chains for the two-state model parameters with decreasing samples in the measurement data sparsity experiment, Fig~\subpanelref{fig:two-comp}{D-E}.
        Each row corresponds to a sparsity level.
        The red boxes indicate samples discarded as burn-in.
    }
    \label{supfig:two-comp-samples-MCMC}
\end{figure}

\begin{figure}[ht!]
    \centering
    \includegraphics[width=\textwidth]{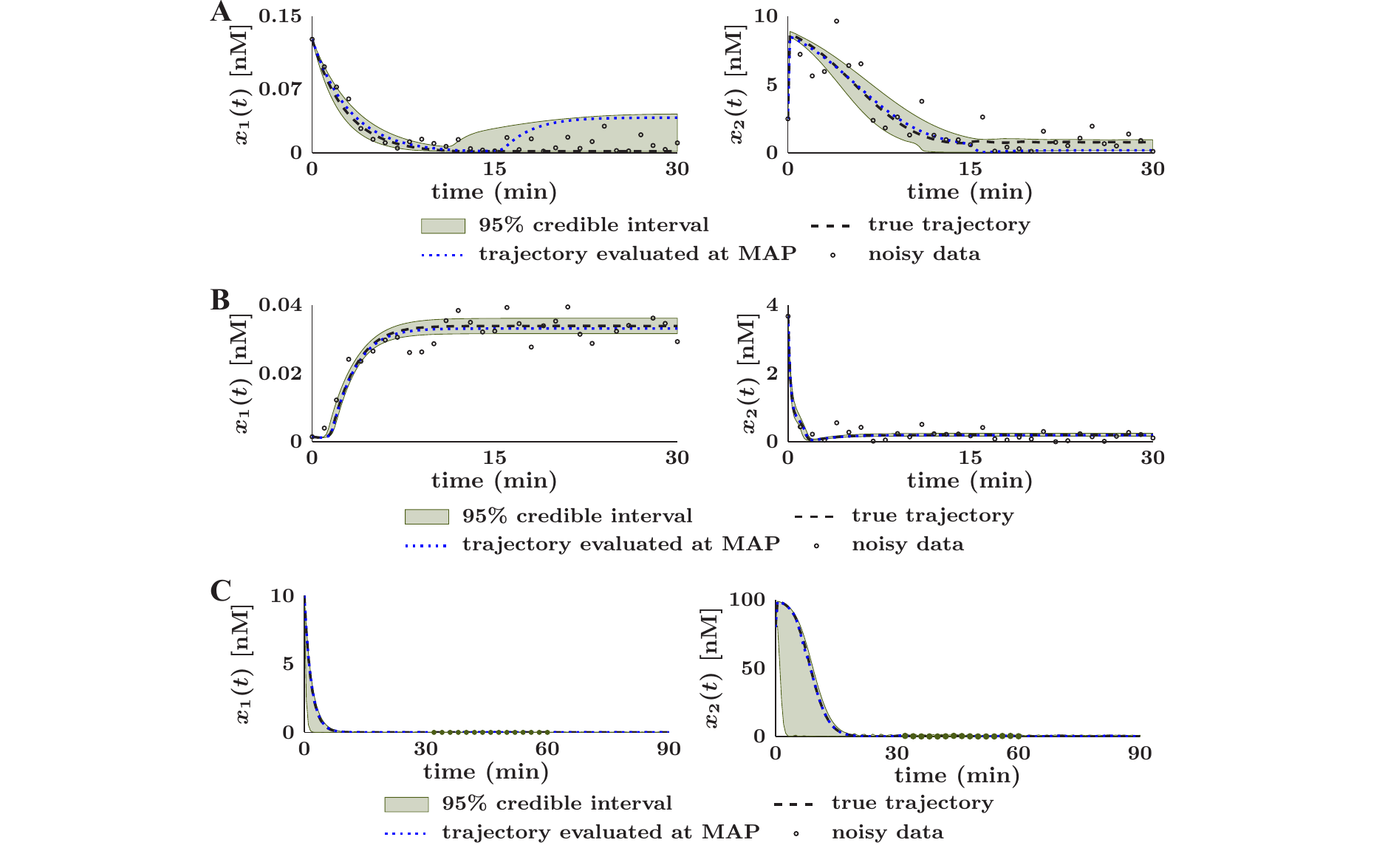}
    \caption{
        Posterior distributions of $x_1(t)$ and $x_2(t)$ for the MAPK model.
        (\textbf{A}) Trajectories for the high steady-state; corresponds to Fig~\subpanelref{fig:mapk-bistable-results}{D}.
        (\textbf{B}) Trajectories for the low steady-state; corresponds to Fig~\subpanelref{fig:mapk-bistable-results}{C}.
        (\textbf{C}) Trajectories for limit cycle oscillations; corresponds to Fig~\subpanelref{fig:mapk-osc-results}{B}.
    }
    \label{supfig:mapk-x1x2-post}
\end{figure}

\begin{figure}[ht!]
    \centering
    \includegraphics[width=\textwidth]{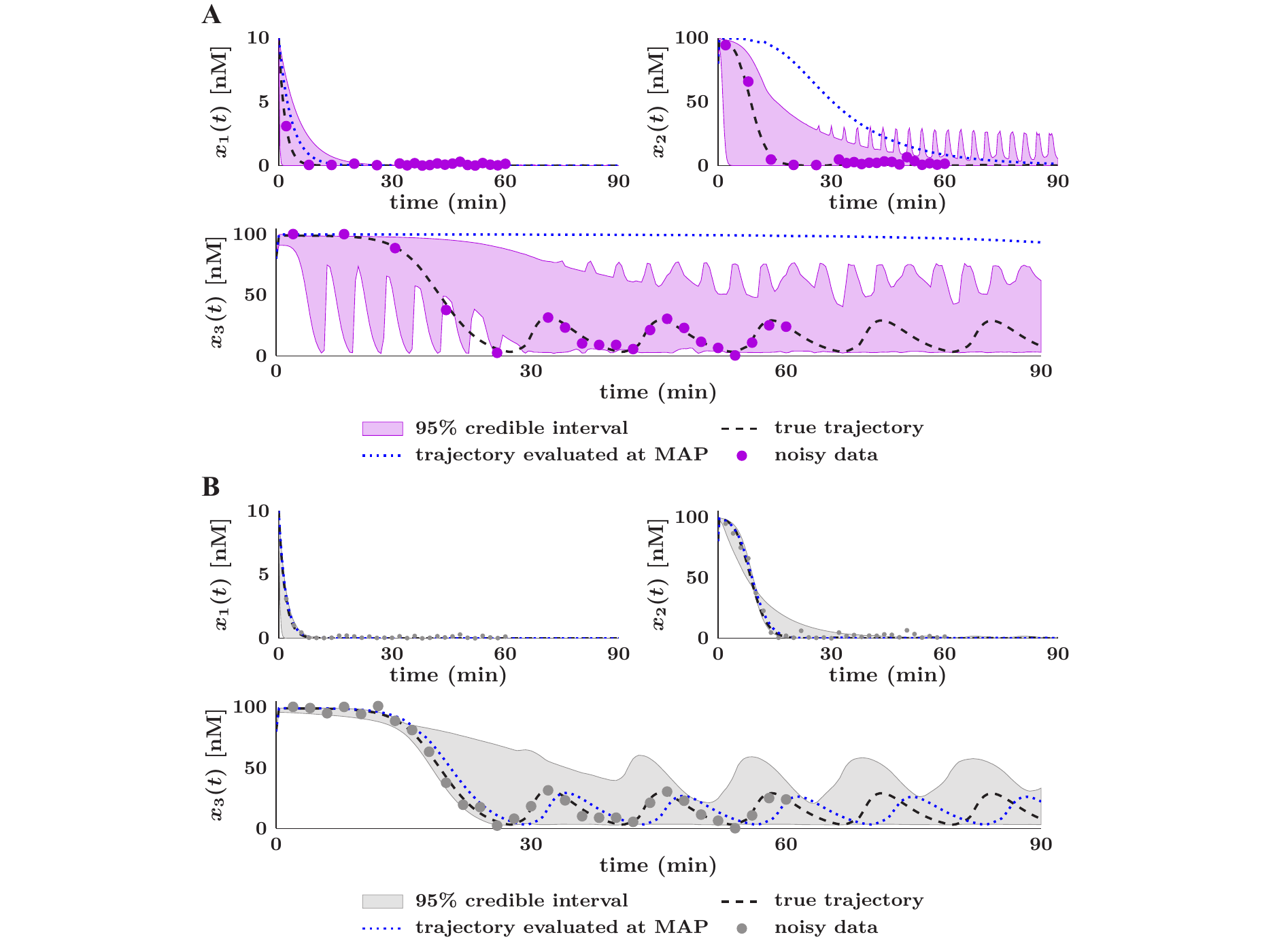}
    \caption{
        Posterior distributions of $x_1(t)$, $x_2(t)$, and $x_3(t)$ for the MAPK model with equidistant and non-equidistant sampled data.
        (\textbf{A}) Distributions for the data with equidistant sampling; corresponds to Fig~\subpanelref{fig:mapk-osc-dataCompare}{C}.
        (\textbf{B}) Distributions for the data with non-equidistant sampling; corresponds to Fig~\subpanelref{fig:mapk-osc-dataCompare}{D}.
    }
    \label{supfig:mapk-newData-dyn-post}
\end{figure}

\begin{figure}[ht!]
    \centering
    \includegraphics[width=\textwidth]{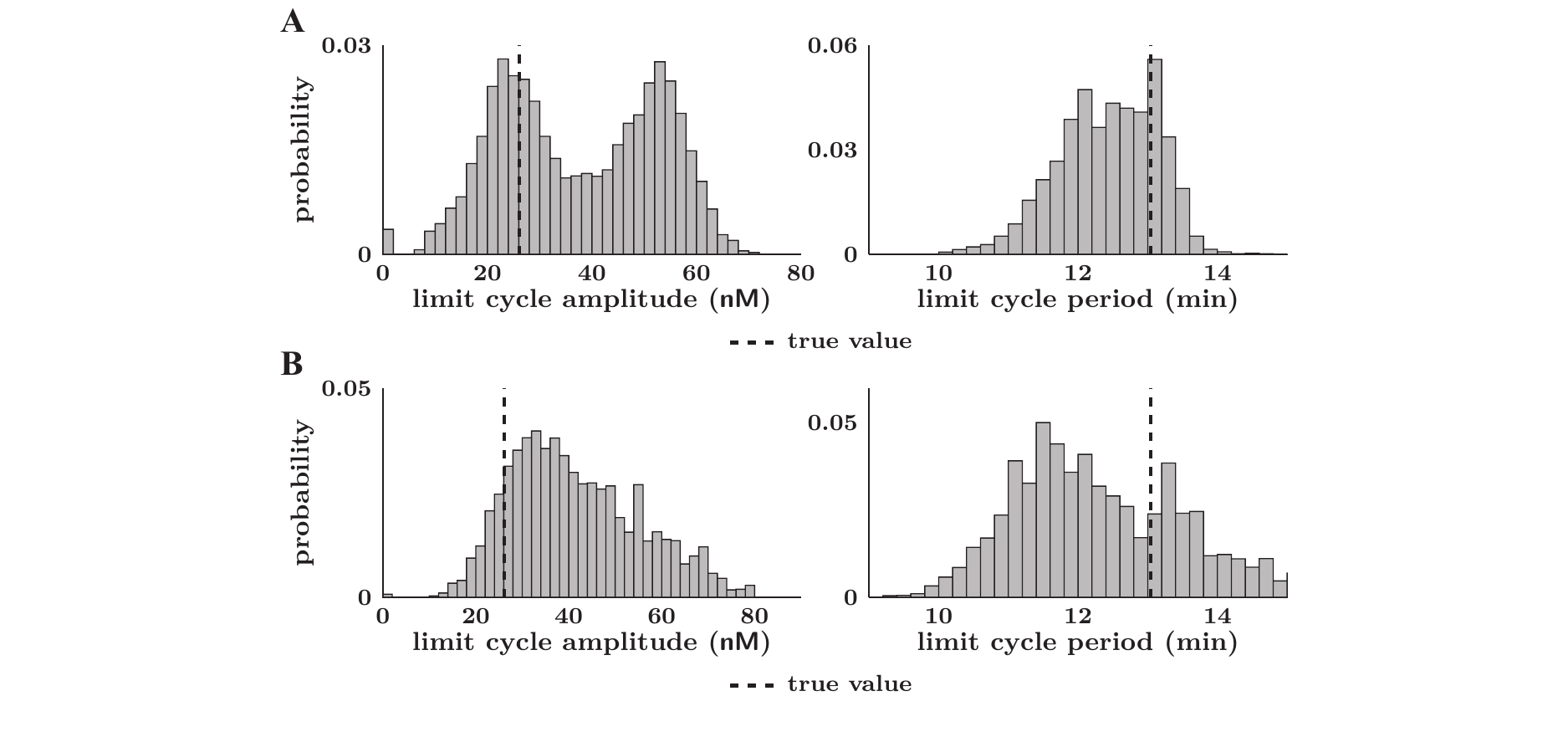}
    \caption{
        Histograms of limit cycle amplitude and period for the oscillating trajectories for the equidistant sampling (\textbf{A}) and non-equidistant sampling (\textbf{B}) data.
        (\textbf{A}) Corresponds to Fig~\subpanelref{fig:mapk-osc-dataCompare}{C} and Supplemental Fig~\subpanelref{supfig:mapk-newData-dyn-post}{A}.
        (\textbf{B}) Corresponds to Fig~\subpanelref{fig:mapk-osc-dataCompare}{D} and Supplemental Fig~\subpanelref{supfig:mapk-newData-dyn-post}{B}.
    }
    \label{supfig:mapk-newData-limit-cycle}
\end{figure}

\begin{figure}[ht!]
    \centering
    \includegraphics[width=\textwidth]{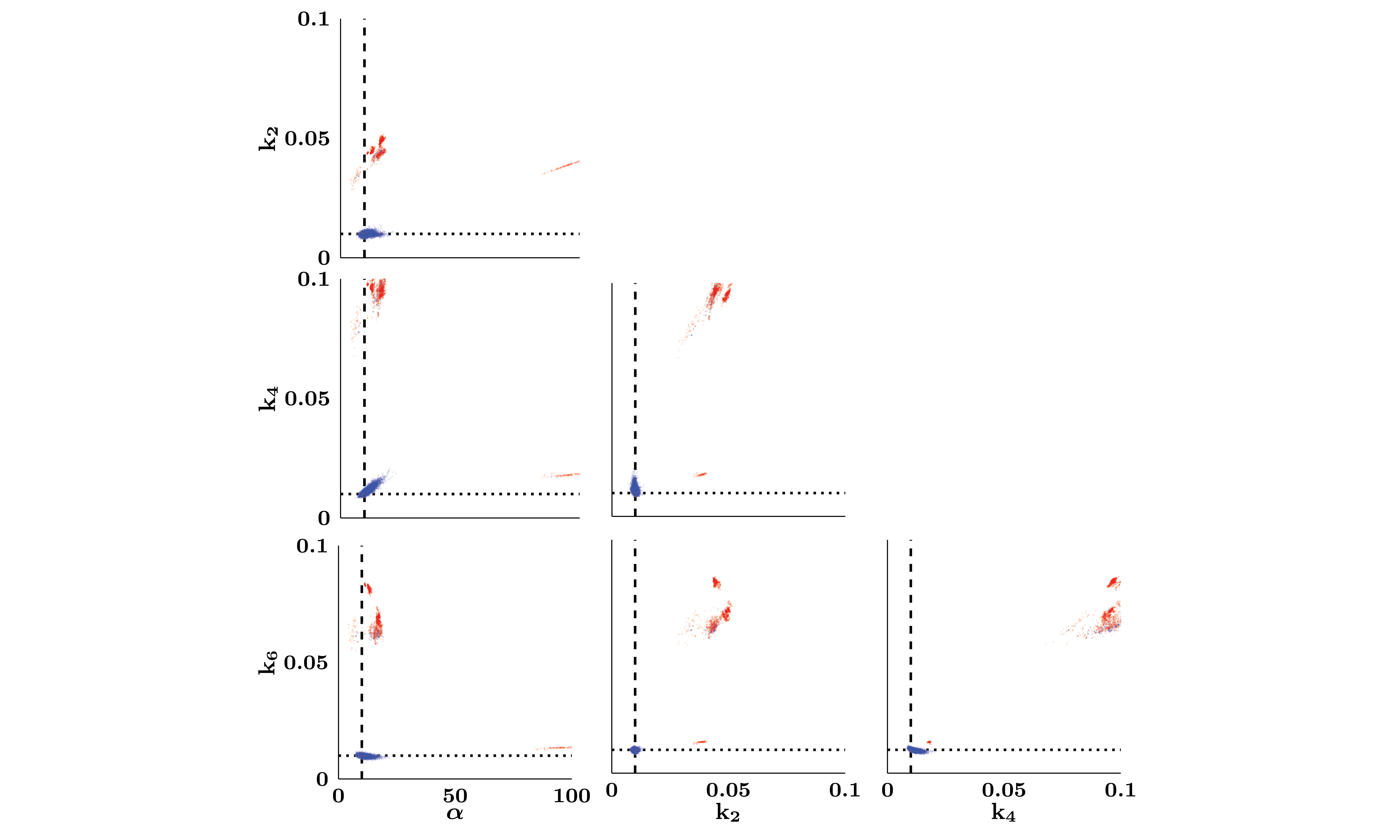}
    \caption{
        Two-dimensional scatter plots between all unique pairings of model parameters estimated from data with oscillations only.
        Each plot has $30,000$ points that are colored according to the nature of the simulated dynamics with that parameter set.
        Simulations with blue points produce limit cycle oscillations, and those with red points produce fixed points.
        Darker regions indicate a higher probability of observing the corresponding parameter values.
    }
    \label{supfig:mapk-oscOnly-scatter}
\end{figure}

\begin{figure}[ht!]
    \centering
    \includegraphics[width=\textwidth]{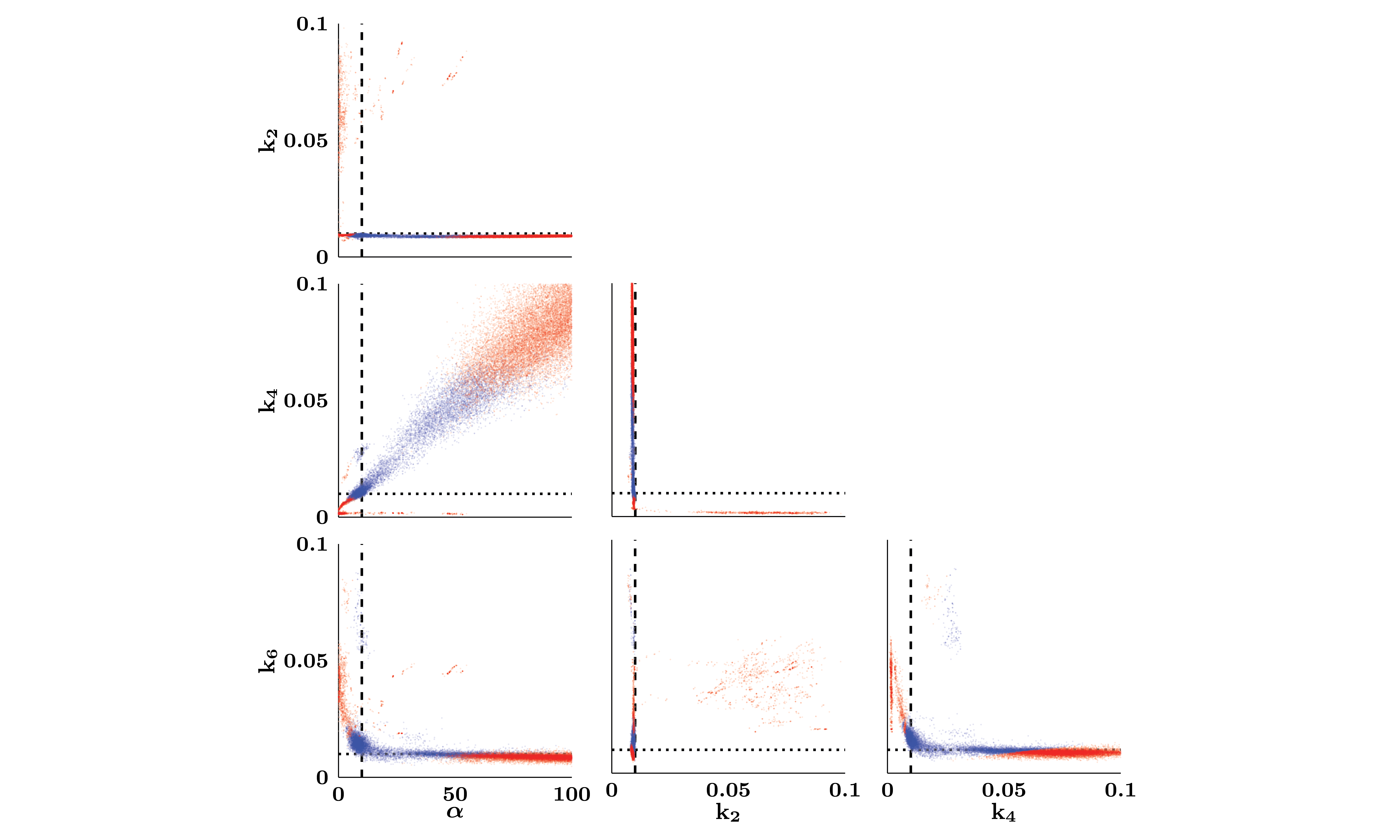}
    \caption{
        Two-dimensional scatter plots between all unique pairings of model parameters estimated from data with equidistant sampling.
        Each plot has $30,000$ points that are colored according to the nature of the simulated dynamics with that parameter set.
        Simulations with blue points produce limit cycle oscillations, and those with red points produce fixed points.
        Darker regions indicate a higher probability of observing the corresponding parameter values.
    }
    \label{supfig:mapk-equidistant-scatter}
\end{figure}

\begin{figure}[ht!]
    \centering
    \includegraphics[width=\textwidth]{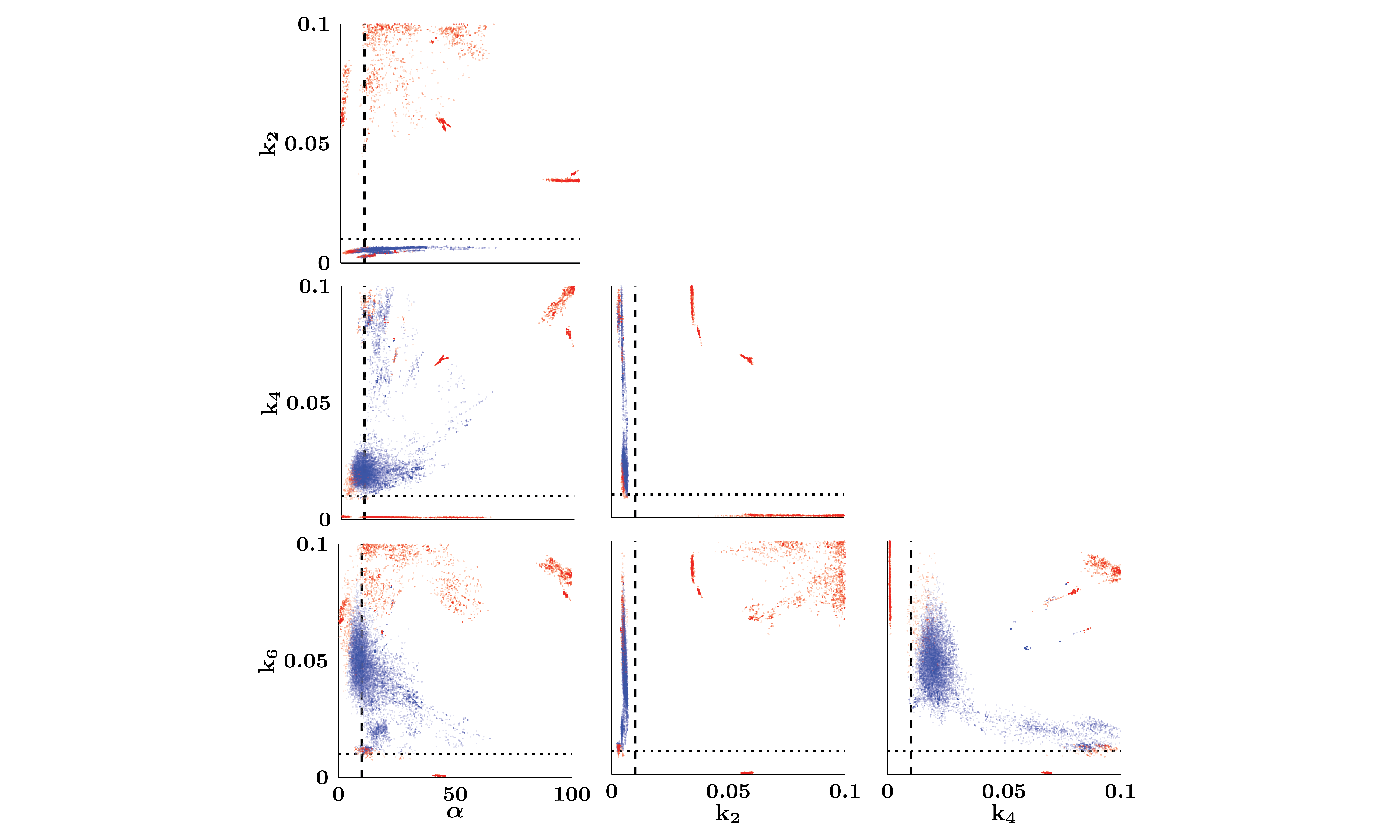}
    \caption{
        Two-dimensional scatter plots between all unique pairings of model parameters estimated from data with non-equidistant sampling.
        Each plot has $30,000$ points that are colored according to the nature of the simulated dynamics with that parameter set.
        Simulations with blue points produce limit cycle oscillations, and those with red points produce fixed points.
        Darker regions indicate a higher probability of observing the corresponding parameter values.
    }
    \label{supfig:mapk-non-equidistant-scatter}
\end{figure}

\begin{figure}[ht!]
    \centering
    \includegraphics[width=\textwidth]{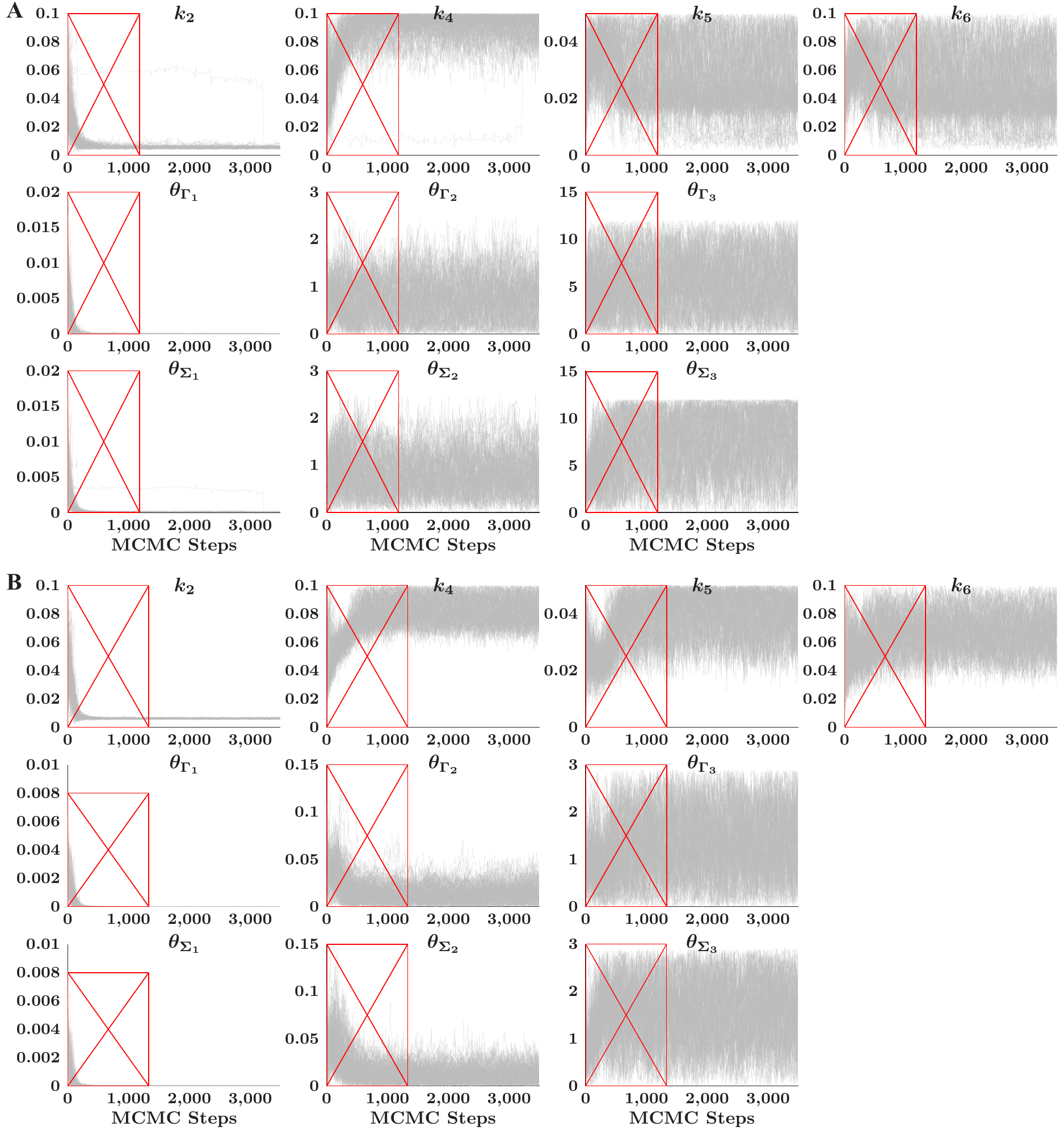}
    \caption{
        Markov chains for the MAPK model parameters and UKF-MCMC noise covariances for bistability.
        (\textbf{A}) Low steady-state, Fig~\subpanelref{fig:mapk-bistable-results}{A}.
        (\textbf{B}) High steady-state, Fig~\subpanelref{fig:mapk-bistable-results}{B}.
        The process noise covariances are $\theta_{\Gamma_i}$ and $\theta_{\Sigma_i}$ are the measurement noise covariances.
        The red boxes indicate samples discarded as burn-in.
        }
    \label{supfig:mapk-bistable-mcmc}
\end{figure}

\begin{figure}[ht!]
    \centering
    \includegraphics[width=\textwidth]{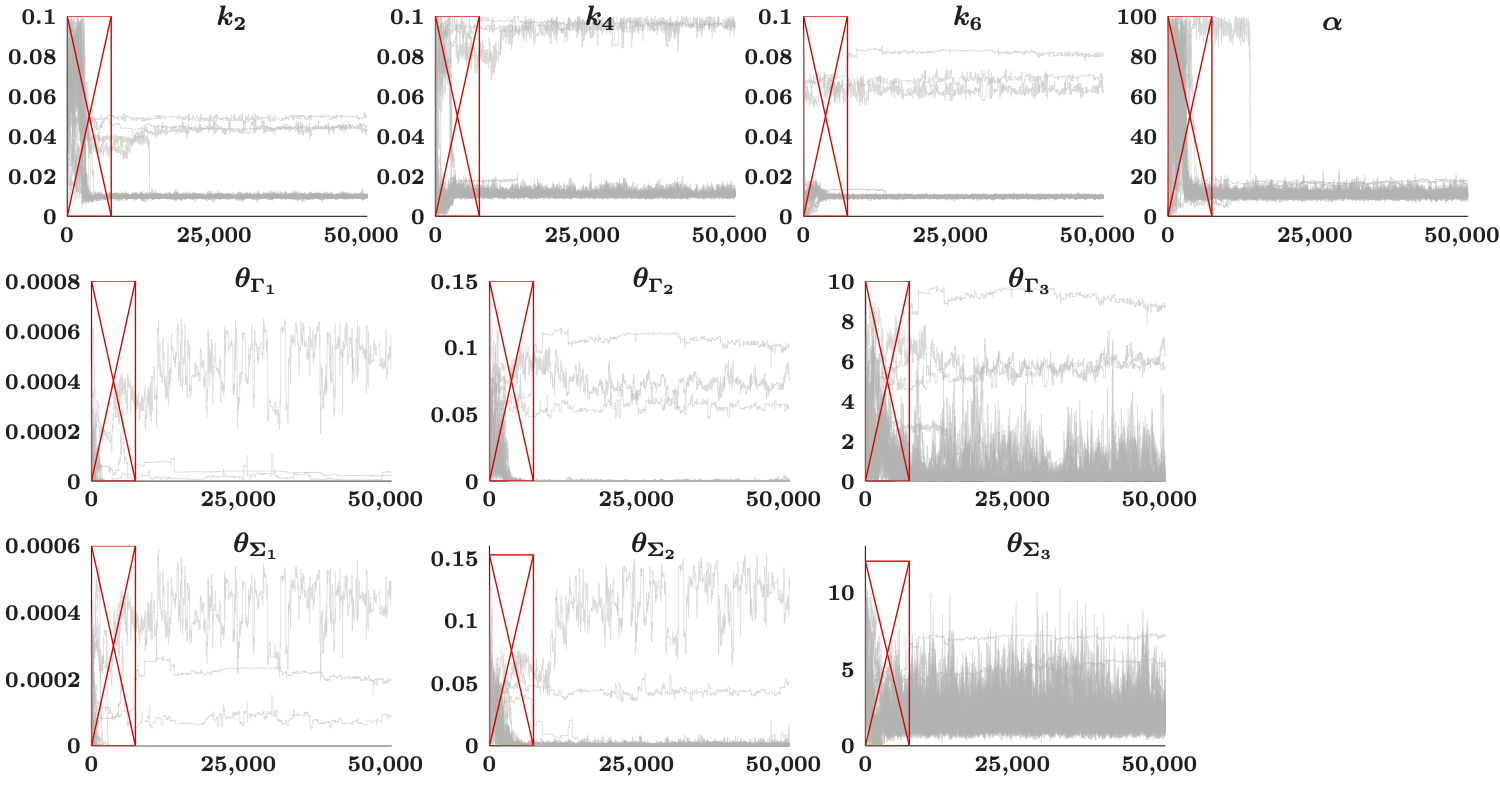}
    \caption{
        Markov chains for the MAPK model parameters and UKF-MCMC noise covariance parameters for limit cycle oscillations with data from oscillations only; corresponds to Fig~\ref{fig:mapk-osc-results} and Fig~~\subpanelref{fig:mapk-osc-dataCompare}{E}
        The process noise covariances are $\theta_{\Gamma_i}$ and $\theta_{\Sigma_i}$ are the measurement noise covariances.
        The red boxes indicate $7,477$ samples per chain discarded as burn-in.
    }
    \label{supfig:mapk-osc-only-mcmc}
\end{figure}

\begin{figure}[ht!]
    \centering
    \includegraphics[width=\textwidth]{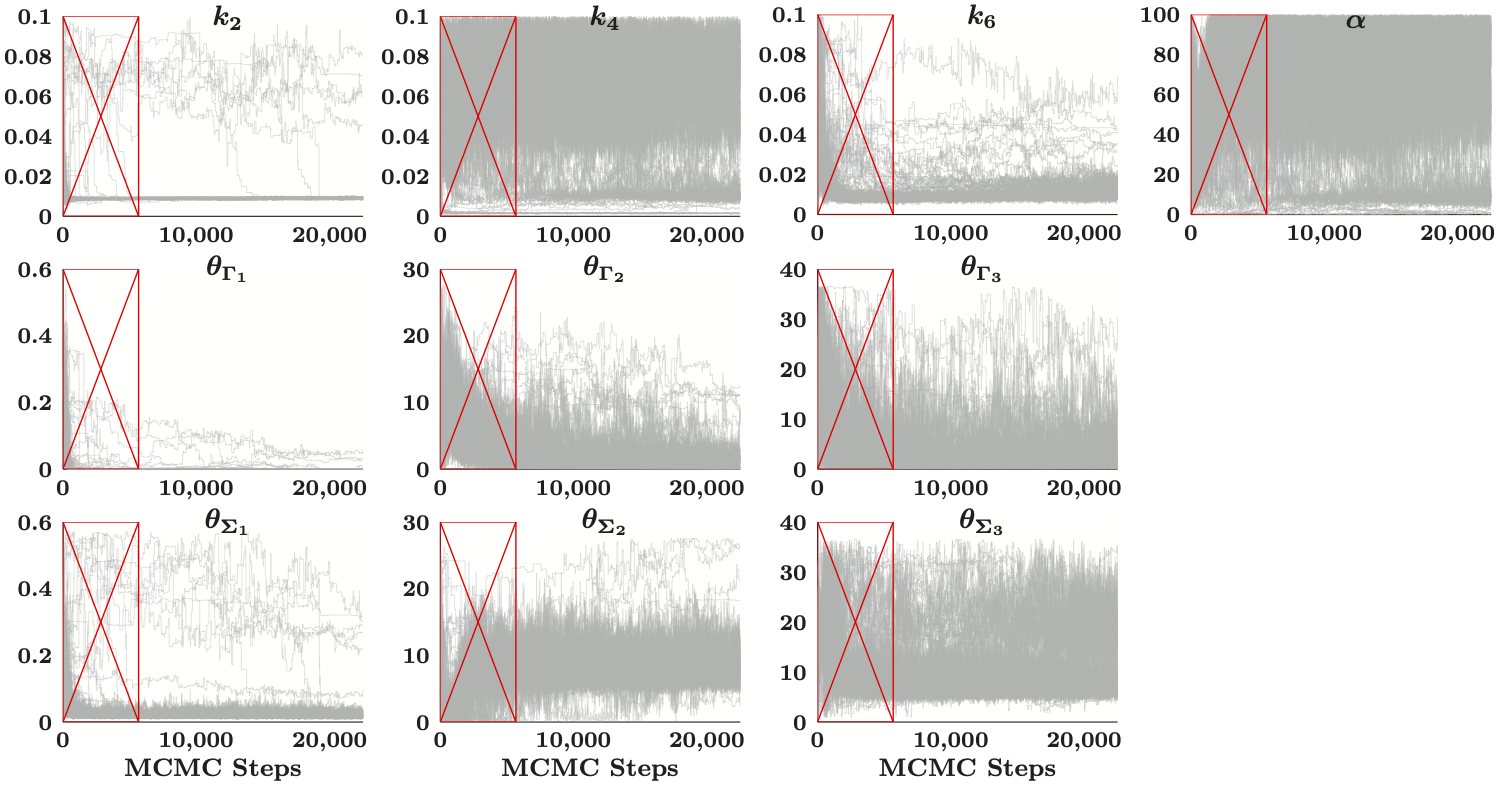}
    \caption{
        Markov chains for the MAPK model parameters and UKF-MCMC noise covariance parameters for limit cycle oscillations estimated from data with equidistant sampling; corresponds to Fig~\subpanelref{fig:mapk-osc-dataCompare}{C}
        The process noise covariances are $\theta_{\Gamma_i}$ and $\theta_{\Sigma_i}$ are the measurement noise covariances.
        The red boxes indicate $5,669$ samples per chain discarded as burn-in.
    }
    \label{supfig:mapk-osc-equidistant-mcmc}
\end{figure}

\begin{figure}[ht!]
    \centering
    \includegraphics[width=\textwidth]{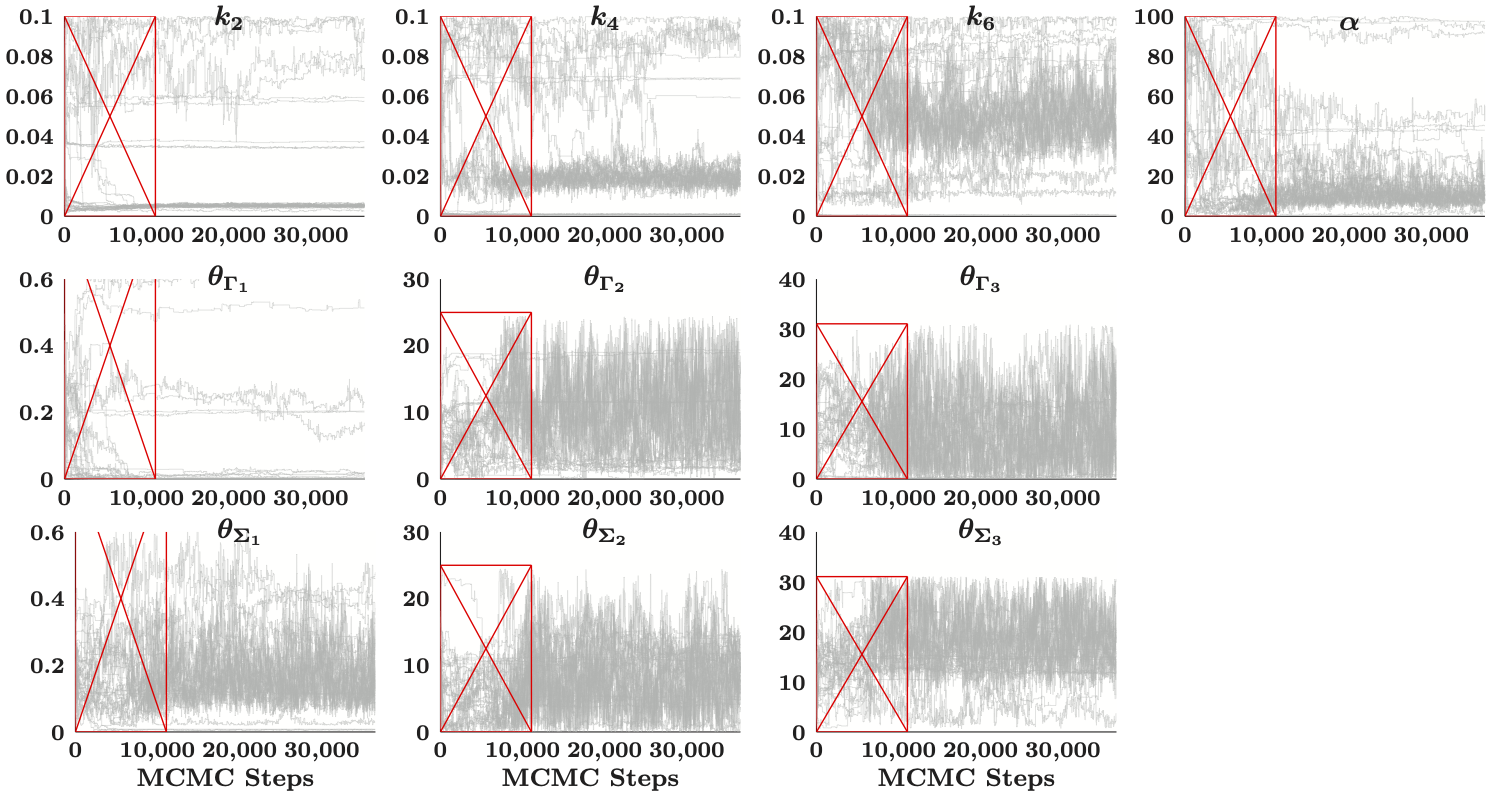}
    \caption{
        Markov chains for the MAPK model parameters and UKF-MCMC noise covariances for limit cycle oscillations estimated from data with non-equidistant sampling; corresponds to Fig~\subpanelref{fig:mapk-osc-dataCompare}{D}
        The process noise covariances are $\theta_{\Gamma_i}$ and $\theta_{\Sigma_i}$ are the measurement noise covariances.
        The red boxes indicate $11,065$ samples per chain discarded as burn-in.
    }
    \label{supfig:mapk-osc-nonequidistant-mcmc}
\end{figure}

\begin{figure}[ht!]
    \centering
    \includegraphics[width=\textwidth]{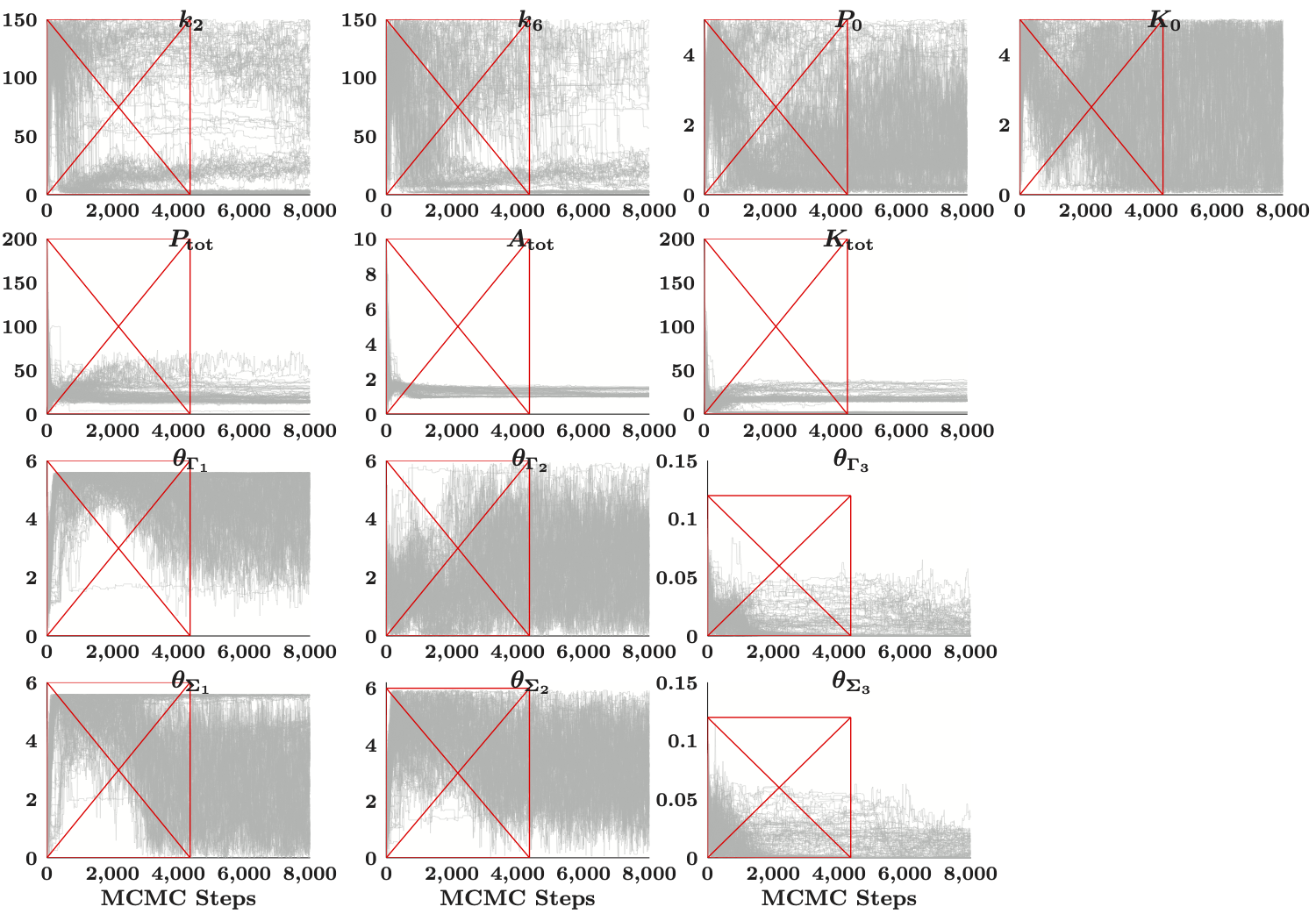}
    \caption{Markov chains for the synaptic plasticity model parameters, Fig~\ref{fig:kinpho-results}.
    The red boxes indicate samples discarded as burn-in.}
    \label{supfig:kinpho-mcmc}
\end{figure}

This section provides supplemental figures for the results presented in Section~3.
Figures~\ref{supfig:two-comp-x2dyn},~\ref{supfig:two-comp-noise-MCMC}, and~\ref{supfig:two-comp-samples-MCMC} correspond to Section~3.1.
Figures~\ref{supfig:mapk-x1x2-post},~\ref{supfig:mapk-bistable-mcmc},~\ref{supfig:mapk-newData-dyn-post},~\ref{supfig:mapk-newData-limit-cycle},~\ref{supfig:mapk-equidistant-scatter},~\ref{supfig:mapk-non-equidistant-scatter},~\ref{supfig:mapk-oscOnly-scatter},~\ref{supfig:mapk-osc-equidistant-mcmc},~\ref{supfig:mapk-osc-nonequidistant-mcmc}, and~\ref{supfig:mapk-osc-only-mcmc} correspond to Section~3.2.
Lastly, Fig~\ref{supfig:kinpho-mcmc} corresponds to Section~3.3.


\newpage


\end{appendix}
\end{document}